\documentstyle[a4p,12pt,twoside,epsfig,amsfonts,amsthm,amsbsy,amssymb,rotating]{article}

\newcommand{\mathlab}[2]{\begin{equation}#1\label{#2}\end{equation}}
\newcommand{\rb}[1]{\raisebox{1.5ex}[-1.5ex]{#1}}
\newcommand{\derr}[2]{\raisebox{-0.7ex}{$\stackrel{\textstyle +#1}{-#2}$}} 
\newcommand{\dderr}[2]{\raisebox{-0.4ex}{$\stackrel{\textstyle \mbox{\scriptsize +#1}}{\mbox{\scriptsize $-$#2}}$}} 
\newcommand{\subsubsubsection}[1]{\vspace*{0.3cm}{\bf #1}\\[0.2cm]}

\newcommand{\clratleft}{0cm}
\newcommand{\clratuup}{-0.5cm}
\newcommand{\clratup}{-0.5cm}
\newcommand{\clratmid}{-0.8cm}
\newcommand{\clratdown}{-0.6cm}
\newcommand{\clratwid}{7.5cm}

\newcommand{\clratlefts}{0cm}
\newcommand{\clratuups}{0cm}
\newcommand{\clratups}{0.2cm}
\newcommand{\clratmids}{-0.2cm}
\newcommand{\clratdowns}{-0.6cm}
\newcommand{\clratwids}{8.2cm}
\newcommand{\clratints}{-0.5cm}

\newcommand{\coneleft}{0cm}
\newcommand{\coneuup}{0cm}
\newcommand{\coneup}{0.3cm}
\newcommand{\conemid}{0cm}
\newcommand{\conedown}{-0.6cm}
\newcommand{\conewid}{7.cm}
\newcommand{\coneint}{0.5cm}

\newcommand{\momleft}{1cm}
\newcommand{\momuup}{-1cm}
\newcommand{\momup}{0cm}
\newcommand{\mommid}{-1.6cm}
\newcommand{\momdown}{-0.5cm}
\newcommand{\momwid}{22.cm}

\newcommand{\fituup}{-0.5cm}

\newcommand{\fitleft}{5.5cm}
\newcommand{\fitwid}{14cm}
\newcommand{\fitmid}{-1.5cm}
\newcommand{\fitcent}{-6cm}
\newcommand{\fitdown}{-1.5cm}

\newcommand{\matchtext}[1]{

\caption[Comparison of fitted $\alpha_s$ values obtained using different
theoretical predictions for #1]
{
Fitted $\alpha_s$ values 
obtained using the different theoretical predictions described in the
text for variable #1 with
the respective experimental (inner) and total (outer) errors. The
$\chi^2/d.o.f.$ values of each fit are given in the rightmost column.}
}

\newcommand{\matchtextt}[1]{

\caption[Comparison of fitted $\alpha_s$ values obtained using different
theoretical predictions for #1]
{Fitted $\alpha_s$ values obtained using the different theoretical predictions described in the text for variable #1, presented as in Fig.~\ref{pr299_10a_fig}. At $\sqrt{s}=$35 GeV, no stable fits could
be obtained using pure ${\cal O}(\alpha_s^2)$ calculations with a free QCD scale factor.}}

\newcommand{\asfitatext}[1]{The central parts of the plots display the 
results of fits of various theoretical predictions (smooth curves) 
in comparison with hadron level measurements (data points) of the four
observables $D_2^D$, $D_2^C$, $N^D$ and $N^C$ at $\sqrt{s}=#1$ GeV.
The error bars on the data represent the total errors, the small
horizontal lines the pure statistical errors which were used in the
fits. Horizontal arrows indicate the respective fit ranges. The 
normalized difference between predictions and the data, $\delta$, is
shown above each plot. The small sections below each plot 
indicate the correction factors $f$ for experimental and hadronization
effects. The hadronization corrections are shown
as lines and shaded bands as described in the text. The experimental
corrections are represented as open triangles and filled circles, respectively,
for the 1982 and 1986 runs.
}

\newcommand{\asfitatextt}[1]{The fit results at $\sqrt{s}=$91 GeV 
are shown in the same
form as in Fig.~\ref{asfitlab_35}.
}

\newcommand{\asfitatexttt}[1]{The fit results at $\sqrt{s}=$189 GeV 
are shown in the same
form as in Fig.~\ref{asfitlab_35}. At this energy, the 
statistical fluctuations of the data points are too large
to draw any meaningful conclusions from the normalized differences between
the predictions and the measurement for $D_2^D$ and $D_2^C$. They are
therefore not shown in the plot.
}

\newcommand{\matchfig}[6]{
\begin{figure}[tbhp]
\vskip #5cm

\centerline{\hspace*{1cm}{\bf\LARGE #3}\hspace*{6.5cm}{\bf\LARGE #4}}
\vspace*{-0.4cm}
\centerline{\hspace*{8.2cm}$\chi^2/d.o.f.$\hspace*{5.8cm}$\chi^2/d.o.f.$}

\centerline{
\hspace*{5.5cm}
\epsfig{file=#1.eps,width=16.cm}\hspace*{-8.5cm}
\epsfig{file=#2.eps,width=16.cm}
}
\vskip -0.5cm

#6

\label{#1_fig}

\end{figure}}

\newcommand{\asfita}[3]{
\begin{figure}[tbhp]

\vskip -1cm

\centerline{{\bf\Huge #1 GeV}}

\vskip 0.5cm

\centerline{
\hspace*{6cm}
\epsfig{file=pr299_#3a.eps,width=14.5cm}
\hspace*{-6cm}
\epsfig{file=pr299_#3b.eps,width=14.5cm}
}

\centerline{
\hspace*{6cm}
\epsfig{file=pr299_#3c.eps,width=14.5cm}
\hspace*{-6cm}
\epsfig{file=pr299_#3d.eps,width=14.5cm}
}

\caption[Fit results for $\alpha_s$ at $\protect\sqrt{s}$=#1 GeV]
{#2
\label{asfitlab_#1}
}

\end{figure}

}

\newcommand{\xmudepfigd}{
\begin{figure}[!t]
\vskip 0cm

\centerline{{\Large $D_2^D$} \hspace*{7cm} {\Large $N^D$}}

\vskip -0.3cm

\centerline{
\hspace*{-4.5cm}
\epsfig{file=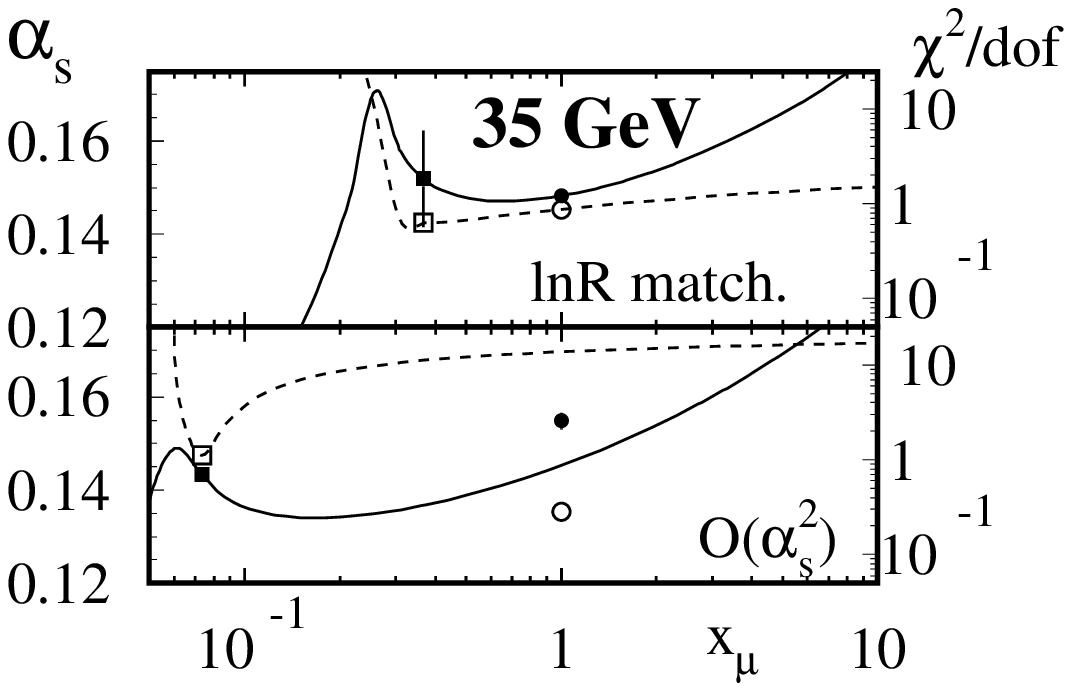,width=14.5cm}}
\vskip -5.8cm
\centerline{
\hspace*{13cm}
\epsfig{file=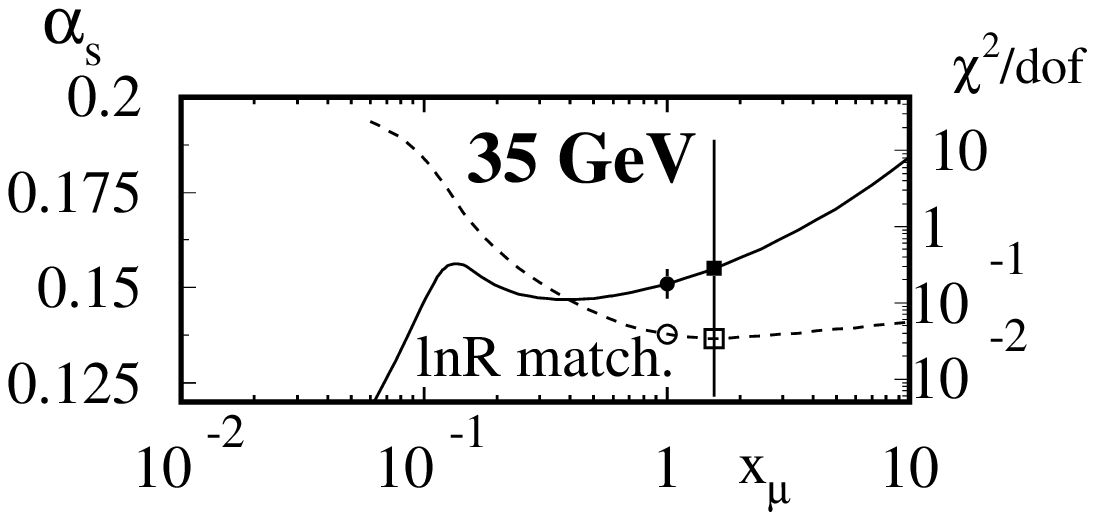,width=14.5cm}
}

\vspace*{0.8cm}

\centerline{
\hspace*{-4.5cm}
\epsfig{file=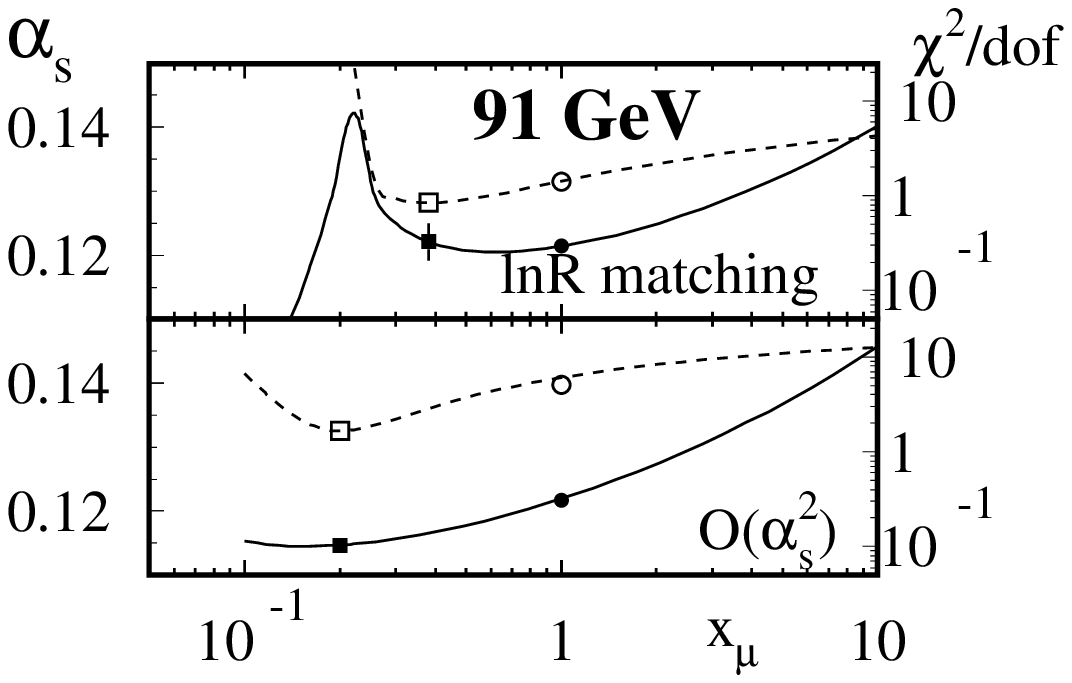,width=14.5cm}}
\vskip -7.3cm
\centerline{
\hspace*{13cm}
\epsfig{file=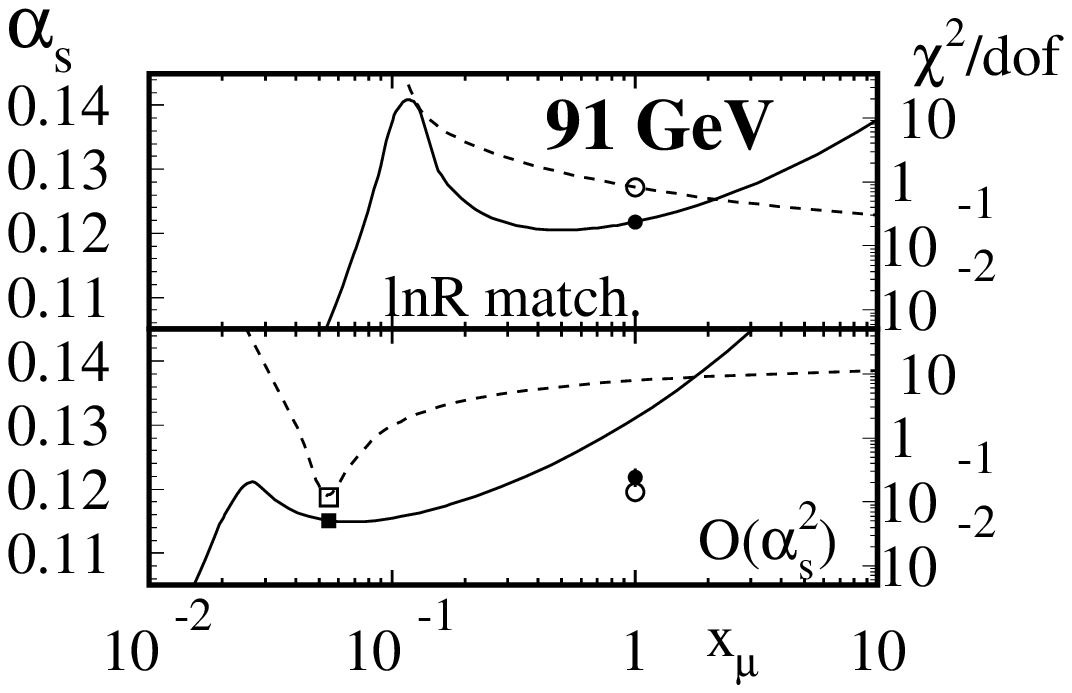,width=14.5cm}
}

\vskip 0.7cm

\centerline{
\hspace*{-4.5cm}
\epsfig{file=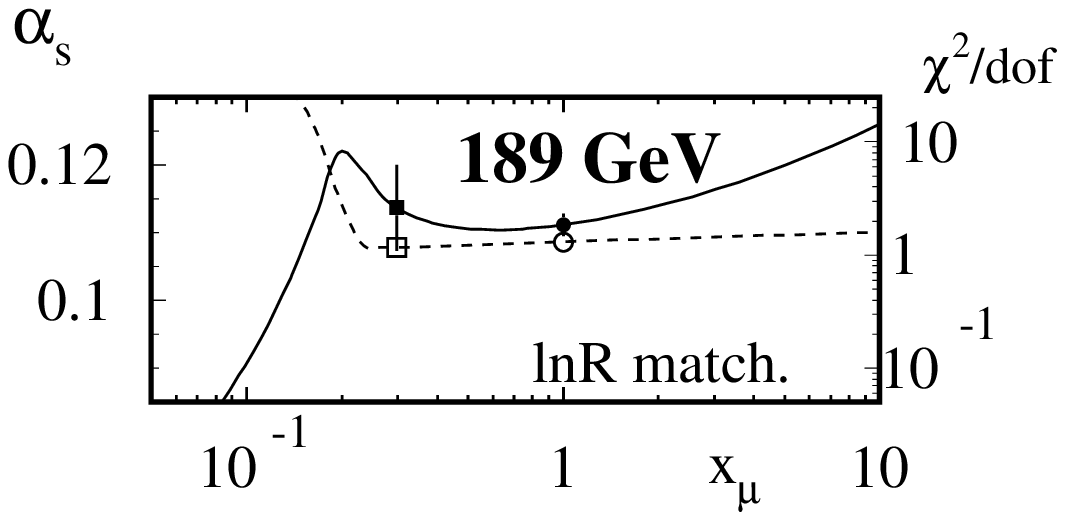,width=14.5cm}}
\vskip -5.7cm
\centerline{
\hspace*{13cm}
\epsfig{file=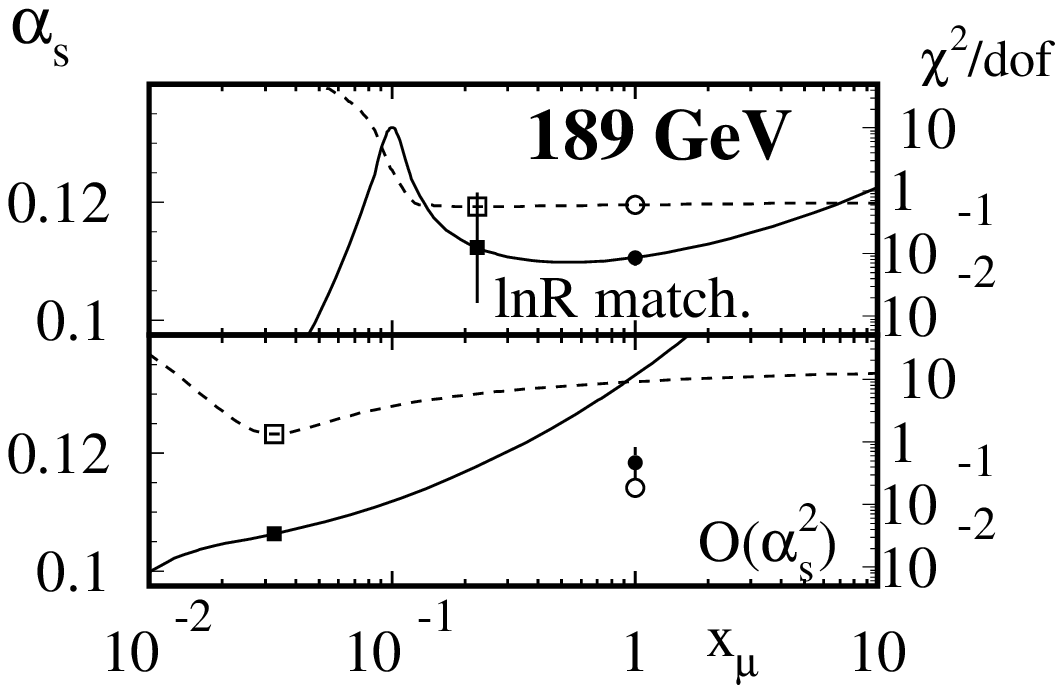,width=14.5cm}
}

\vskip 0cm

\vskip -0.8cm

\caption[QCD scale dependence of $\alpha_s$ fits]{
Sensitivity of fit results to the scale $x_{\mu}$.
The solid and dashed lines are, respectively, the fitted $\alpha_s$ and
the corresponding $\chi^2/d.o.f.$ 
as a function
of the QCD
scale $x_{\mu}$ for observables $D_2^D$ (left) and $N^D$ (right). 
Filled and empty squares denote $\alpha_s$ and
$\chi^2/d.o.f.$ for a fitted scale; circles have the same meaning for
a fixed scale of 1. For $D_2^D$ at $\sqrt{s}=189$ and $N^D$ at 
35 GeV, results are presented only for
the $\ln{R}$-matching scheme. In all other cases, fits of the
pure ${\cal O}(\alpha_s^2)$ predictions are also presented in the lower partitions
of the plots.\label{xmudep_lab}}
\end{figure}
}

\begin{document}

\begin{titlepage}

\begin{center}
{\large EUROPEAN ORGANISATION FOR NUCLEAR RESEARCH}
\end{center}
\bigskip
\begin{flushright}
CERN-EP/99-175 \\ 13th December 1999
\end{flushright}
\bigskip
\begin{center}
  {\huge\bf\boldmath
    \center{QCD Analyses and Determinations}
    \vspace*{-0.3cm}
    \center{of $\alpha_s$ in ${\rm e}^+{\rm e}^-$ Annihilation}
    \center{at Energies between 35 and 189 GeV}
  }
\end{center}
\bigskip\bigskip\bigskip
\begin{center}
{\LARGE
   The  JADE (*) and the OPAL (**) Collaboration
}
\vspace{0.25cm}
\end{center}
\bigskip\bigskip

  \centerline{\bf Abstract:}
  We employ data taken by the JADE and OPAL experiments for an integrated
  QCD study in hadronic ${\rm e}^+{\rm e}^-$ 
  annihilations at c.m.s. energies ranging from 35 GeV through 189 GeV.
  The study is based on jet-multiplicity related observables.
  The observables are obtained 
  to high jet resolution scales
  with the JADE, Durham, Cambridge and cone jet finders, and compared with
  the predictions of various QCD and Monte Carlo models. 
  The strong coupling strength, $\alpha_s$, 
  is determined at each energy by fits of 
  ${\cal O}(\alpha_s^2)$ calculations, as well as matched
  ${\cal O}(\alpha_s^2)$ and NLLA predictions, to the data. 
  Matching schemes are compared, and the
  dependence of the results on the choice of the renormalization scale is investigated.
  The combination of the results using matched predictions gives 
  $$\alpha_s(M_{Z^0})=0.1187\derr{0.0034}{0.0019}.$$
  The strong coupling is also obtained, at lower precision,  
  from ${\cal O}(\alpha_s^2)$
  fits of the c.m.s. energy
  evolution of some of the observables.
  A qualitative comparison is made between the data and 
  a recent MLLA prediction
  for mean jet multiplicities.

\bigskip\bigskip\bigskip
\begin{center}

\bigskip

\vspace*{0.5 cm}
  {\large To be submitted to European Physical Journal C}
\end{center}
\end{titlepage} 

\begin{center}{\Large     (*) The JADE Collaboration
}\end{center}
P. Pfeifenschneider$^{ 14}$, O. Biebel$^{ 14,  i}$, P.A. Movilla Fern\'andez$^{ 14,  i}$
and the members of the former JADE Collaboration who are fully listed in Ref. \cite{naroska}
\bigskip\bigskip\bigskip\bigskip

\begin{center}{\Large     (**) The OPAL Collaboration
}\end{center}\bigskip
\begin{center}{
G.\thinspace Abbiendi$^{  2}$,
K.\thinspace Ackerstaff$^{  8}$,
P.F.\thinspace Akesson$^{  3}$,
G.\thinspace Alexander$^{ 22}$,
J.\thinspace Allison$^{ 16}$,
K.J.\thinspace Anderson$^{  9}$,
S.\thinspace Arcelli$^{ 17}$,
S.\thinspace Asai$^{ 23}$,
S.F.\thinspace Ashby$^{  1}$,
D.\thinspace Axen$^{ 27}$
% ...
}\end{center}\bigskip
\bigskip
$^{  1}$School of Physics and Astronomy, University of Birmingham,
Birmingham B15 2TT, UK
\newline
$^{  2}$Dipartimento di Fisica dell' Universit\`a di Bologna and INFN,
I-40126 Bologna, Italy
\newline
$^{  3}$Physikalisches Institut, Universit\"at Bonn,
D-53115 Bonn, Germany
\newline
% ...
$^{  8}$CERN, European Organisation for Particle Physics,
CH-1211 Geneva 23, Switzerland
\newline
$^{  9}$Enrico Fermi Institute and Department of Physics,
University of Chicago, Chicago IL 60637, USA
\newline
% ...
$^{ 16}$Department of Physics, Schuster Laboratory, The University,
Manchester M13 9PL, UK
\newline
$^{ 17}$Department of Physics, University of Maryland,
College Park, MD 20742, USA
\newline
% ...
$^{ 22}$Department of Physics and Astronomy, Tel Aviv University,
Tel Aviv 69978, Israel
\newline
$^{ 23}$International Centre for Elementary Particle Physics and
Department of Physics, University of Tokyo, Tokyo 113-0033, and
Kobe University, Kobe 657-8501, Japan
\newline
% ...
$^{ 27}$University of British Columbia, Department of Physics,
Vancouver BC V6T 1Z1, Canada
\newline
% ...


\section{Introduction}
The renormalized
strong coupling strength $\alpha_s$ of quantum chromodynamics (QCD)
is predicted to depend upon the momentum transfer of the interaction
under study. It is desirable to perform tests of QCD at
different values of this momentum scale. 
These tests are best performed using uniform experimental
conditions since systematic uncertainties may thereby be
reduced.
Until recently, 
rather limited sets of definite c.m.s. energies were available for
QCD analyses in ${\rm e}^+{\rm e}^-$ collisions under consistent conditions.
Since the start of the
``LEP 2'' program in 1995, the c.m.s.
energy of the LEP collider at CERN has been increased in several steps
from its original values close to 91.2 GeV,
allowing QCD analyses over a wide range
of high c.m.s. energies. However, the inclusion of measurements at lower 
c.m.s. energies 
is important as well, 
because QCD becomes more strongly dependent on the energy scale towards
lower energies, and tests of the theory will be most significant here.\par
This paper presents QCD tests using 
${\rm e}^+{\rm e}^-$ annihilations into 
hadrons (so-called multihadronic events) 
from $\sqrt{s}=35$ GeV to 189 GeV.
Data recorded at the OPAL experiment at LEP are analyzed in combination
with data from the JADE experiment at the PETRA collider at DESY, where 
${\rm e}^+{\rm e}^-$
collisions were studied from 1978 to 1986 at lower c.m.s. energies. 
The OPAL and JADE detectors
are similar in construction, and we have tried to keep
the experimental procedure in
both 
analyses as similar as possible.\par
The observables used are exclusively based on the multiplicities of hadronic
jets, defined 
using standard techniques. In the first part of the present work we present
measurements of a large variety of such observables and compare them with
several Monte Carlo predictions. The measurements are then employed
to determine the strong coupling strength
$\alpha_s$ and to test the QCD prediction
for the momentum transfer dependence, i.e. the ``running,'' of $\alpha_s$.
We compare the results from different
types of matched ${\cal O}(\alpha_s^2)$ and NLLA 
predictions as well as pure ${\cal O}(\alpha_s^2)$ predictions
and study the dependence of the fit results on the renormalization scale.
The individual results are combined into a final value for $\alpha_s(M_{Z^0})$.
Finally, a recent MLLA calculation for the mean jet multiplicity is compared
with our measurements without extracting a value for $\alpha_s$.
\begin{table}[!t]

\begin{center}
\begin{tabular}{||c|c|c|c|c||} \hline\hline
 &\hfill $\sqrt{s}$ \hfill ($\langle\sqrt{s}\rangle$)\hfill & period of & integrated & number of \\
 &    [GeV]  &  data taking & luminosity [pb$^{-1}$]  & selected events\\ \hline\hline
{\bf J}  & \hfill (34.6) & 1982 & 37.5 & 8721 \\ \cline{3-5}
{\bf A} & \rb{34.5--35.5}\hfill (35.0) & 1986 & 92.3 & 20793 \\ \cline{2-5}
{\bf D} &  &  & & \\ 
{\bf E} & \rb{43.4--44.3}\hfill \rb{(43.7)} & \rb{1984/85} & \rb{30.3} & \rb{4110}\\ \hline \hline
 & 91.2 & 1994 & 34 & 1508031 \\ \cline{2-5}
 & 130 &  & 2.6 & 144 \\  \cline{2-2}\cline{4-5}
{\bf O} & 136 & \rb{1995} & 2.5 & 140 \\  \cline{2-5}
{\bf P} & 130 &  & 2.6 & 179 \\  \cline{2-2}\cline{4-5}
{\bf A} & 136 & \rb{1997} & 3.3 & 167 \\  \cline{2-5}
{\bf L} & 161 & 1996 & 10.0 & 281 \\  \cline{2-5}
 & 172 & 1996 & 10.42 & 224 \\  \cline{2-5}
 & 183 & 1997 & 55.22 & 1082 \\  \cline{2-5}
 & 189 & 1998 & 186.3 & 3300 \\  \cline{2-5} \hline\hline

\end{tabular}
\end{center}

\caption[Data Samples]{
\label{datatablab}
Overview of the data samples used in this analysis. The quoted
numbers of selected events refer to the standard 
selection so described in
Sect. \ref{datsamplab}. 
For the c.m.s. energies of the JADE experiment, 
the mean value $\langle\sqrt{s}\rangle$, averaged over the selected
events, is given in parentheses. 
The small difference between the event
number at $\sqrt{s}=35$ GeV and that given in
\cite{ourjade} is due to a somewhat more stringent definition of ``long
tracks'' (see table \ref{trackqualtablab}) in this analysis.
}

\end{table}

\section{The Experiments}
The ${\rm e}^+{\rm e}^-$ storage ring PETRA (see e.g. \cite{petra})
was operated for physics measurements from 1978 until 1986 at c.m.s. energies ranging
from 12 GeV to 46.7 GeV. Extensive energy
scans were made between these values. This analysis uses data samples recorded
at energies of $\sqrt{s}=35$ and 44 GeV. The precise ranges in c.m.s. energy
and the luminosities are given in 
Table \ref{datatablab}.\par
The LEP ${\rm e}^+{\rm e}^-$ collider at CERN began operation in 1989 at 
$\sqrt{s}\approx91$ GeV, 
i.e. around the mass of the $Z^0$ boson. Since Fall 1995, the energy has been increased
in steps from 130 GeV, 136 GeV, 161 GeV, 172 GeV, 183 GeV to 189 GeV
in 1998, the last year for which we include data in this study. The luminosities
recorded by the OPAL experiment at each of these energies are also listed
in Table \ref{datatablab}.\par
Descriptions of the JADE and OPAL detectors can be found in \cite{naroska,jadesel}
and \cite{opaldet}, respectively. Apart from the dimensions, the detectors
are very similar in their construction. Both are multi-purpose devices with a large
solid angle coverage. 
Table \ref{detparamlab} summarizes some detector parameters. The $r$, $\phi$ and $z$
coordinates refer to a cylindrical coordinate system with the origin lying
in the center of the
detector and the $z$ axis pointing along the incoming electron beam direction.\par
The detector components
primarily used in this analysis are the tracking systems and the
electromagnetic calorimeters. 
The main parts of the tracking systems of both detectors are drift chambers built
with a ``jet chamber'' geometry.
The relative resolution of the transverse track momentum
can be parameterized as $\sigma(p_t)/p_t=\sqrt{A_t^2+(B_t\cdot p_t\left[GeV\right])^2}$. The
values of $A_t$ and $B_t$ for JADE and OPAL are given in Table \ref{detparamlab}.\par
The electromagnetic calorimeters (ECAL) of JADE and OPAL are arrays of
lead-glass blocks. The relative 
resolution of the electromagnetic cluster energy $E$ is
$\sigma(E)/E=A_c+B_c/\sqrt{E\left[GeV\right]}$. The parameters $A_c$ and $B_c$ 
are given in Table \ref{detparamlab} for both detectors.

\begin{table}[!t]

{\footnotesize

\begin{center}
\begin{tabular}{||c|lcr||c|c||} \hline\hline
& \multicolumn{3}{|c||}{{\bf Parameter}} & {\bf JADE} & {\bf OPAL} \\ \hline
 & \multicolumn{3}{|l||}{overall length} & 8 m & 12 m \\
\rb{{\bf Dimensions}} &  \multicolumn{3}{|l||}{overall height} & 7 m & 12 m \\ \hline
 & & \multicolumn{2}{r||}{length} & 2.4 m & 4 m\\
 & \rb{dimension} & \multicolumn{2}{r||}{outer radius} & 0.8 m & 1.85 m \\ \cline{2-6}
 & \multicolumn{2}{|l}{transv. momentum} & \multicolumn{1}{|r||}{$A_t$} & 0.04 & 0.02 \\ 
 & \multicolumn{2}{|l}{resolution $\sigma(p_t)/p_t$} & \multicolumn{1}{|r||}{$B_t$} & 0.018 & 0.0015 \\ \cline{2-6}
 & \multicolumn{2}{|l}{spatial} & \multicolumn{1}{|r||}{$r-\phi$} & 
180 $\mu$m (110 $\mu$m) & 135 $\mu$m \\ 
 & \multicolumn{2}{|l}{resolution} & \multicolumn{1}{|r||}{$z$} & 1.6 cm & 
4.5---6 cm\\ \cline{2-6}
 & \multicolumn{3}{|l||}{double hit resolution} & 
7.5 mm (2 mm) & 2.5 mm\\ \cline{2-6}
{\bf Tracking} &  \multicolumn{3}{|l||}{gas composition} & &  \\
{\bf system}  & \multicolumn{3}{|l||}{argon/methane/isobutane}  & \rb{88.7\%/8.5\%/2.8\%} & \rb{88\%/9.4\%/2.6\%} \\ \cline{2-6}
 & \multicolumn{3}{|l||}{gas pressure} &  4 bar  & 4 bar \\ \cline{2-6}
 & \multicolumn{3}{|l||}{max. no. of hits}  & 48 & 159 \\
 & \multicolumn{3}{|l||}{reachable in} &  $0.83\cdot4\pi$  & $0.73\cdot4\pi$  \\ \cline{2-6}
 & \multicolumn{3}{|l||}{at least 8 hits}  & & \\
 & \multicolumn{3}{|l||}{reachable in}  & \rb{$0.97\cdot4\pi$} & \rb{$0.98\cdot4\pi$} \\ \cline{2-6}
 & \multicolumn{3}{|l||}{magnetic field}  & 0.48 T & 0.435 T \\ \hline
 & \multicolumn{2}{|l}{energy} & \multicolumn{1}{|r||}{$A_c$} & 0.015 & 0.002 \\
 & \multicolumn{2}{|l}{resolution $\sigma(E)/E$} & \multicolumn{1}{|r||}{$B_c$} & 0.04 & 0.063 \\ \cline{2-6}
 & \multicolumn{3}{|l||}{solid angle coverage}    & 90\% & 98\% \\ \cline{2-6}
 & \multicolumn{3}{|l||}{angular resolution}      & 7 mrad & 2 mrad \\ \cline{2-6}
 &  & \multicolumn{2}{|r||}{radial extent} & 1---1.4 m & 2.5---2.8 m \\
{\bf Electromagnetic} &  & \multicolumn{2}{|r||}{length} & 3.6 m & 7 m \\
{\bf calorimetry} & barrel & \multicolumn{2}{|r||}{polar angle covered} & $32^{\circ}$--$148^{\circ}$ & $36^{\circ}$--$144^{\circ}$ \\
 & & \multicolumn{2}{|r||}{radiation depth} & 12.5$X_0$ (15.7$X_0$) & 24.6$X_0$ \\
 & & \multicolumn{2}{|r||}{granularity} & 8.5$\times$10 cm$^2$ & 10$\times$10 cm$^2$ \\ \cline{2-6}
 & & \multicolumn{2}{|r||}{outer radius} & 0.9 m & 1.8 m \\
 & endcap & \multicolumn{2}{|r||}{polar angle covered} & 
$11^{\circ}$--$32^{\circ}$/$148^{\circ}$--$169^{\circ}$
& $11^{\circ}$--$36^{\circ}$/$144^{\circ}$--$169^{\circ}$ \\
 &  & \multicolumn{2}{|r||}{radiation depth} & 9.6$X_0$ & 22$X_0$ \\
 & & \multicolumn{2}{|r||}{granularity} & 14$\times$14 cm$^2$ & 9$\times$9 cm$^2$ \\ \hline\hline
\end{tabular}
\end{center}

}

\caption[Parameters of the JADE and OPAL detectors]{
Parameters of the JADE and OPAL detectors are compiled from \cite{naroska,jadesel,opaldet,otmar}.
Wherever a second number is given in parentheses, it 
refers to a later period of detector operation. The significance of the
parameters $A_t$, $B_t$, $A_c$ and $B_c$ is described in the text.
\label{detparamlab}
}

\end{table}

\section{Data Samples and Multihadronic Event Selection}
\label{datsamplab}
In the JADE part of the analysis we use ``detector level''
information from
data and Monte Carlo samples 
as described in \cite{ourjade}, obtained from
(measured or simulated) detector signals before any corrections
are applied.
We use data from the 1982 and 1986 runs at $\sqrt{s}=35$ GeV and 
the 1985 runs at 44 GeV, where
Monte Carlo samples including detector level information are 
available. It was shown in previous reanalyses of JADE 
data \cite{ourjade,otherjade} that measurements made in 1986 and before
could be reproduced using these data and Monte Carlo samples.\par
At $\sqrt{s}=M_{Z^0}$, we 
use the complete OPAL run of 1994 which
is the largest available homogeneous run without changes in the OPAL experimental set-up.\par
From the first runs at higher c.m.s. energies in 1995 we combine
the two samples at $\sqrt{s}=130$ 
and 136 GeV, weighting them with their statistical errors. 
Runs in 1997 at the same two c.m.s. energies 
are used as well and subjected to the
same treatment. 
The runs from 1995 and
1997 are analyzed independently of each other since 
changes of the detector were made between the two dates. In both 
cases, results are quoted at $\sqrt{s}=133$ GeV.
For each of the following energy steps we use
the full event samples recorded by OPAL 
and quote results for the respective
averaged c.m.s. energies.\par
In order to reject lepton pairs 
and two-photon collisions, we apply standard
selections from the two experiments. The selection used for the JADE part of the
analysis is the same as described in \cite{ourjade} and \cite{bethkejade},
and is closely related to that of OPAL, taking into account the difference
in c.m.s. energies. 
For the OPAL analysis we apply the criteria given in
\cite{tkmh}. Table \ref{tkmhtablab} contains a comparative list of the
most important
cuts used for the JADE and the OPAL analyses. 
We shall refer to this set of cuts as 
the ``preselection''.\par
At c.m.s. energies above $M_{Z^0}$, photon radiation in the initial state becomes
a significant source of background. In order to reject such ``ISR events'' we
determine the total hadronic mass $\sqrt{s'}$ of an event following a procedure
based on that described in \cite{opal161} which takes possible multiple photon
radiation into account. We require events to have 
$\sqrt{s}-\sqrt{s'}<10\;\mbox{GeV}$. For systematics studies we apply 
alternatively a 
combination of cuts on the visible energy and missing momentum of the event and on
the energy of an isolated photon candidate \cite{opal133}. 
We shall refer to the former procedure as the ``invariant
mass'' selection and to the latter as the ``energy balance'' selection.\par
\begin{table}[!t]
{\footnotesize

\begin{center}
\begin{tabular}{||c||c||c|c||} \hline\hline
Reaction to be &  &  &  \\
suppressed & \rb{Cut variable} & \rb{JADE} & \rb{OPAL} \\ \hline \hline
2-lepton & & $\ge3$ long tracks & \\
events   & \rb{$n_{ch}$} & and $\ge4$ central tracks & \rb{$\ge7$} \\[0.2cm]
 & $n_{cal}$ & --- & $\ge7$ \\[0.2cm]
& & $>3.0$ GeV (barrel) & \\
& \rb{$E_{shw}$} & or $>0.4$ GeV (per endcap) & \rb{---} \\ \hline

2-photon & $E_{vis}/\sqrt{s}$ & $>0.5$ & $>0.1$  \\
events   & $p_{bal}$ & $<0.4$ & $<0.6$  \\ 
         & $p_{miss}$ & $0.3\cdot\sqrt{s}$ & ---\\ \hline

other & $|\cos{\theta_T}|$ & $<0.8$ & $<0.9$  \\ \hline\hline

\end{tabular}
\end{center}

}

\caption[Preselection cuts at JADE and OPAL]{
\label{tkmhtablab}
Comparison of the preselection cut values
against two-lepton and two-photon events 
for the JADE and OPAL experiments. 
$n_{ch}$, $n_{cal}$ and
$E_{shw}$ denote, respectively, the number of charged tracks and calorimeter clusters and
the total shower energy deposited in the calorimeters. $E_{vis}$ is the total visible
energy, $p_{miss}$ the total missing momentum
and $p_{bal}$ the momentum sum in $z$ direction, normalized by $E_{vis}$.
Both experiments also apply cuts on the polar angle of the thrust axis, $\theta_T$. 
The terms ``long'' and ``central tracks'' are
explained in Table \ref{trackqualtablab} in Appendix~\ref{qualitylab}.
}

\end{table}

At $\sqrt{s}=161$ GeV and above, the production of $W^{\pm}$ (and later also $Z^0$) pairs 
with hadronic decays
is another source of background. At these higher energies, we 
reject such reactions by 
dividing each event into hemispheres 
using the plane perpendicular to the thrust axis 
\cite{thrust}. We denote the heavier and lighter 
invariant mass of each hemisphere, normalized with the visible energy, by
$M_H$ and $M_L$ \cite{jetmass}.
Events with weak boson pairs
decaying hadronically usually have larger hemisphere
invariant masses.
We apply the cut $M_H/2+M_L<0.35$ 
and refer to this as the ``jet mass selection''.
An alternative selection method has
been used in previous OPAL analyses (e.g. \cite{mymontp}): the event is 
resolved into four jets using the Durham jet finder, and
the ${\cal O}(\alpha_s^2)$ QCD matrix element 
for a four-parton final state
is calculated using the jet four-momenta $p_1$...$p_4$. The value of
the matrix element is used as a cut
variable.
This method will
be called the ``event weight selection'' and serves as a systematic check. 
The jet mass and event weight selections have very similar 
performance when applied to a sample of Monte Carlo events from PYTHIA 
\cite{pythia} (for
multihadronic events) and GRC4F \cite{grc4f} (for all relevant four-fermion processes) 
which have been passed through the preselection and the invariant 
mass selection.\par
Table \ref{datatablab} lists the numbers of events which we select at the individual c.m.s. energies
using the 
standard selections.

\section{Measurement of Jet Fractions}
\label{jetmeaslab}
We present measurements of jet-multiplicity related quantities 
for various jet finders at c.m.s. energies of 35 through 189 GeV 
and compare them with
predictions of several Monte Carlo models.
The measured quantities 
are corrected for
effects of limited detector resolution and acceptance, as well as for
inefficiencies of 
the selection and ISR, 
i.e. they are presented at the ``hadron level,'' which is
understood to include all charged and neutral
particles emerging after all intermediate particles with
lifetimes below $3\cdot 10^{-10}$ s have decayed.
\subsection{General Analysis Procedure}
\subsubsection{Reconstruction of Single Particles}
\label{singpartlab}
For both experiments, the
measurements are based on 
charged tracks and electromagnetic calorimeter clusters.
From these
the four-vectors of single particles are reconstructed
according to techniques which are
quite similar in both the
JADE and the OPAL part of the analysis.
After imposing the quality criteria for charged tracks and electromagnetic calorimeter
clusters described in Appendix~\ref{qualitylab}, each accepted track is regarded
as a charged particle having the measured three-momentum
of the track and the mass of a charged pion.
If a track can be linked to a particular ECAL cluster an estimate is
made of the energy which a charged pion would 
deposit in the calorimeter; this amount is
subtracted from the energy of the cluster. If the entire cluster energy
is used up by such subtractions, the cluster is discarded; otherwise, the remaining
energy is assumed to have been deposited by an additional neutral particle, and a zero-mass 
four-vector is constructed from this energy and the position of the cluster.\par
The set of four-vectors from each selected event
is then passed to the jet algorithms. At those
c.m.s. energies where initial state photon radiation is large ($\sqrt{s}=133$ GeV
and higher), the entire system of vectors is boosted into its own
rest frame before the algorithms are applied.

\subsubsection{Correction for Experimental Effects}
\label{detcorrlab}
At the c.m.s. energies of the JADE experiment (35 and 44 GeV), corrections
rely on existing Monte Carlo samples with full detector simulation,
taking into account the changes of the detector with time. The detector
level Monte Carlo samples available for this analysis were generated using 
the JETSET 6.3 program \cite{jetset63} with the standard parameter set
of the JADE collaboration \cite{jadesetting}
and including 
photon radiation in the initial state.
The Monte Carlo samples were shown in previous publications
\cite{ourjade,otherjade} to describe the 
data well. They were processed in the same way as the
data, i.e. subjected to the same selection cuts and single particle 
reconstruction procedures.\par
To obtain the corresponding hadron level information we have run
the JETSET 6.3 program to generate 
events at both c.m.s. energies, using
the same parameter settings, but
without initial state photon radiation. By dividing the hadron level by the
detector level prediction,
binwise multiplicative correction factors were determined for all observables
and then
applied to the data.\par 
A similar binwise multiplicative correction procedure is applied to the OPAL data
at $\sqrt{s}=91$ GeV. The correction factors were determined from two distinct 
Monte Carlo samples with detector simulation, 
generated by JETSET 7.4 \cite{jetset74} and 
HERWIG 5.9 \cite{herwig}.
The parameter settings
for both generators 
are described in \cite{opal161,opaljettune}.
Initial and final state photon radiation is
included in both cases.\par
For c.m.s. energies of 133 GeV and above, the JETSET Monte Carlo is replaced by
PYTHIA 5.7 \cite{pythia}, which has a more accurate
modelling of initial state photon radiation.
In addition, versions
5.8d (at $\sqrt{s}=161$ GeV) and 5.9 (otherwise) 
of HERWIG 
were used to generate multihadronic events.
The correction procedure for ISR and detector effects
applied at energies of 133 GeV and above was the same as 
at the lower c.m.s. energies.\par
At $\sqrt{s}=161$ GeV and above, 
all relevant four-fermion final states 
were generated by the GRC4F generator 
\cite{grc4f} and subjected
to detector simulation. 
The background predicted by GRC4F for each observable 
is {\it subtracted} at the detector level
before the multiplicative corrections for residual ISR background and
detector effects are applied.

\subsubsection{Determination of Systematic Errors}
\label{systlab}
To assess the size of systematic uncertainties inherent to the analysis
procedure, the entire analysis was repeated with variations of the
selection, of the correction
Monte Carlo generators and of the detector components used.
For each variation, 
the deviation of the final result from
the standard measurement is taken as a systematic uncertainty. 
All systematic uncertainties 
are added quadratically to yield the total systematic error 
for every bin of each variable.\par
The influence of the detector components used for the
single particle reconstruction (tracking chambers and electromagnetic calorimeter)
is estimated by repeating the analysis using only charged tracks.
As a consistency check, where there are data sets from two different run
periods ($\sqrt{s}=35$ GeV or 133 GeV), the influence of changes in the detectors carried
out between the two dates has been investigated, but no effects were found.\par
A systematic
variation of the selection mechanism 
is done at all energies by
tightening the cut on the thrust axis from $|\cos{\theta_T}|<0.9$ to $|\cos{\theta_T}|<0.7$. 
In addition, at 
$\sqrt{s}=35$ GeV
and 44 GeV, the cut variations described in \cite{ourjade} are performed:
The cut on the total missing three-momentum $p_{miss}$ 
is either 
      tightened from $0.3\cdot\sqrt{s}$ to $0.25\cdot\sqrt{s}$ or
      removed;
the upper limit on the energy balance in beam direction, $p_{bal}$, is either tightened from 0.4 to 
      0.3 or removed;
the cut on the normalized visible energy
      $E_{vis}/\sqrt{s}$ is varied from 0.5 to 0.55 and 0.45;
a minimum of 7 rather than 3 ``long tracks'' (as 
defined in Appendix~\ref{qualitylab})
is demanded.
For the analysis at $\sqrt{s}=91$ GeV, no further cross-checks aside from the variation
of the thrust axis cut are done.
At $\sqrt{s}=133$ GeV and above, 
the influence of the standard invariant mass selection is tested
by replacing it by the energy balance selection, and
at c.m.s. energies of 161 GeV and higher, the analysis is repeated using the 
event weight
selection rather than the standard jet mass selection.\par
A variation of the correction Monte Carlo is not possible at JADE energies because
no detector level samples aside from the ones used are currently available.
At all other energies ($\sqrt{s}\ge91$ GeV), the analysis is repeated using HERWIG
for the determination of the corrections instead of
JETSET or PYTHIA. The effect of using the
EXCALIBUR generator \cite{excalibur} rather than GRC4F 
for the background has also been investigated. The deviation from the main
result induced by this change was found to be negligible throughout and is
therefore not added as an additional uncertainty.\par
\label{sssslab} Any systematic variation of the analysis procedure involving changes in the number
of events entering the measurement of some numerical value (e.g. some bin) 
will necessarily generate a
statistical deviation from the standard value. 
This effect 
becomes
significant at c.m.s. energies with low statistics, i.e. at
$\sqrt{s}=133$ GeV, 161 GeV and 172 GeV.
To obtain a more 
realistic estimate of the systematic errors at these c.m.s. energies
a number of subsamples are created from the 
respective multihadronic Monte Carlo sample, all of which contain on average
the number of events
corresponding to the integrated luminosity of the data (cf. \cite{opal161}). Each of these subsamples is subjected
to the same analysis procedure as the data.
The standard deviation $\sigma$ of each separate contribution to
the systematic error over all subsamples 
is determined for all measured values and regarded as the statistical
component of the systematic error. If
$\sigma$ is smaller than the corresponding systematic 
error contribution $\delta$ measured from the data, the
latter is reduced to $\sqrt{\delta^2-\sigma^2}$. 
Otherwise, the respective contribution to the systematic error is
dropped completely. The procedure is performed separately for each contribution
to the systematic error before they are added. 
The number of Monte
Carlo subsamples used is 16 for the 1995 run at 133 GeV and 30 in all other cases.\par
In order to further reduce fluctuations of
the errors between measured values in adjacent bins
due to low statistics, the overall systematic errors in all except the extreme bins of
each quantity shown are
averaged over each three adjacent values 
of that quantity. The systematic 
errors of the first and the last bin are subsequently set to the
averaged errors of the three first and three last bins, respectively.
This method of averaging systematic errors is performed
at all c.m.s. energies.

\subsection{Description of the Measured Jet Fractions}
\label{jetresultlab}
In the following, measurements 
are shown for 
three representative c.m.s. energy values: 35, 91 and 189 GeV.
All quantities are plotted versus the
parameter(s) of the respective jet algorithm. Error bars represent the quadratic
sum of systematic and statistical uncertainties. The latter include effects of
limited statistics of the data and of the correction
Monte Carlo samples. 
Listings of the numerical values, including
those for the c.m.s. energies which are not shown in the figures, 
are to appear in the Durham data base \cite{databas}.\par
In each plot, the measurements are compared with hadron level predictions from
the generators
PYTHIA 5.722, HERWIG 5.9, ARIADNE 4.08 \cite{ariadne} 
and COJETS 6.23 \cite{cojets},
representing different kinds of parton shower and hadronization modelling. The parameter
settings for ARIADNE and COJETS are those from \cite{ariadne} and \cite{cojetsset}.
The predictions are based on samples of multihadronic events without
initial state photon radiation, generated independently of those used 
for the corrections. In the ARIADNE samples, initial states generated by PYTHIA 
were subjected to the ARIADNE parton shower simulation, in which
the hadronization modelling was taken over from JETSET 7.408. 
\subsubsection{{\boldmath$n$}-Jet Fractions}
\begin{figure}[tbhp]
\vskip \clratuup
\centering{\Huge JADE Scheme}

\vskip 1cm

\begin{center}
\vskip \clratup
\hskip \clratleft
\epsfig{file=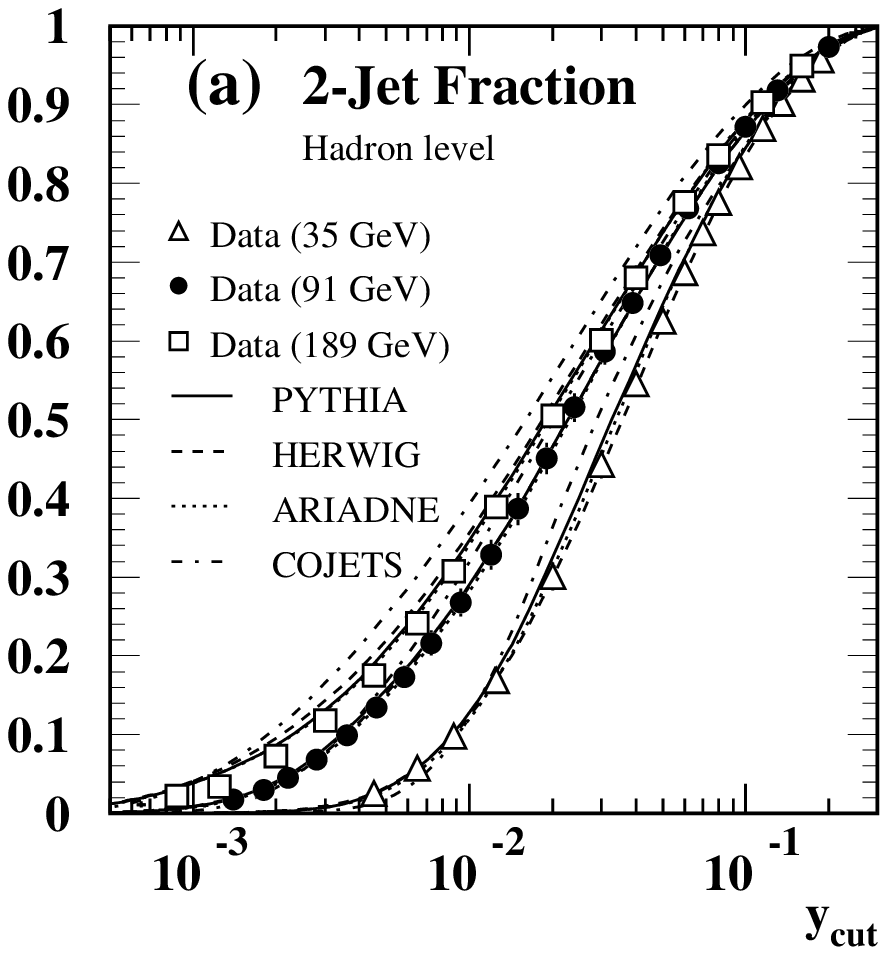,width=\clratwid}
\epsfig{file=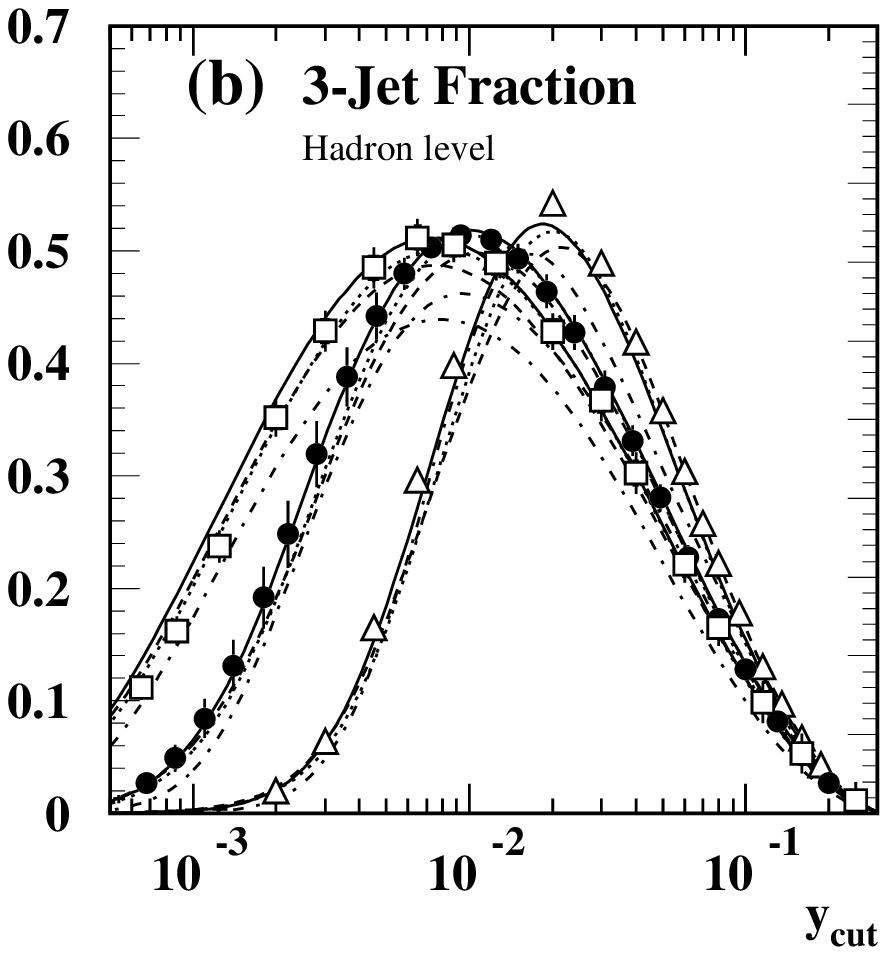,width=\clratwid}
\end{center}

\vspace*{\clratmid}

\begin{center}
\hskip \clratleft
\epsfig{file=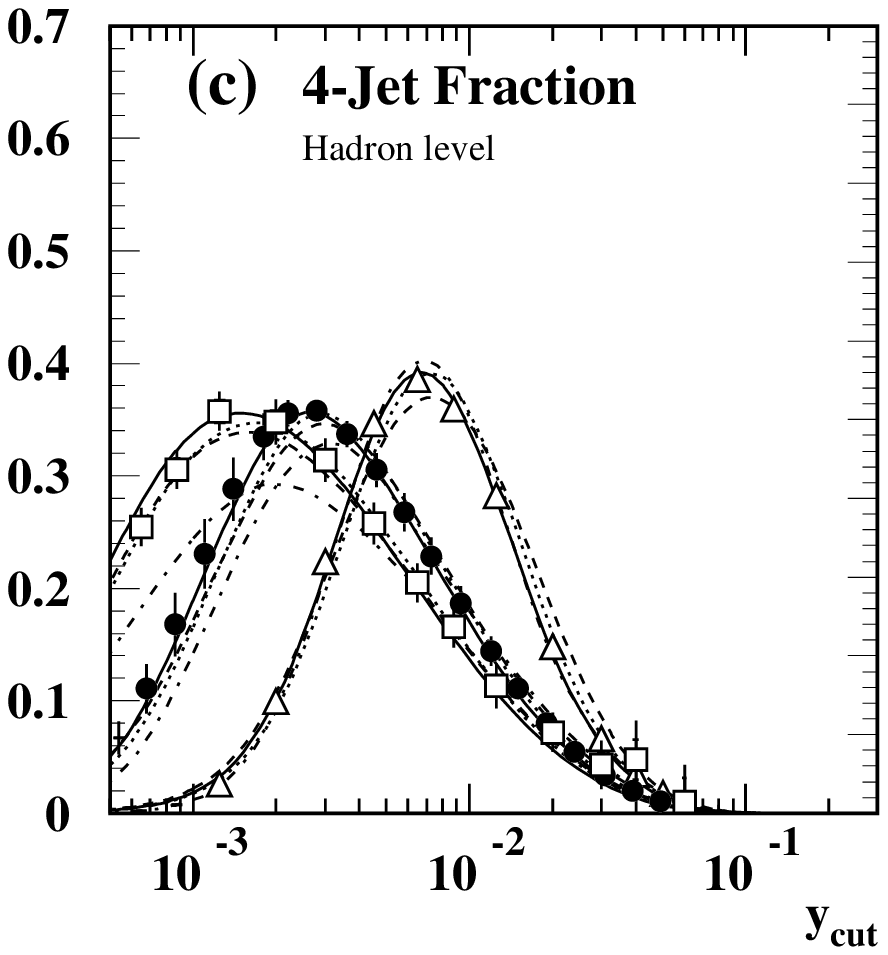,width=\clratwid}
\epsfig{file=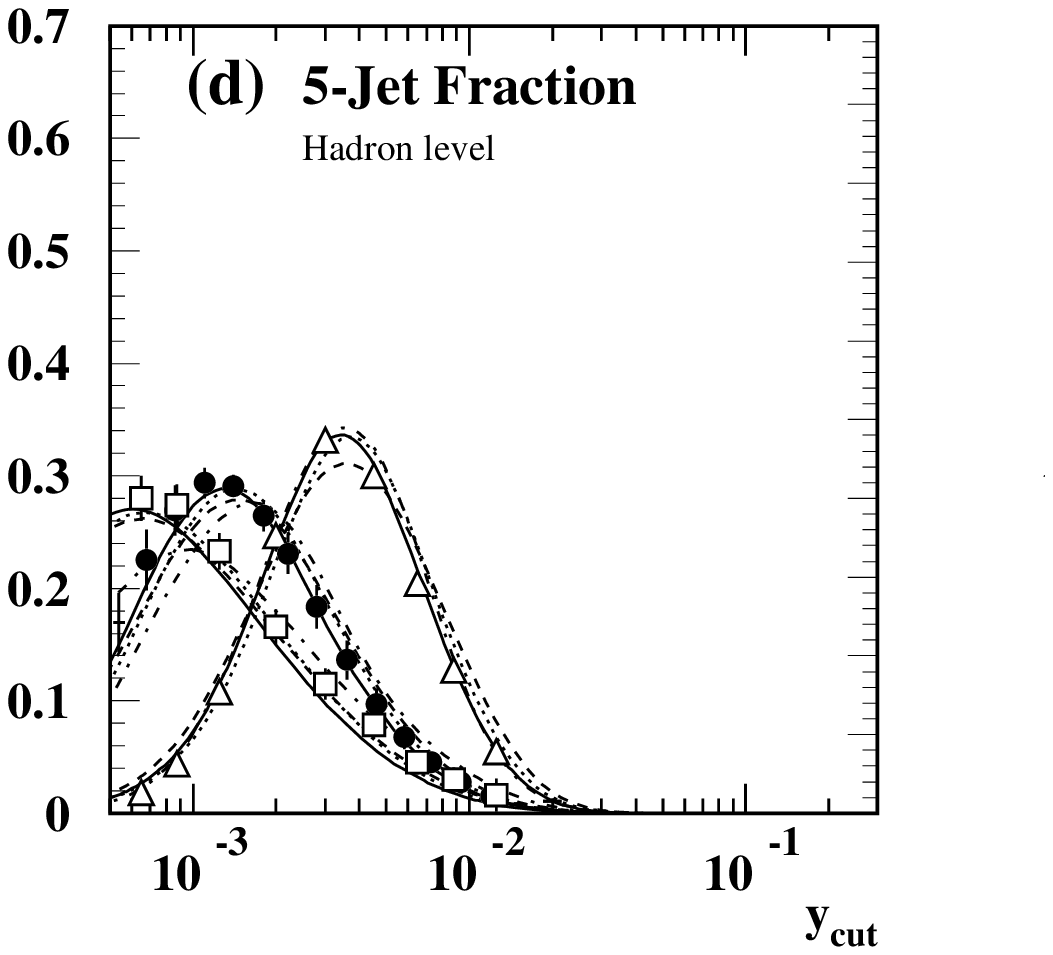,width=\clratwid}
\end{center}

\vspace*{\clratmid}

\begin{center}
\hskip \clratleft
\centerline{
\epsfig{file=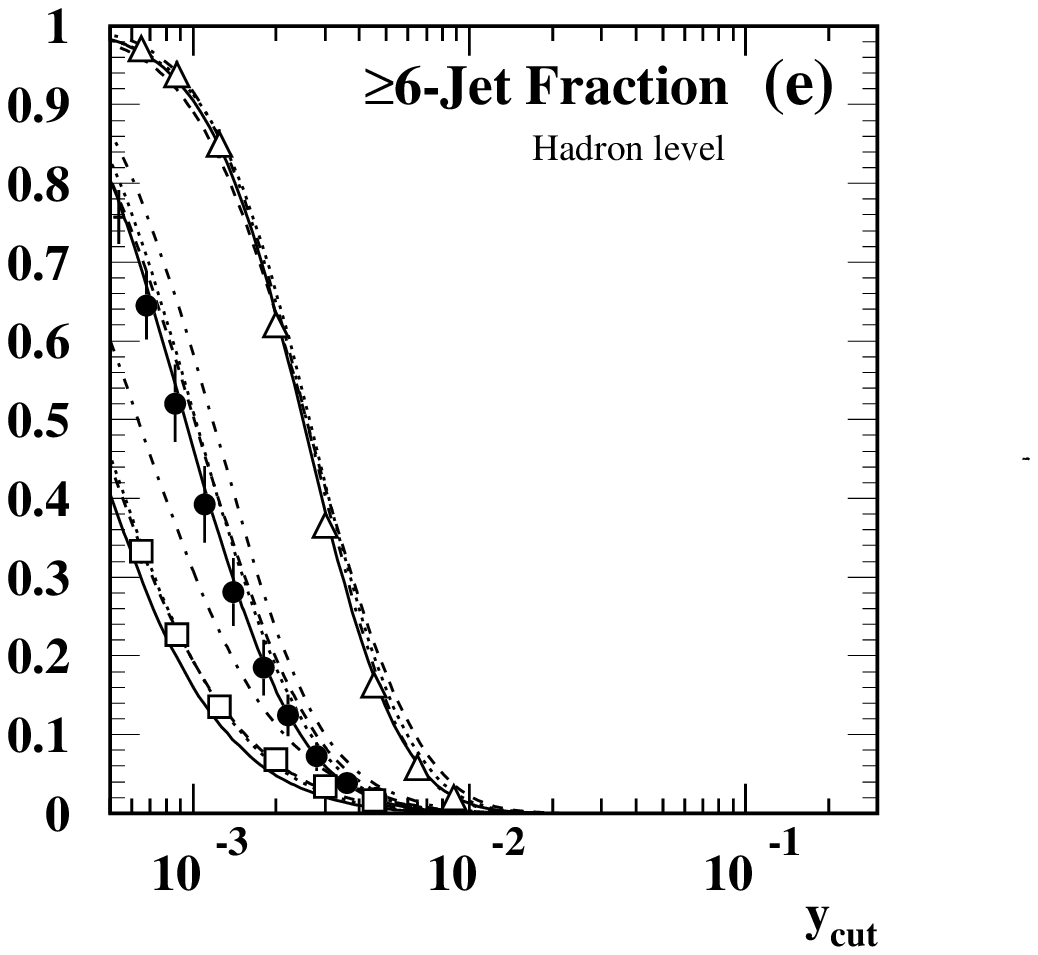,width=\clratwid}
}
\end{center}

\vspace*{\clratdown}

\caption[Jet fractions (JADE scheme) at $\protect\sqrt{s}$=35, 91 and 189 GeV]
{The fractions of events with 2, 3, 4, 5 and more 
than 5 jets as obtained by the JADE jet algorithm are plotted versus 
$y_{cut}$ at 35, 91 and 189 GeV. 
Data points are the measurements by the JADE and OPAL experiments. 
The error bars represent the total errors. 
Predictions by PYTHIA, HERWIG,
ARIADNE and COJETS are shown using lines of different styles. 

\label{e0ratefig}}

\end{figure}

\begin{figure}[tbhp]
\vskip -1cm

\begin{sideways}
\begin{minipage}[b]{\textheight}{

\vskip \clratuups
\centering{\Huge Durham Scheme}
\vspace*{\clratups}

\centerline{
\hskip \clratlefts
\epsfig{file=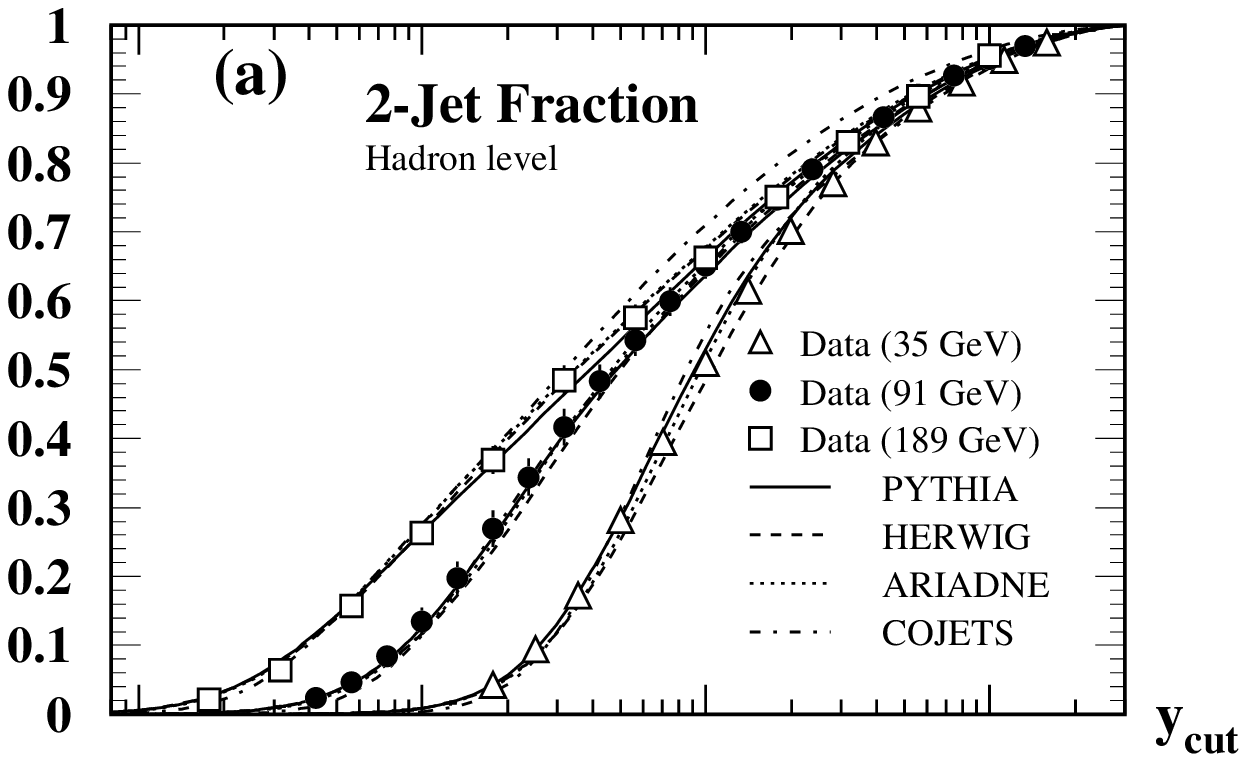,width=\clratwids}\hspace*{\clratints}
\epsfig{file=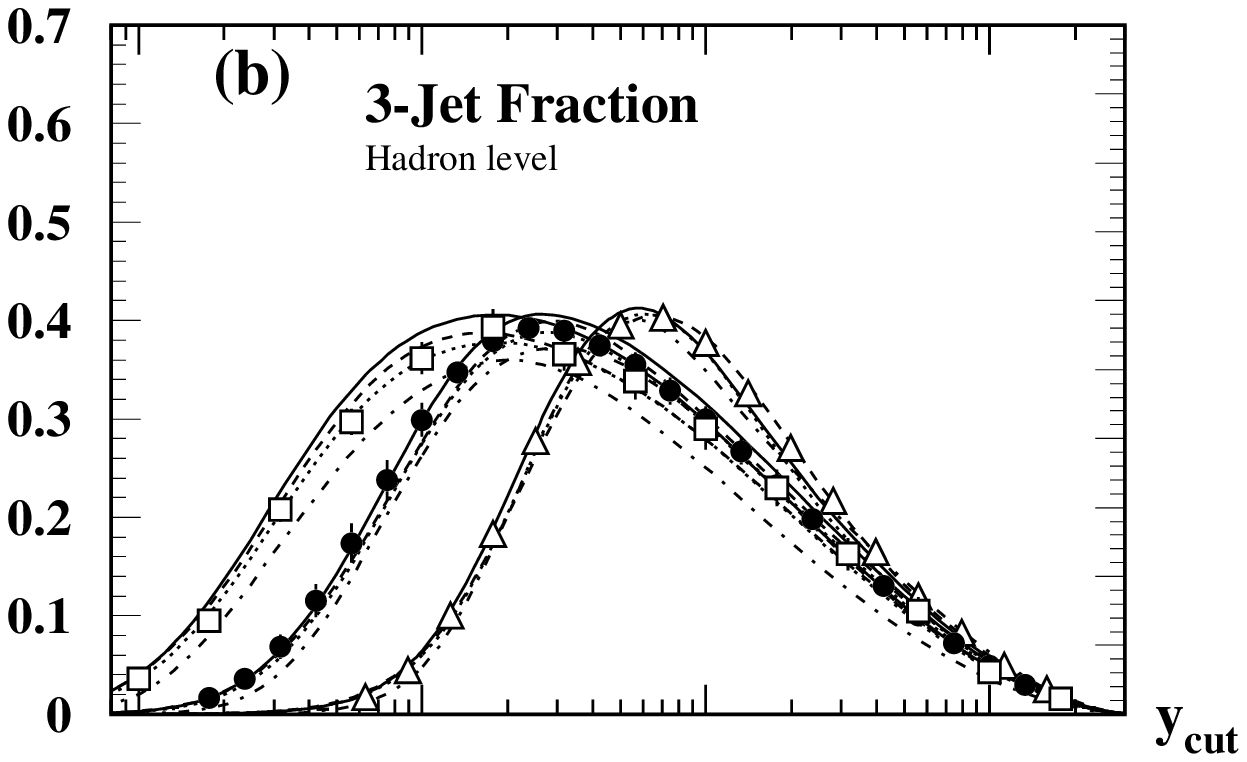,width=\clratwids}\hspace*{\clratints}
\epsfig{file=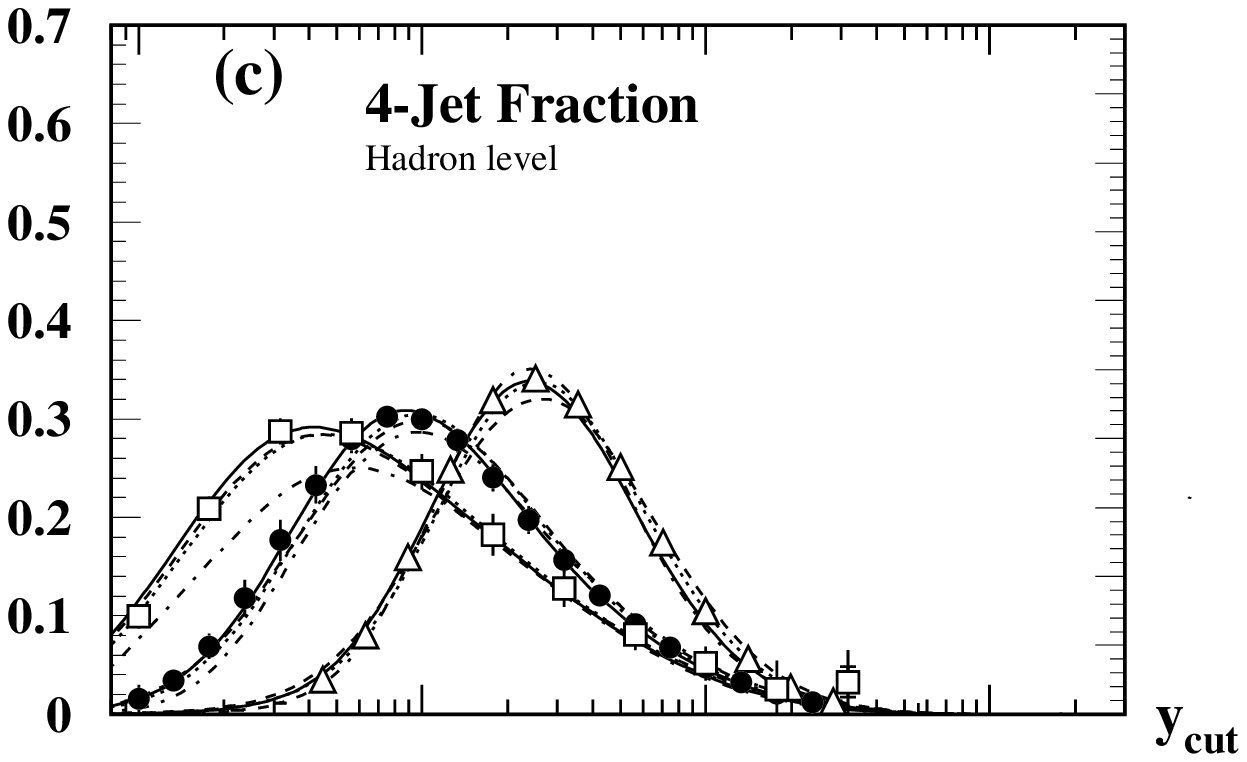,width=\clratwids}\hspace*{\clratints}
}
\vskip -1.2cm

\centerline{
\hskip \clratlefts
\epsfig{file=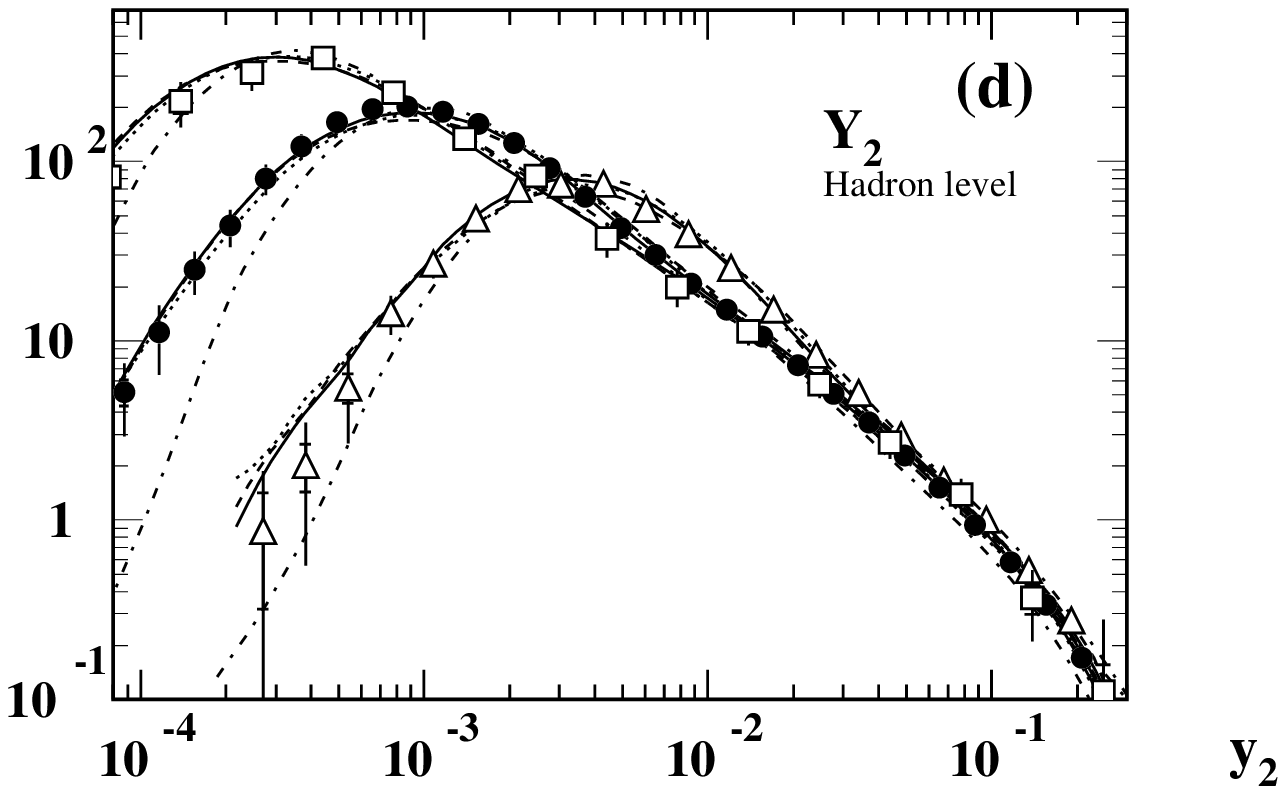,width=\clratwids}\hspace*{\clratints}
\epsfig{file=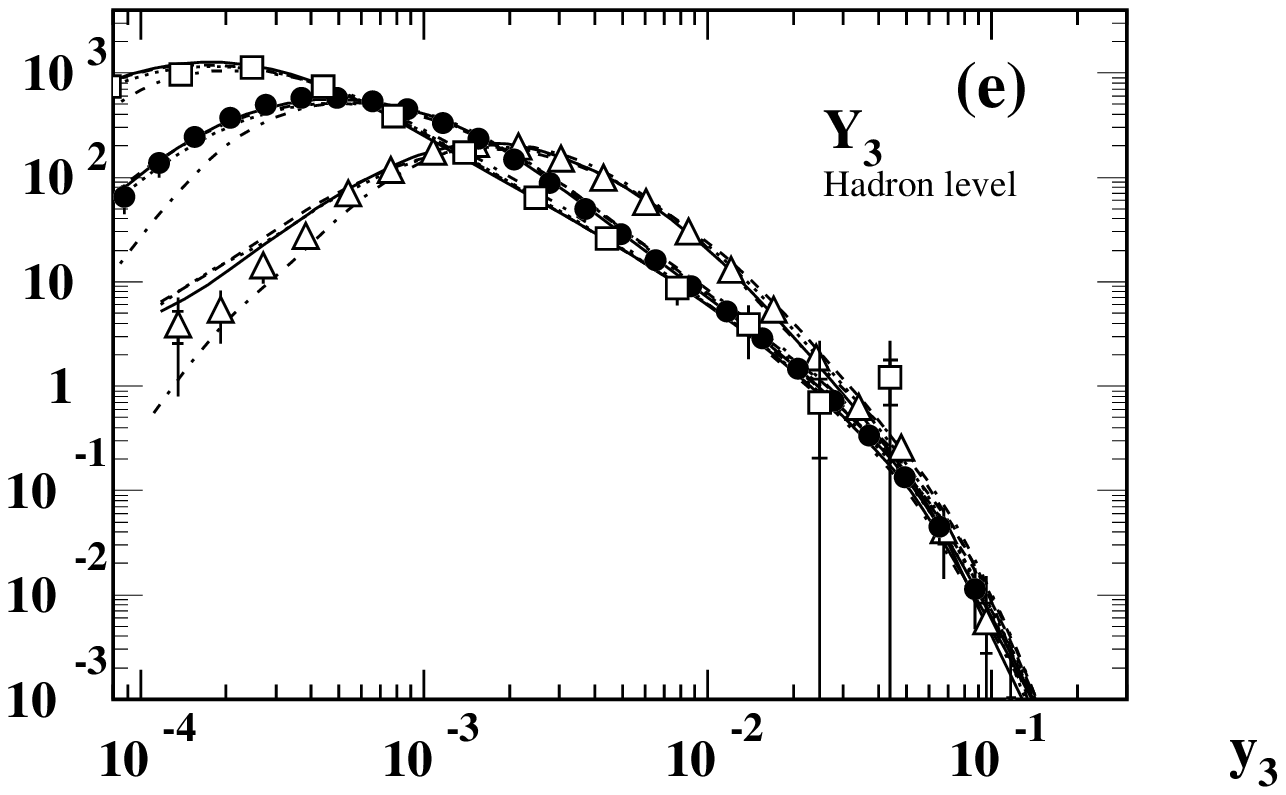,width=\clratwids}\hspace*{\clratints}
\epsfig{file=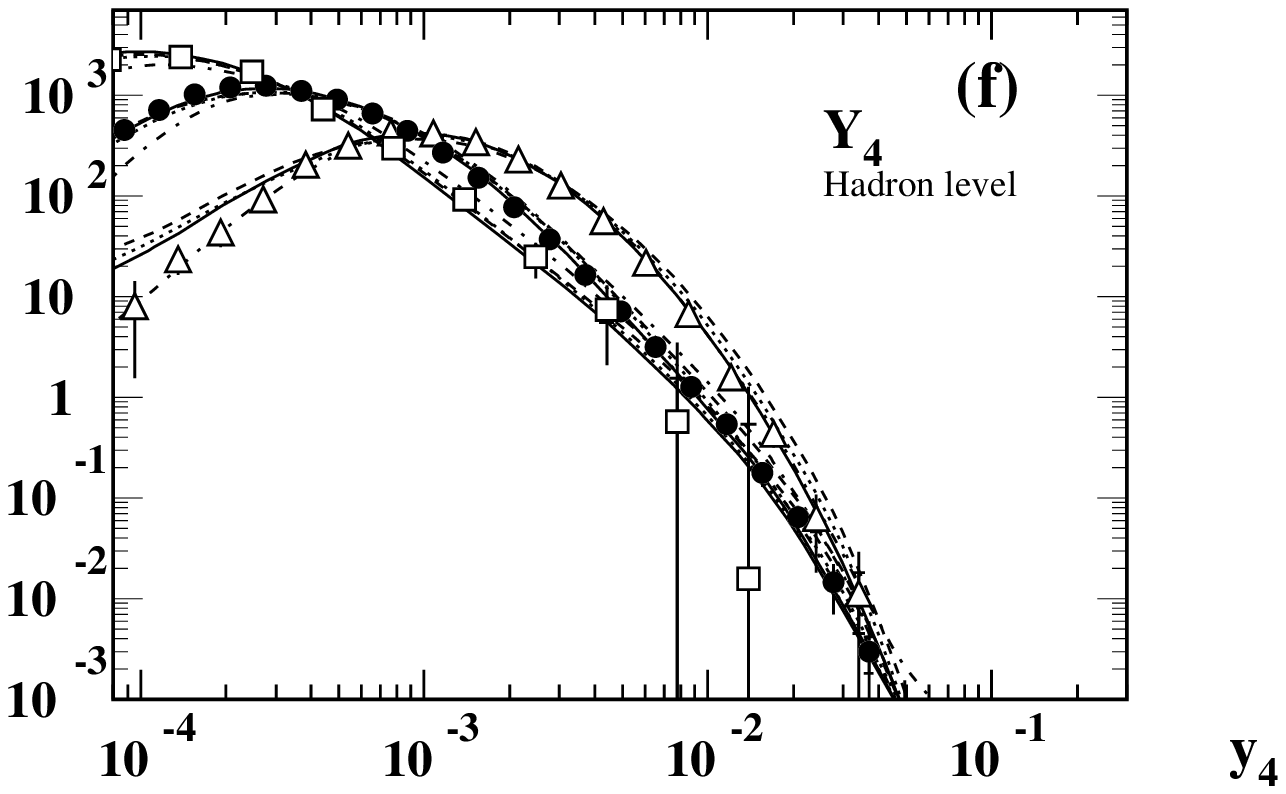,width=\clratwids}\hspace*{\clratints}
}

\vskip\clratmids

\centerline{
\hskip \clratlefts
\epsfig{file=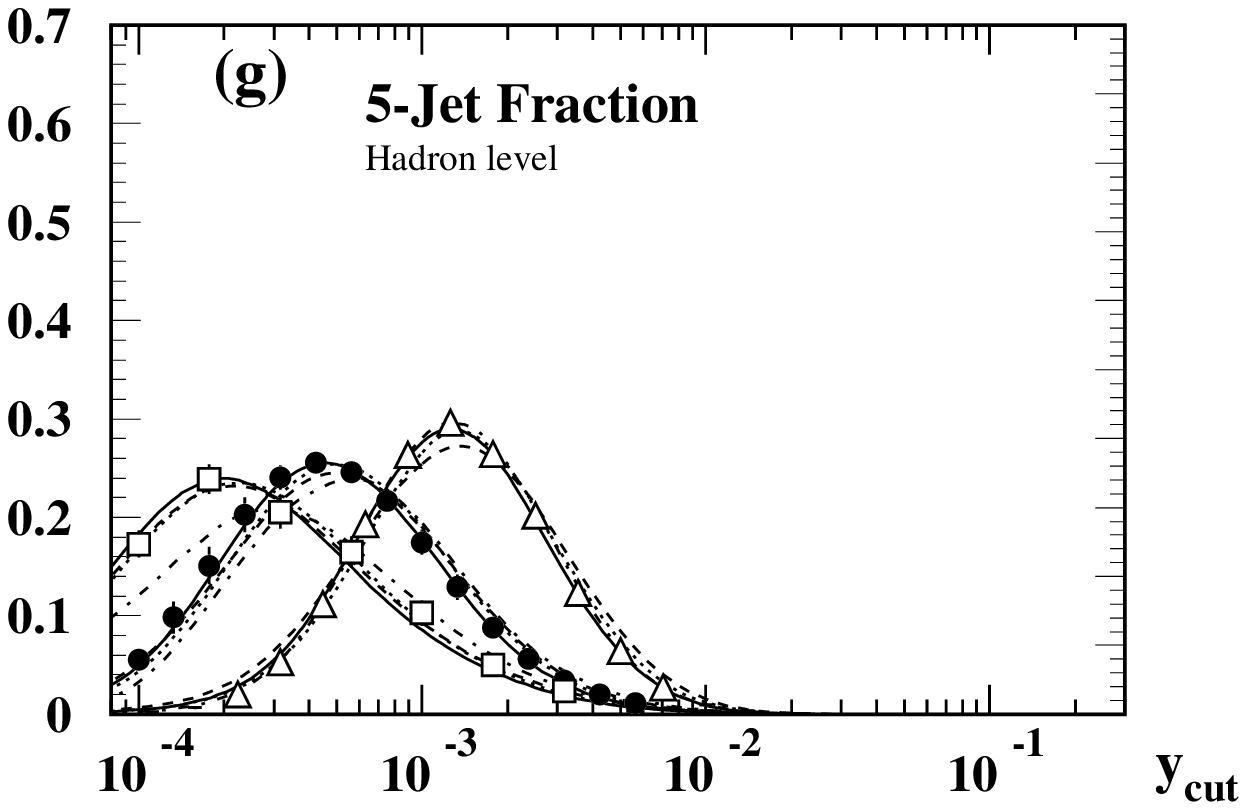,width=\clratwids}\hspace*{\clratints}
\epsfig{file=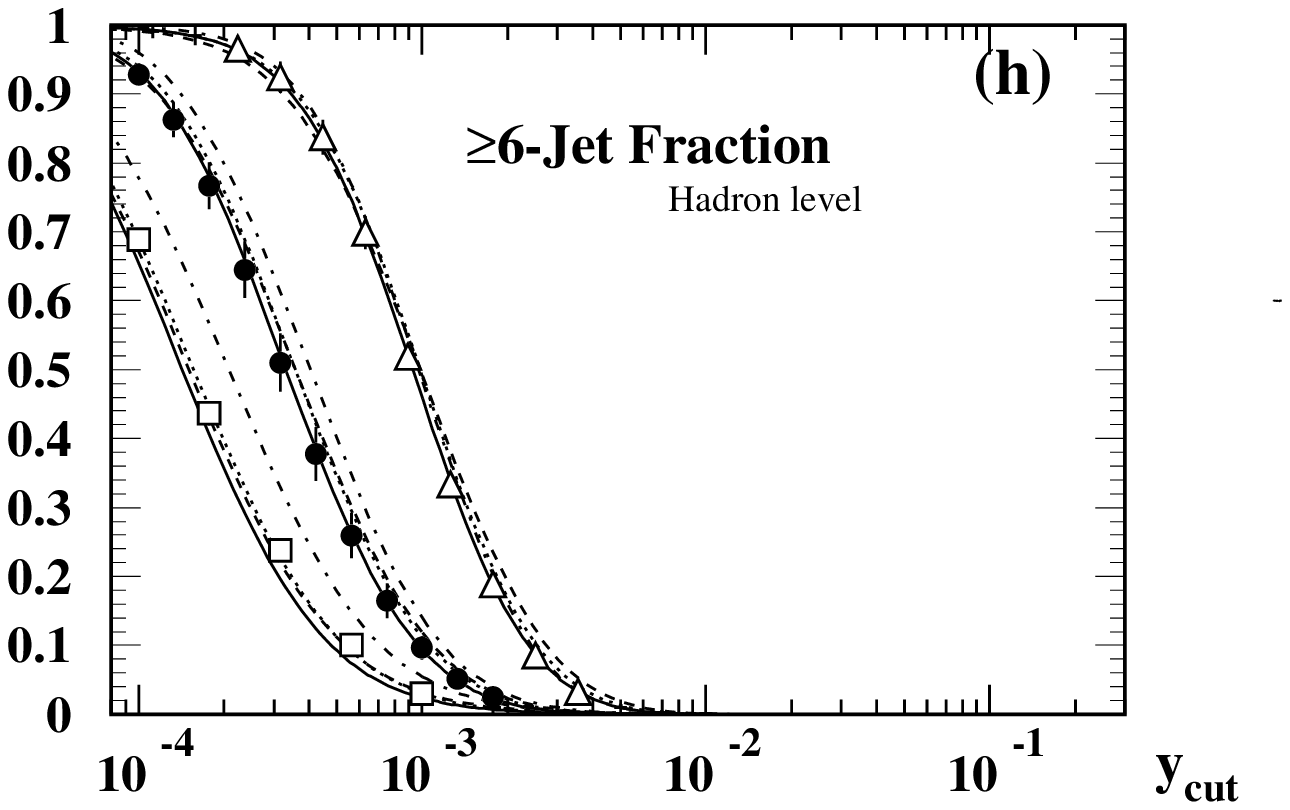,width=\clratwids}\hspace*{\clratints}
}

\vskip \clratdowns

\caption[Jet fractions and $y_n$ distributions (Durham scheme) at $\protect\sqrt{s}$=35, 91 and 189 GeV]
{In plots (a), (b), (c), (g) and (h),
the fractions of events with 2, 3, 4, 5 and more 
than 5 jets as obtained by the Durham scheme are plotted 
versus 
$y_{cut}$ at 35, 91 and 189 GeV. 
Diagrams (d) through (f) show the distributions in $y_n$ 
(see Sect. \ref{ydistlab}) corresponding to (a), (b) and (c).
The presentation is the same as in Fig.~\ref{e0ratefig}.
Small horizontal ticks
indicate the size of the statistical errors. 
\label{dratefig}
}

}
\end{minipage}

\end{sideways}

\end{figure}

\begin{figure}[tbhp]

\vskip -1cm

\begin{sideways}
\begin{minipage}[b]{\textheight}{

\vskip \clratuups
\centering{\Huge Cambridge Scheme}
\vspace*{\clratups}

\centerline{
\hskip \clratlefts
\epsfig{file=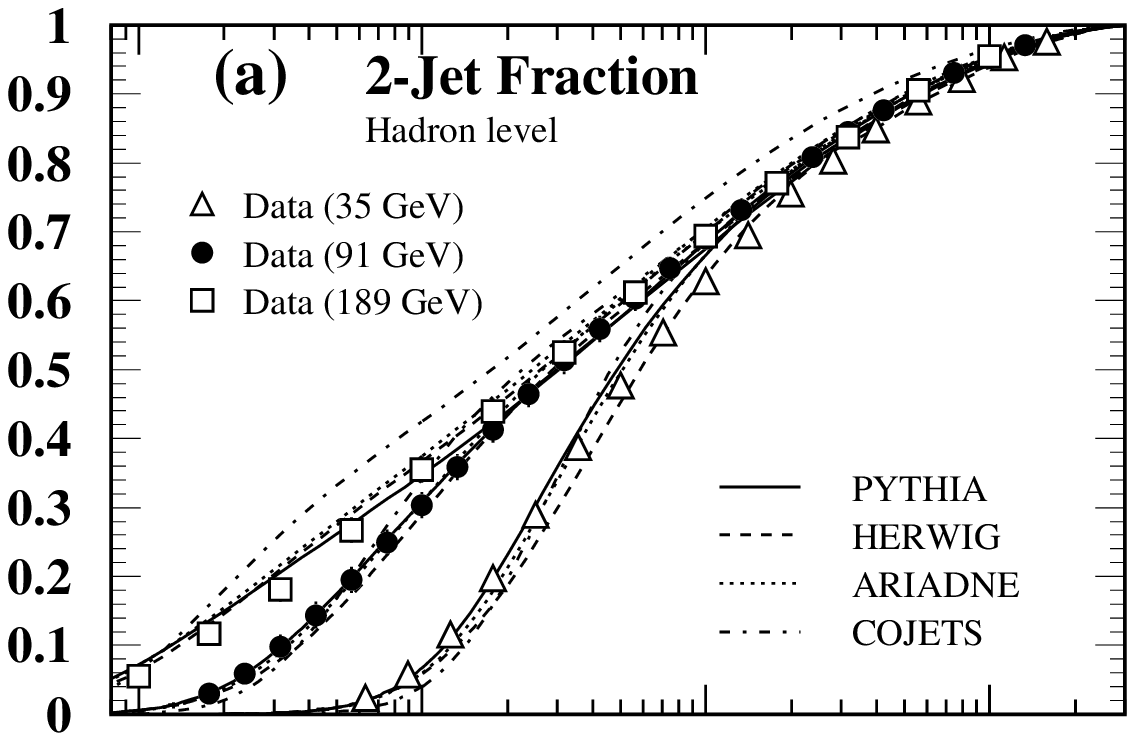,width=\clratwids}\hspace*{\clratints}
\epsfig{file=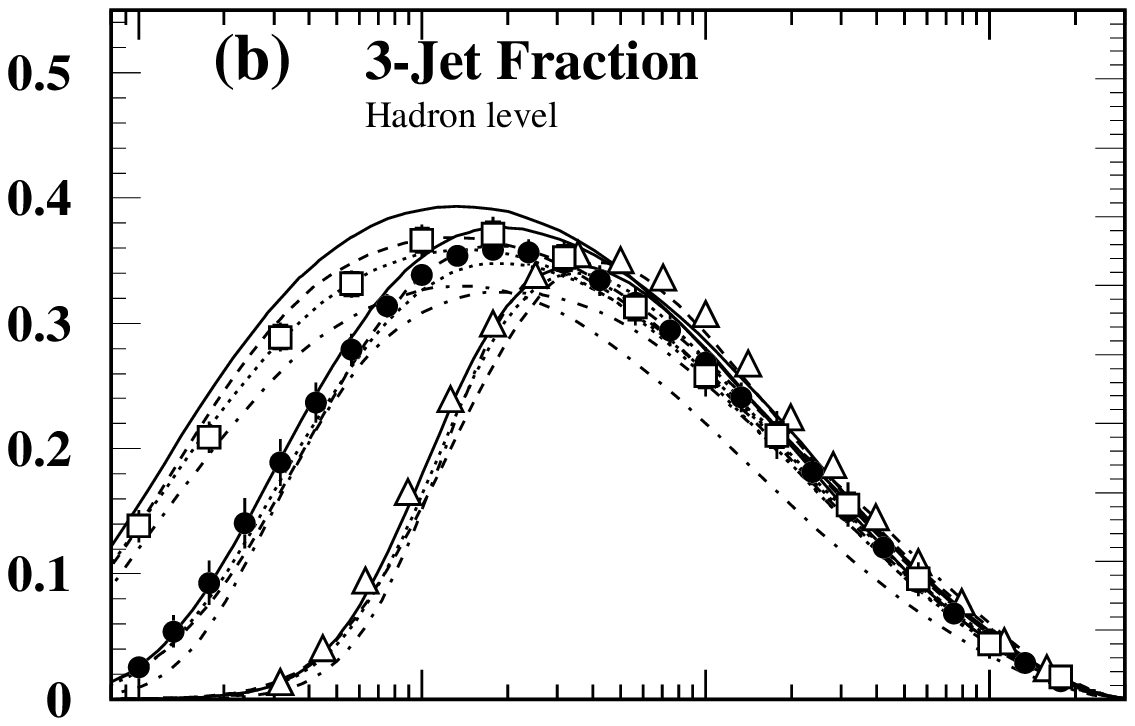,width=\clratwids}\hspace*{\clratints}
\epsfig{file=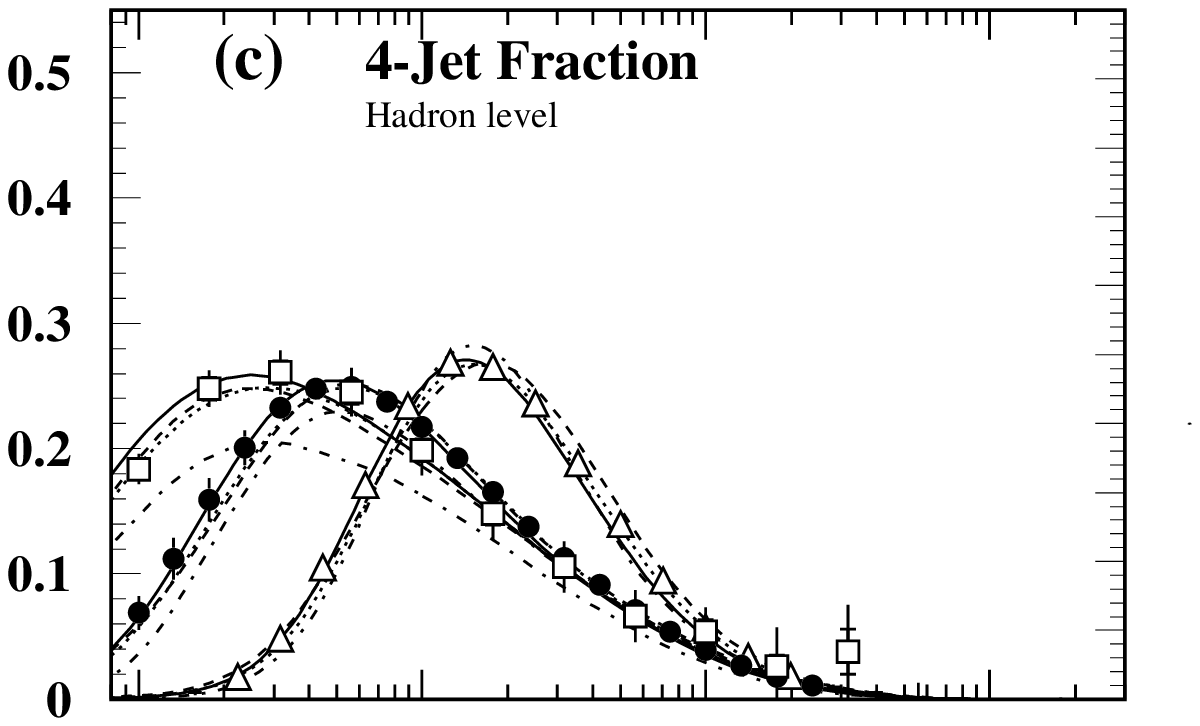,width=\clratwids}\hspace*{\clratints}
}
\vskip -1.2cm

\centerline{
\hskip \clratlefts
\epsfig{file=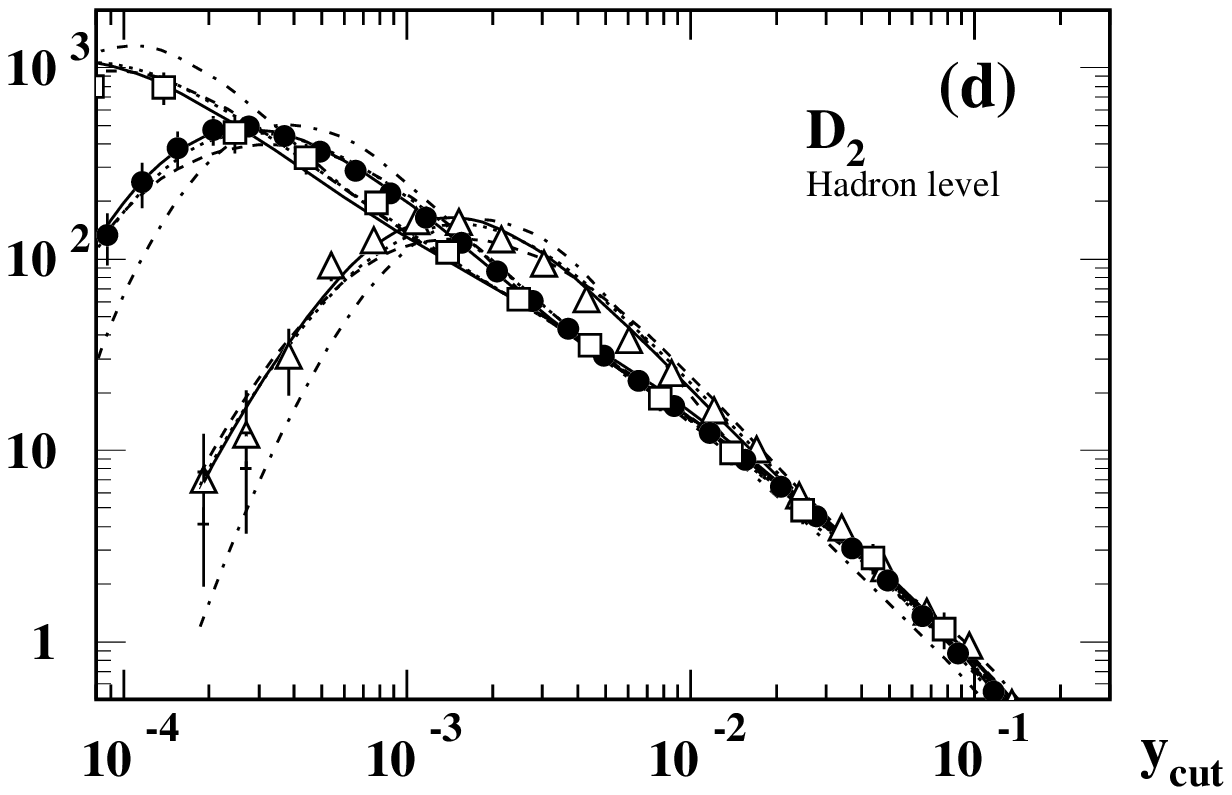,width=\clratwids}\hspace*{\clratints}
\epsfig{file=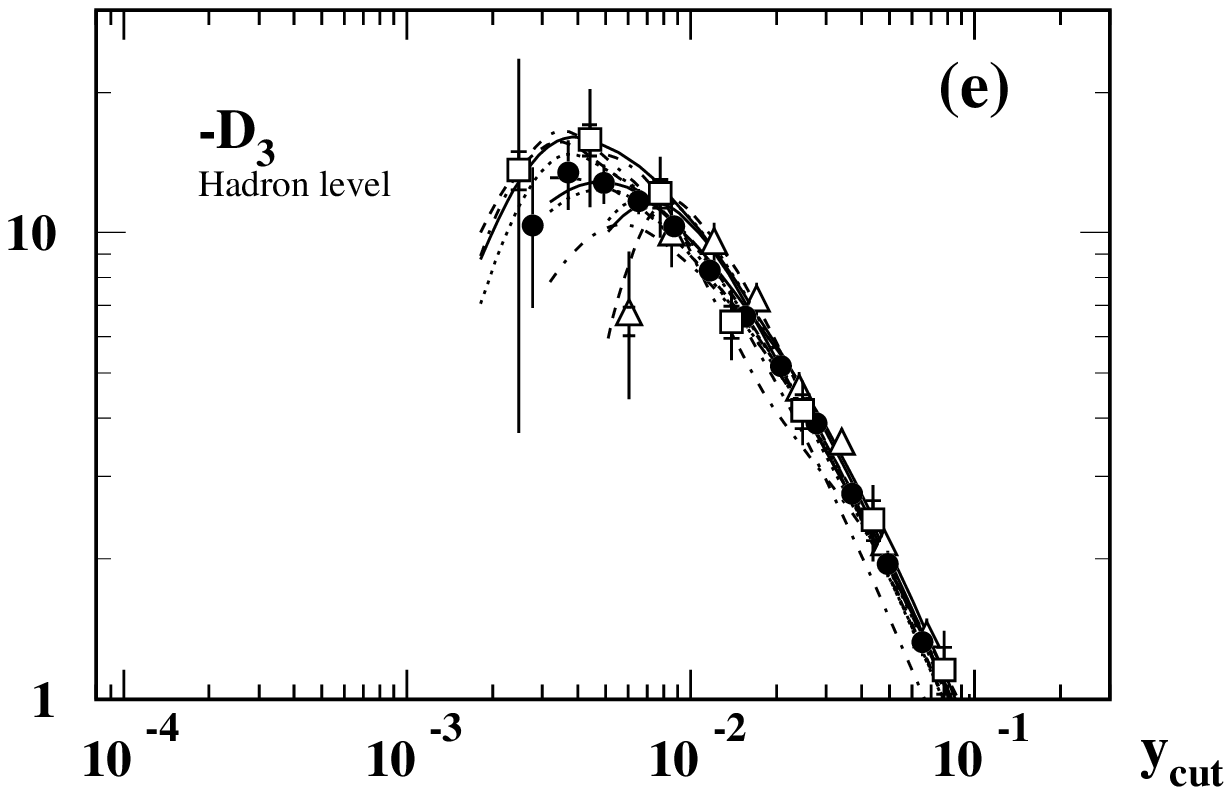,width=\clratwids}\hspace*{\clratints}
\epsfig{file=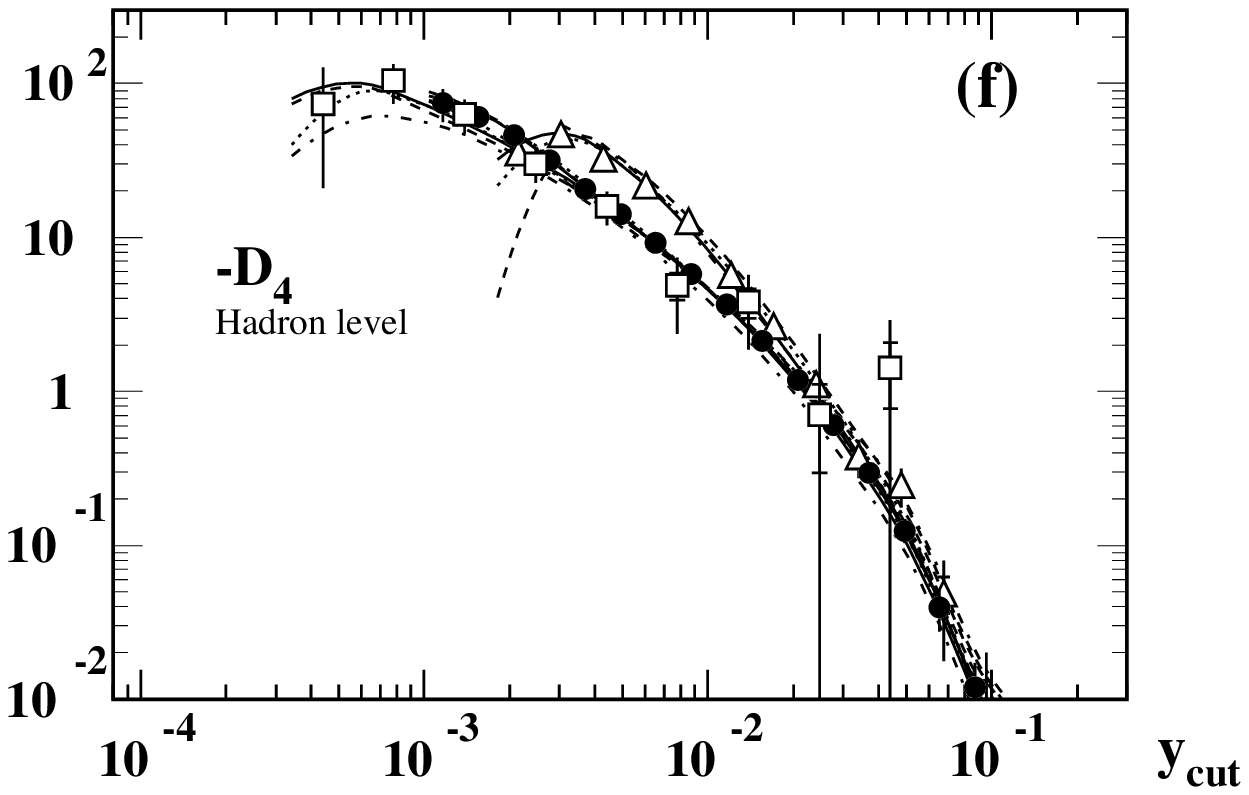,width=\clratwids}\hspace*{\clratints}
}

\vskip\clratmids

\centerline{
\hskip \clratlefts
\epsfig{file=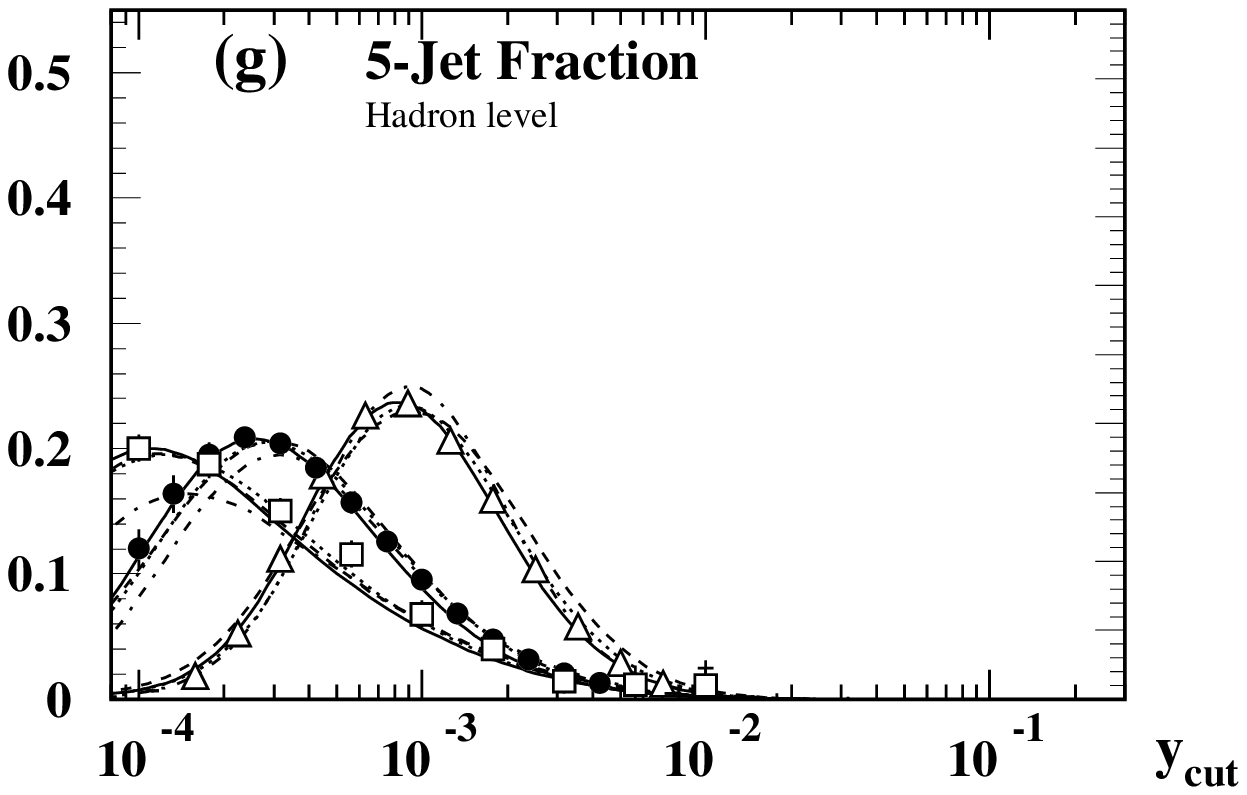,width=\clratwids}\hspace*{\clratints}
\epsfig{file=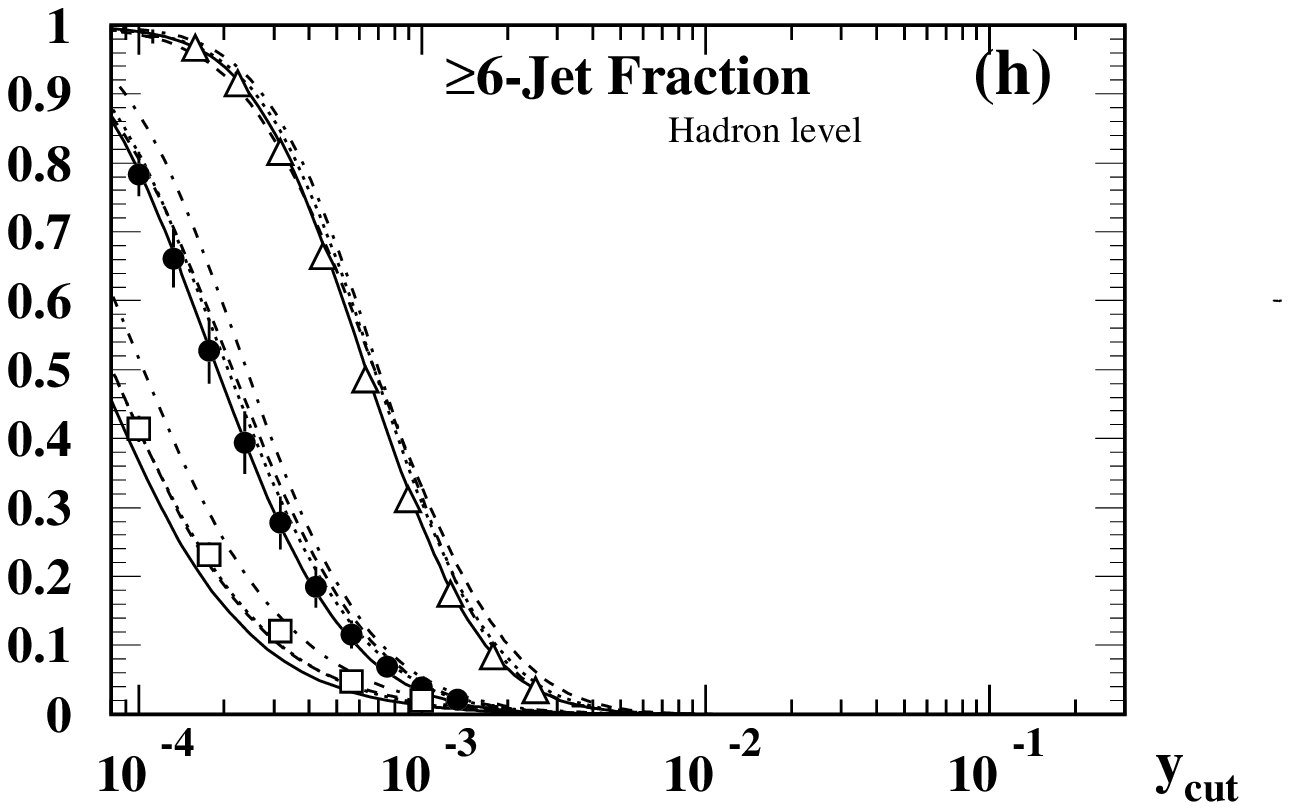,width=\clratwids}\hspace*{\clratints}
}

\vskip \clratdowns
\caption[Jet fractions (Cambridge scheme) at $\protect\sqrt{s}$=35, 91 and 189 GeV]
{In plots (a), (b), (c), (g) and (h), jet fractions as obtained by the 
Cambridge scheme are plotted 
versus 
$y_{cut}$ at 35, 91 and 189 GeV
as in Fig.~\ref{dratefig}. Diagrams (d) through (f)
show the differential jet fractions $D_n$ (see section \ref{ydistlab})
corresponding to (a), (b) and \nolinebreak (c).
\label{cratefig}
}
}
\end{minipage}

\end{sideways}

\end{figure}

We analyze $n$-jet fractions, $R_n$, from the JADE \cite{jadealgo}, 
Durham \cite{dalgo} 
and Cambridge \cite{calgo} jet finders, which reconstruct hadron jets based on different
resolution parameters $y_{cut}$ and different procedures to recombine unresolvable jets.
Measurements for $n=2$ through $n=5$ and $n\ge6$ are shown versus $y_{cut}$ 
in Fig.~\ref{e0ratefig},
and in the upper and lower plots of Figs.~\ref{dratefig} and \ref{cratefig},
respectively. 
For each variable, the measurements at the three c.m.s. energies are 
combined in one plot to make changes with energy 
more visible. The same ranges in
abscissa and ordinate were chosen for all c.m.s. energies within one scheme.
We also use a cone jet finder.\par
At the highest c.m.s. energies, systematic effects due to hadronization are
smallest. At $\sqrt{s}=35$ 
GeV and 44 GeV no HERWIG samples were available for the
evaluation of the model uncertainties. Thus the apparent systematic errors
are largest for the 91 GeV sample.
The predictions of 
PYTHIA, HERWIG and ARIADNE 
are similar and lie mostly within the uncertainties of the measurements over the 
range in $y_{cut}$ studied. 
It is impossible to choose a clear best generator from these three.
Since 
measurements of jet fractions for neighboring $y_{cut}$ values are
partially based on the same events, the single bins of each jet fraction
are strongly correlated.\par
The predictions of COJETS are also in many cases in agreement with the data.
However, especially at higher c.m.s. energies 
($\sqrt{s}\gtrsim133$ GeV), 
too many
jets are predicted in the low $y_{cut}$ regions. This is likely to be a consequence of
the neglect of gluon coherence in the COJETS generator which may lead
to an increased number of soft gluons between jets
and therefore to high jet multiplicities. 
Coherence effects have already been
observed by OPAL \cite{coherence}.\par
Scaling violations of both the Monte Carlo predictions and the 
measurements with $\sqrt{s}$
are visible for all three schemes in form of a 
reduction of the jet fractions for $n\ge3$ (and a corresponding increase of $R_2$) 
at higher $\sqrt{s}$,
as is expected from the
running of $\alpha_s$ according to QCD. 
The resolution parameters
$y_{cut}$ are defined in such a way that 
no scaling violations would be expected
for jet rates if $\alpha_s$ did not run.\par
\begin{figure}[tbhp]

\begin{sideways}
\begin{minipage}[b]{\textheight}{

\vskip \coneuup
\centering{\Huge Cone Algorithm}
\vspace*{\coneup}

\centerline{
\hskip \coneleft
\epsfig{file=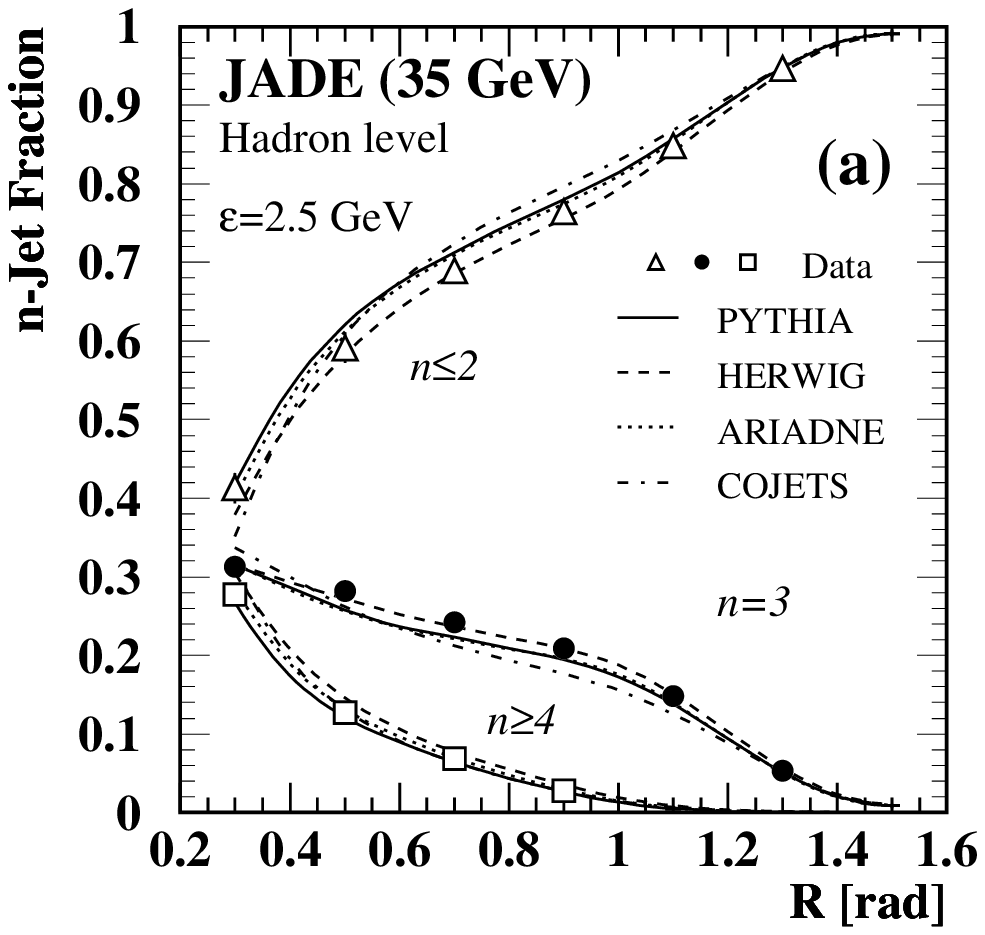,width=\conewid}\hspace*{\coneint}
\epsfig{file=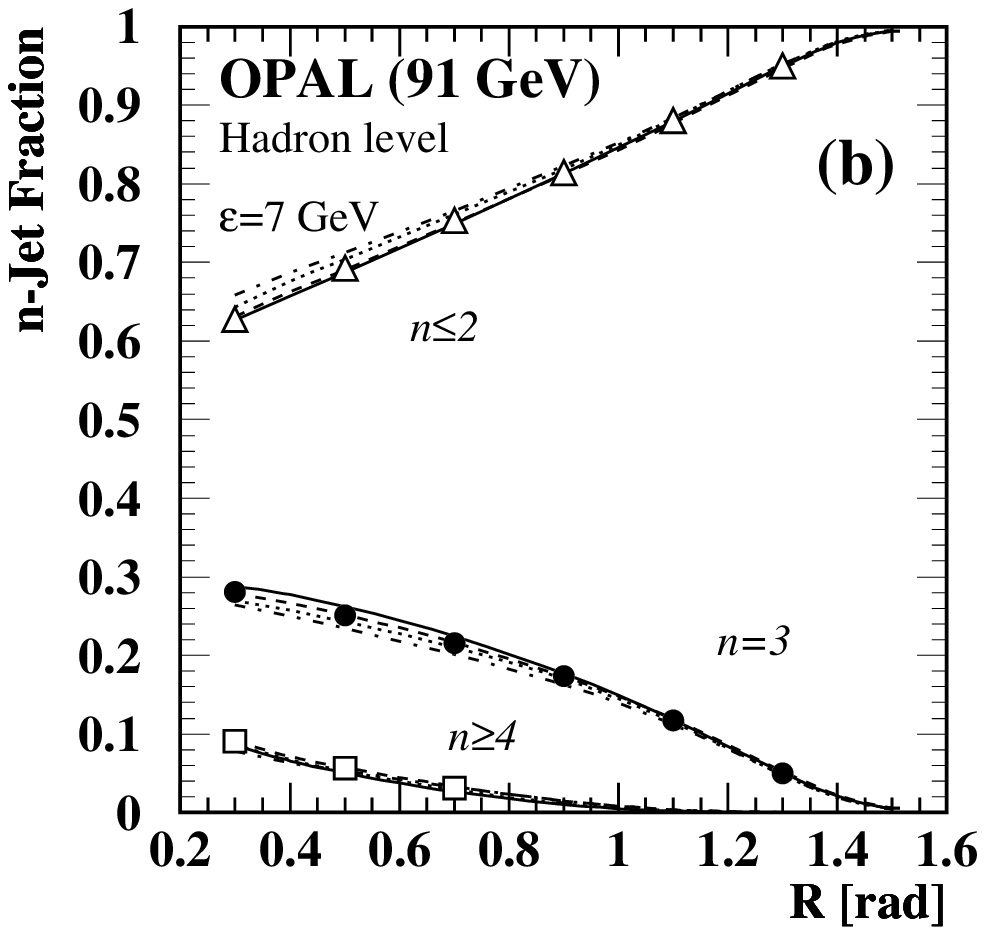,width=\conewid}\hspace*{\coneint}
\epsfig{file=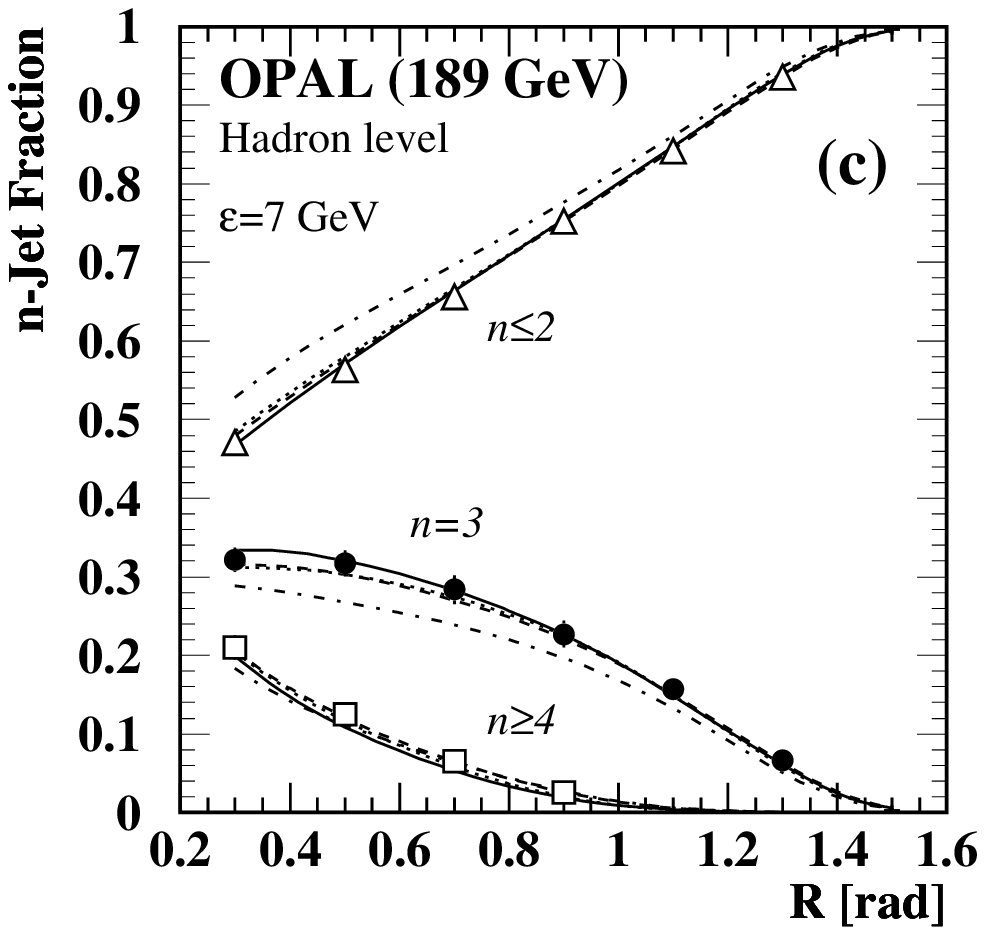,width=\conewid}\hspace*{\coneint}
}

\vskip\conemid

\centerline{
\hskip \coneleft
\epsfig{file=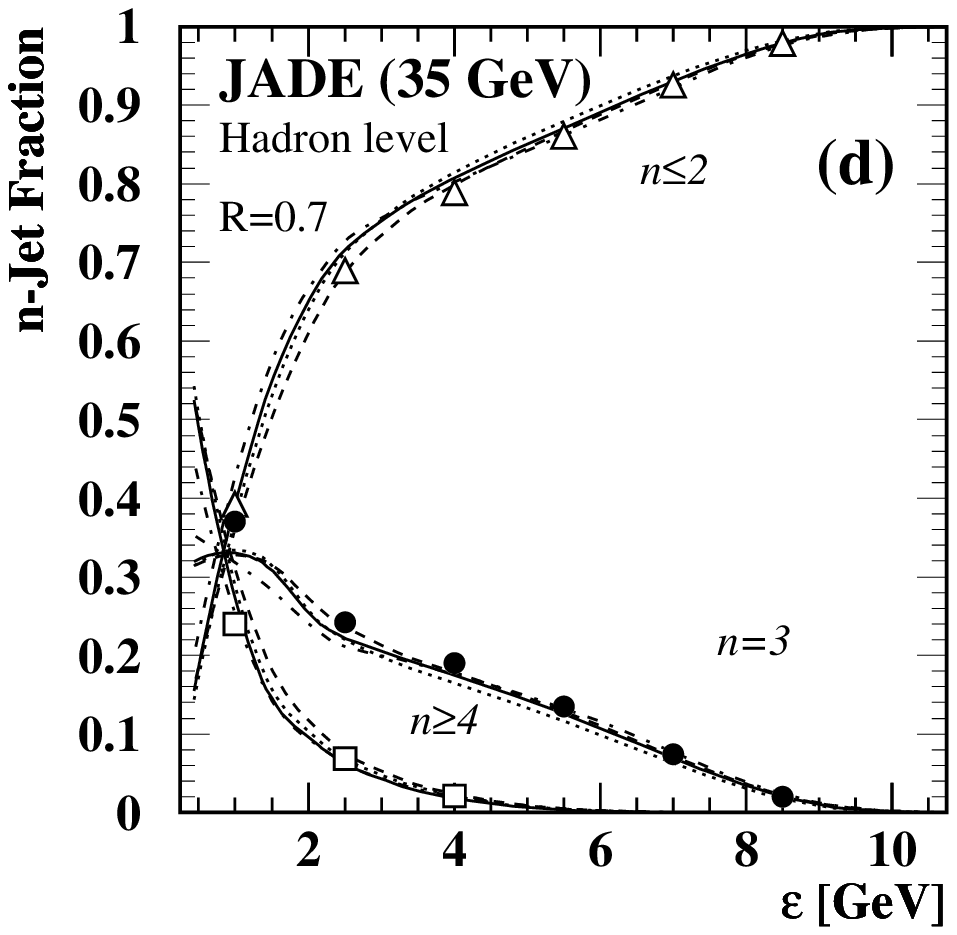,width=\conewid}\hspace*{\coneint}
\epsfig{file=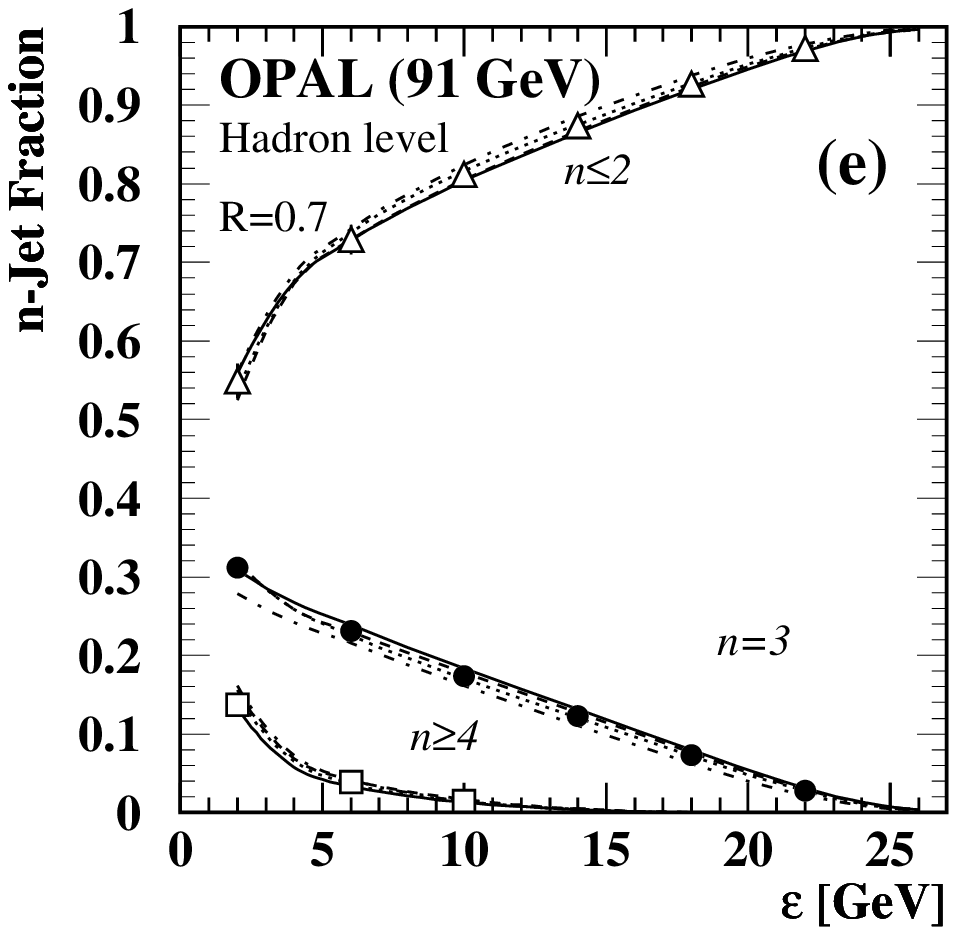,width=\conewid}\hspace*{\coneint}
\epsfig{file=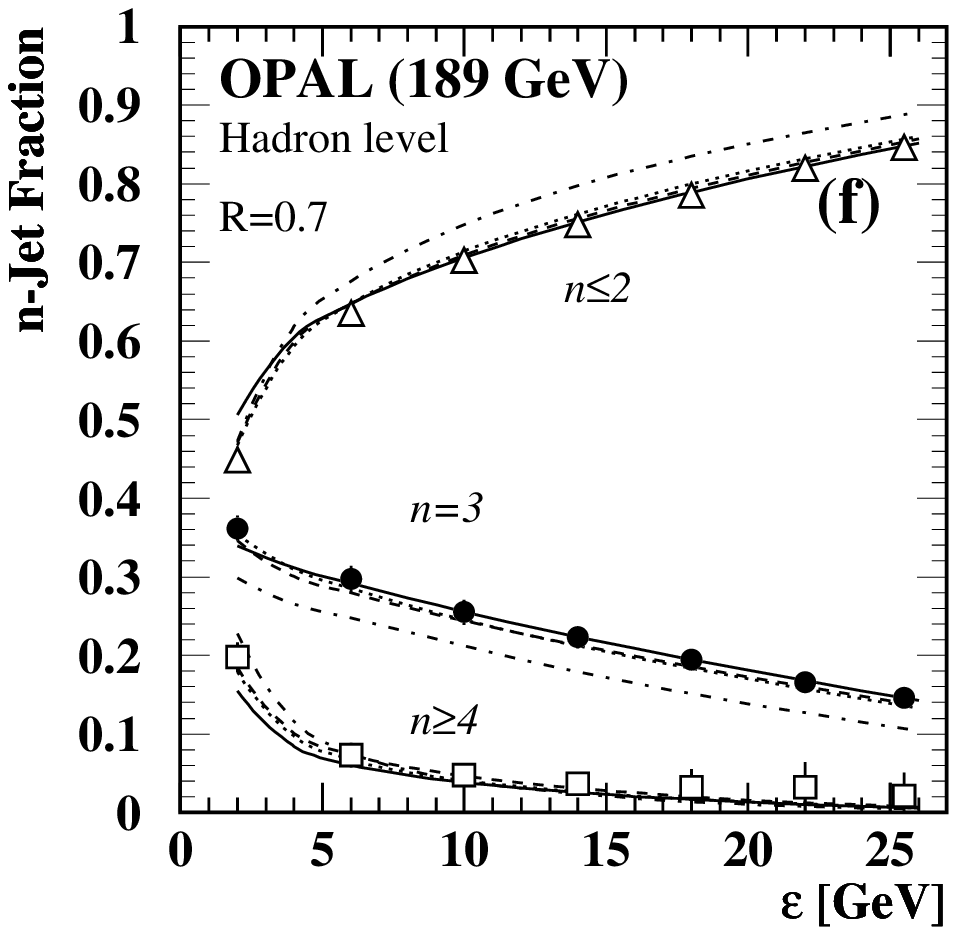,width=\conewid}\hspace*{\coneint}
}

\vskip \conedown

\caption[Jet fractions (cone algorithm) at $\protect\sqrt{s}$=35, 91 and 189 GeV]
{The fractions of events with 2 or less, 3 and more 
than 3 jets as obtained by the cone algorithm are plotted in the upper row
versus the cone half angle $R$
at 35, 91 and 189 
GeV, the jet energy cut-off $\epsilon$ being fixed at 2.5 GeV for $\sqrt{s}$=35 GeV
or 7 GeV otherwise. The lower plots show
the corresponding fractions versus $\epsilon$ with $R$ kept fixed at 0.7 rad.
Data and Monte Carlo predictions are presented in the same form
as in figures \ref{e0ratefig} through \ref{cratefig}.
\label{coneratefig}
}

}
\end{minipage}

\end{sideways}

\end{figure}

Fig.~\ref{coneratefig} displays
$n$-jet fractions obtained with a cone jet finder described in \cite{opalcone}, which 
reconstructs jets within cones of half angle $R$ having a minimum energy $\epsilon$.
Because of the explicit energy cut inherent in the algorithm, results of
the jet finding will depend strongly on the total energy $E_{vis}$ 
of the input particles.
In order to
remove this dependence,
the cone algorithm is run with the {\it scaled} energy
cut $\epsilon'\equiv\epsilon\cdot E_{vis}/\sqrt{s}$.
In the upper part of the figure, jet fractions are plotted
versus $R$ for $n\le2$, $n=3$ and $n\ge4$ at a fixed
minimum jet energy $\epsilon$. A value of $\epsilon=7$ GeV is chosen for
$\sqrt{s}=91$ GeV and above. At $\sqrt{s}=35$ GeV and 44 GeV, where typical
jet energies are considerably lower, $\epsilon$ is set to 2.5 GeV and 3 GeV.
The same jet fractions are shown against $\epsilon$ in the lower part of the
figures, where $R$ is kept fixed at 0.7 rad.\par
As in the case of the clustering algorithms, the predictions of all Monte
Carlo programs except that of COJETS are similar for all c.m.s. energies.
Again, COJETS deviates from them and from the data above
$\sqrt{s}=91$ GeV.

\subsubsection{{\boldmath$y_n$} Distributions and Differential {\boldmath$n$}-Jet Fractions}
\label{ydistlab}
In the context of the JADE and Durham schemes,
we shall denote by $y_n$ the value of $y_{cut}$ at which a
particular event switches from a $n$-jet to an $n+1$-jet configuration\footnote{Strictly this definition is only reasonable provided that
the $y_n$ fall off monotonically with $n$ which is 
not always the case for the resolution measures considered here.}.
Any one of the quantities $y_n$ can be
regarded as an event shape variable. We write differential distributions in $y_n$ as
\mathlab{Y_n\equiv\frac{1}{\sigma}\frac{\mbox{d}\sigma}{\mbox{d}y_n},}{yndef}
where $\sigma$ denotes the total hadronic cross section. In comparisons with theory, statistical
correlations between the values at different parameter settings have
to be taken into account. Such correlations are certainly present in $n$-jet fractions since
the measurements at neighboring parameter values contain partially the same events.
In this respect, the differential distributions $Y_n$ 
are more convenient
quantities
since they contain each event
only once. In the middle row of 
Fig.~\ref{dratefig}
we present distributions $Y_n$ as obtained from the Durham scheme for 
$n=2$ through \nolinebreak 4.\par
The Cambridge scheme employs a more complicated procedure for the
jet reconstruction than the JADE and Durham schemes, involving two
distinct resolution measures.
There is therefore no counterpart of the quantities $y_n$ in the 
Cambridge scheme which would allow the interpretation described above 
(see e.g. \cite{cambopal,jetcomp}). We show instead, in the middle row of Fig.~\ref{cratefig}, the corresponding ``differential
$n$-jet fractions,''
\mathlab{D_n\equiv\frac{1}{\sigma}\frac{\mbox{d}\sigma_n}{\mbox{d}y_{cut}},}{dndef}
with $\sigma_n$ being the cross
section for the production of $n$ jets
determined from
explicit binwise differentiation of the jet fractions. 
The relation $D_n=Y_n-Y_{n-1}$, and in particular $D_2=Y_2$, holds for
conventional cluster algorithms like the JADE or Durham scheme, as long
as the condition specified in footnote $^1$ is satisfied.\par
Both for the $y_n$ distributions and the differential jet fractions,
corrections and error calculation were carried out independently of the 
measured jet fractions $R_n$.
All plots show that the Monte Carlo predictions are almost 
indistinguishable and describe the data well except for COJETS. 
None of the models is
ruled out by the data on the basis of these quantities.
There are scaling violations of the differential jet fractions as there
were in the jet fractions.

\subsubsection{Mean Jet Multiplicities}
\label{jetmeanlab}
\begin{figure}[!t]
\vskip \coneuup

\vspace*{\coneup}

\centerline{
\hskip \coneleft
\epsfig{file=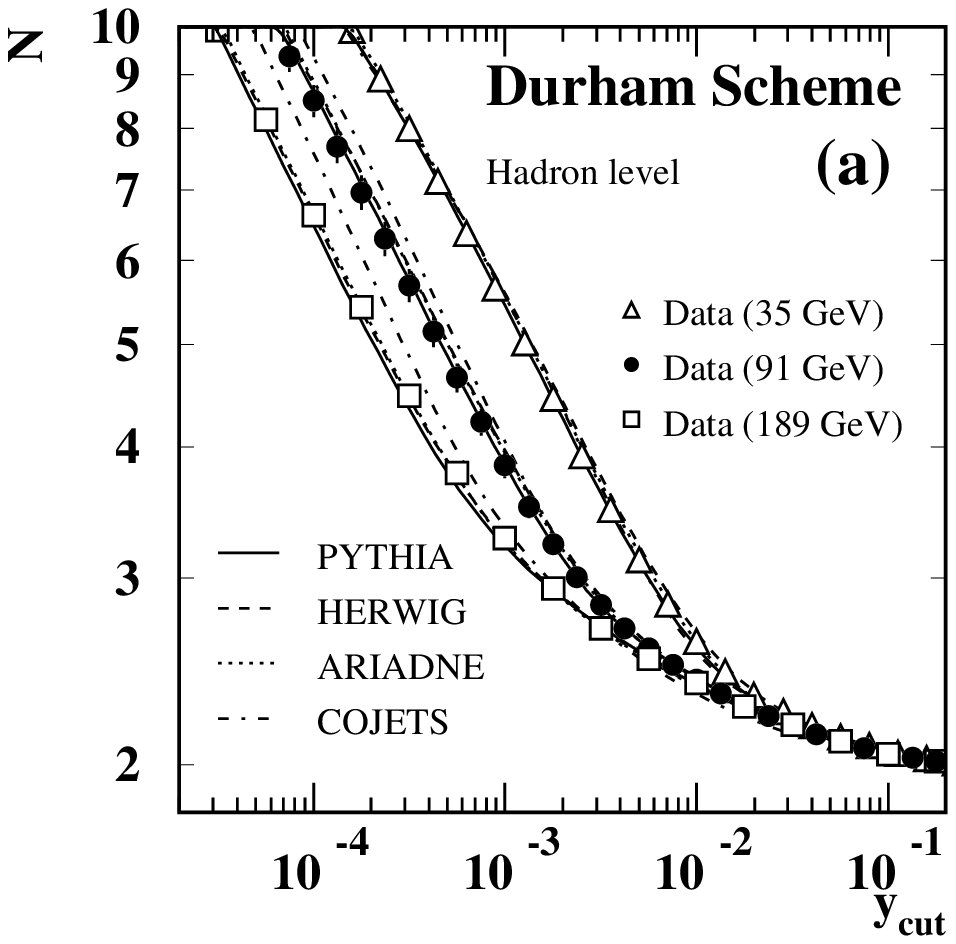,width=\conewid}\hspace*{\coneint}
\epsfig{file=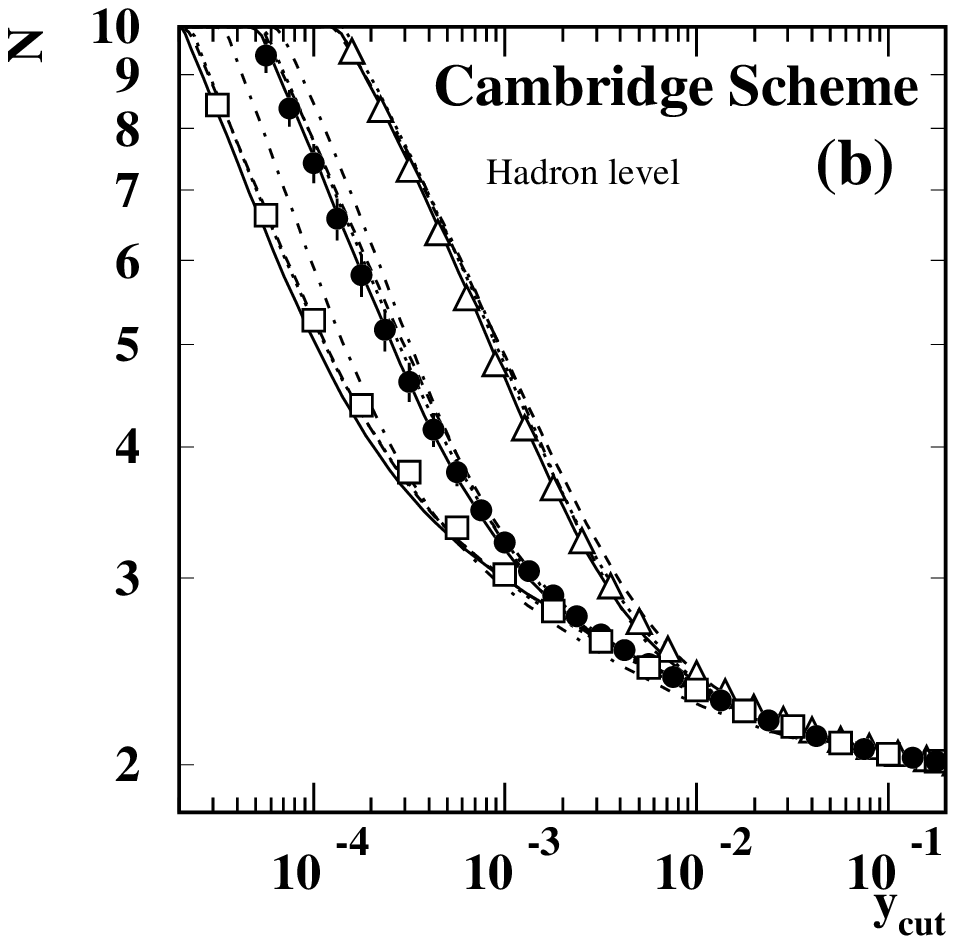,width=\conewid}\hspace*{\coneint}
}

\vskip \conedown

\caption[Mean jet multiplicities (Durham and Cambridge schemes) at $\protect\sqrt{s}$=35, 91 and 189 GeV]
{The mean jet multiplicities as obtained by the Durham (a) and Cambridge 
schemes (b) are plotted versus
$y_{cut}$ at 35, 91 and 189 GeV. 
Data and Monte Carlo predictions are presented in the same form
as in the previous figures.
\label{jetmeanfig}
}

\end{figure}

Another relevant quantity in the context of QCD tests is the 
mean jet multiplicity $N$ defined by
\mathlab{N\equiv\frac{1}{\sigma}\sum_n n\sigma_n,}{ndef}
for which various
theoretical calculations
exist. Measurements of $N$ versus $y_{cut}$ are presented for the Durham and
Cambridge schemes
in Fig.~\ref{jetmeanfig}. The most complete theoretical predictions for
this observable exist for these two schemes 
(cf. Appendix~\ref{matchtheolab}).\par
The decrease of the average number of jets at given $y_{cut}$ 
with rising c.m.s. energy, as 
predicted by the energy-dependence of the strong coupling, is clearly visible.
Differences between the Durham and Cambridge results
are predicted by all
Monte Carlo programs and confirmed by the data.
At and above $\sqrt{s}=133$ GeV, the Durham scheme resolves 
systematically more jets than the Cambridge scheme.\par
All Monte Carlo predictions except that of
COJETS are almost identical and lie within the error bars of the measurements.
The COJETS prediction overshoots the data at $\sqrt{s}=91$ GeV and
higher energies in regions of low $y_{cut}$, i.e. in regions of large
jet multiplicities, consistent with the explanation suggested in the previous
section and with OPAL
results from the analysis of charged particle based quantities 
\cite{opal161,opal133}. 

\subsubsection{Mean values of $y_n$}
\begin{figure}[tbhp]
\vskip \momuup

\vskip \momup

\centerline{
\hspace*{8cm}
\epsfig{file=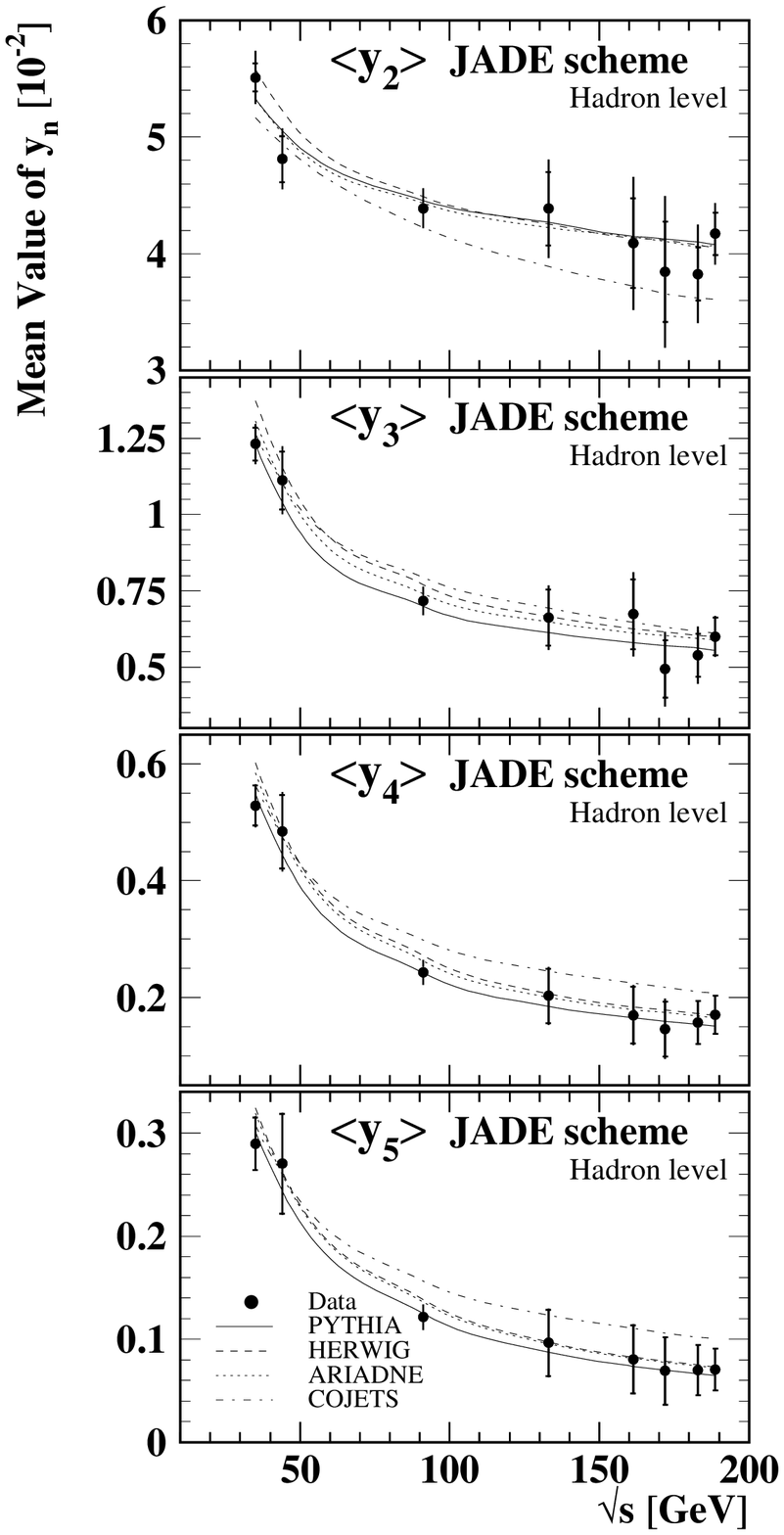,width=18cm}
\hspace*{-10cm}
\epsfig{file=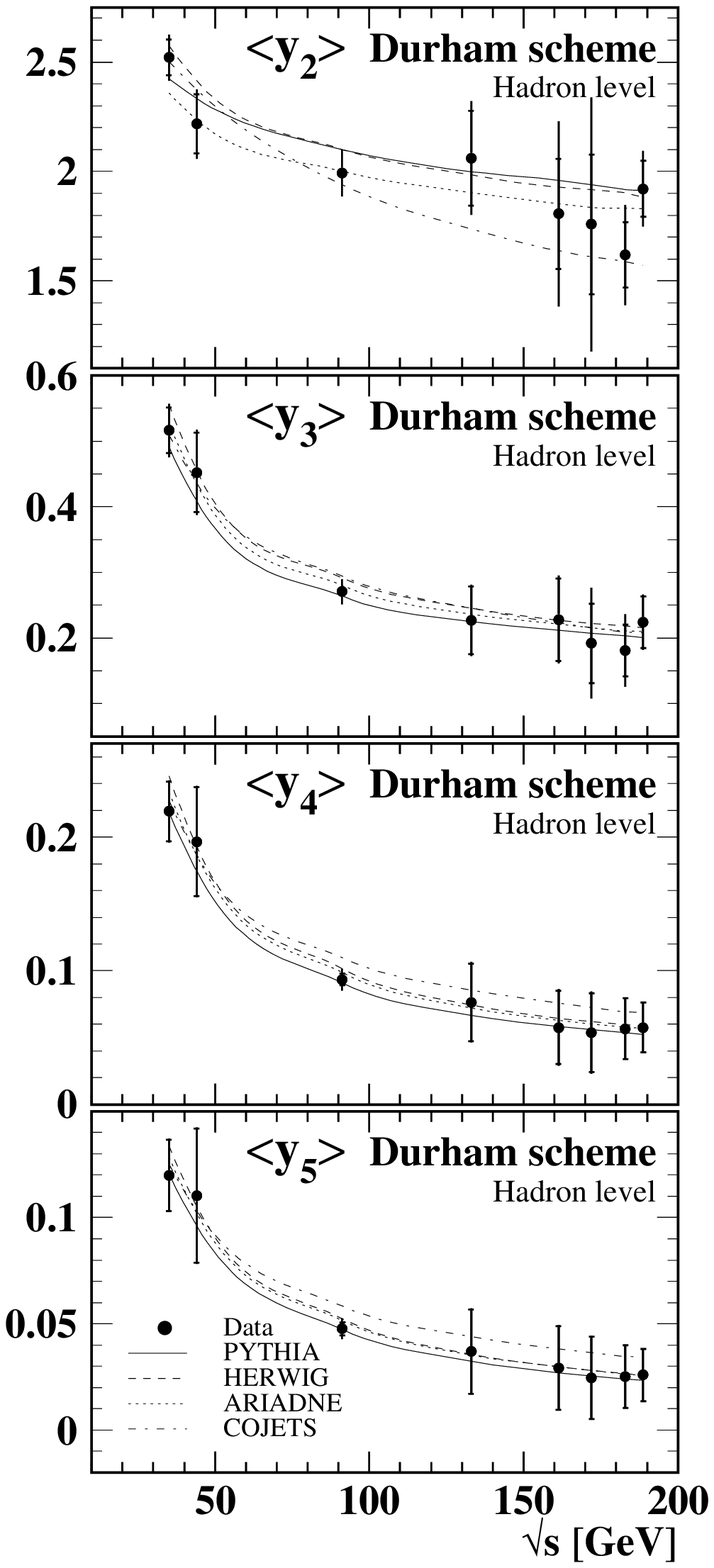,width=18cm}
}

\vspace*{0cm}

\caption[Mean values of $y_n$ vs. $\protect\sqrt{s}$ (JADE and Durham schemes)]
{The average of the observables $y_n$ 
is plotted versus the c.m.s. energy
$\sqrt{s}$ for 
the JADE (left) and Durham schemes (right). 
Data points are the measurements by the JADE and OPAL 
experiments. The error bars represent the total errors. Small 
horizontal lines indicate the size of the statistical errors.
The curves are spline functions connecting the Monte Carlo predictions at the
single c.m.s. energies.
\label{momfiglab}}

\end{figure}

Finally, we consider the mean values of the $y_n$ distributions given by
\mathlab{\langle y_n\rangle\equiv\frac{1}{\sigma}\int y_n\frac{\mbox{d}\sigma}{\mbox{d}y_n}\mbox{d}y_n}{momdef}
Measurements for $n=2$ through 5 for the JADE and Durham schemes are plotted 
in Fig.~\ref{momfiglab}, against the c.m.s. energy.
Hadron level predictions of PYTHIA, HERWIG, ARIADNE and COJETS at each of
the eight c.m.s. energies are also shown.
To facilitate
qualitative comparisons, the respective predicted values at each c.m.s.
energy are connected by spline
functions to guide the eye. 
All generators are found in almost equal agreement with the data, except for
COJETS.\par
Higher moments of the observables $y_n$ have also been investigated. They
have large uncertainties and are not presented.

\subsubsection{Summary}
The measurements of jet-multiplicity related observables show 
scaling violations. PYTHIA, HERWIG and ARIADNE describe the data well, but 
COJETS predicts
too many jets at and above $\sqrt{s}=91$ GeV.

\section{Tests of Quantum Chromodynamics}
\label{qcdtestslab}

\subsection{Procedure for {\boldmath$\alpha_s$} Determinations}
\label{fixasfitslab}

\subsubsection{General Description}
\label{genfixlab}
All jet-multiplicity based quantities considered in this analysis 
can be expressed as a
power series in the
strong coupling strength
$\alpha_s$, where the coefficients of the powers of $\alpha_s$ 
depend on the 
observable.
The approximate 
methods used here 
for the calculation of such series, pure ${\cal O}(\alpha_s^2)$ predictions
and matched ${\cal O}(\alpha_s^2)$ and next-to-leading logarithmic
predictions, are given in
Appendix~\ref{theocalclab}.
The coupling $\alpha_s$ will be regarded as unknown quantity, and the
tests of perturbative QCD will consist of $\chi^2$ fits of the predictions
to the measurements over ranges in $y_{cut}$
with $\alpha_s$ being fitted. These result in a determination
of $\alpha_s$ at each c.m.s. energy.\par
All fits are performed at the hadron level, using only
the statistical errors for the calculation of $\chi^2$. The error
returned by the fit, i.e. the amount by which the fitted parameters
may be varied without increasing $\chi^2$ by more than 1,
defines the statistical errors of the fit results.
Statistical correlations between bins in $y_{cut}$ are taken into
account in the definition of $\chi^2$. At 
c.m.s. energies where the statistics are sufficient (i.e. 
$\sqrt{s}$=35 GeV, 44 GeV and 91 GeV) we determine the correlation
matrix by
subdividing the data sample, and the
corresponding Monte Carlo sample used for the correction from detector
to hadron level (see section \ref{detcorrlab}), into 
$N$ independent subsamples and carrying out the entire
analysis for each subsample. We choose $N=30$ at $\sqrt{s}=35$ GeV, 
$N=15$ at $\sqrt{s}=44$ GeV and $N=80$ at $\sqrt{s}=91$ GeV.
At the higher c.m.s. energies we determine the correlations
from Monte Carlo subsamples.
At $\sqrt{s}=133$, 161 and 172 GeV, 
30 subsamples are used. At
the two higher c.m.s. energies, we use 50 subsamples each.\par
For the fits, we use only those observables for which there exist 
matched predictions of ${\cal O}(\alpha_s^2)$ and NLLA calculations, 
i.e. the 2-jet fractions and the mean jet multiplicities as measured
using the Durham and Cambridge schemes (see Appendix~\ref{theocalclab}). 
Fits are performed to 
the {\it differential} 2-jet fraction $D_2$ rather than the 2-jet
fraction itself. 
Throughout the remainder of this chapter,
the fitted variables shall be denoted $D_2^D$, $D_2^C$, $N^D$ and $N^C$
where upper indices $D$ and $C$ stand for the ``Durham'' and
``Cambridge'' scheme.
One aim of the analysis is
a comparison between the different types of calculations. The functional 
expressions to be used in the fits are given for the various matching schemes in
(\ref{bla}), (\ref{blaa}), (\ref{blaaa}) and
(\ref{blaaaa}) of Appendix~\ref{theocalclab}. 
Furthermore, fits of the pure ${\cal O}(\alpha_s^2)$ predictions
are carried out using (\ref{oas2r2}) and (\ref{oas2mean}). 
The parameter to
be varied is always $\alpha_s$ as appearing in the equations. The
renormalization scale factor $x_{\mu}\equiv\mu/\sqrt{s}$ 
is either kept fixed at $x_{\mu}=1$
or fitted simultaneously. Since the matched predictions are more complete
than pure second-order calculations, one may expect an increased need for
an adapted scale for the latter.

\subsubsection{Hadronization Effects}
\label{hadronizlab}
In order to study the influence of hadronization on the quantities to
be considered, we compare the predictions of PYTHIA 5.722, HERWIG 5.9
and ARIADNE 4.08 before and after the hadronization step. Each generator
represents a different model for either hadronization or the 
partonic state. The generated
event samples are the same as the ones used for the hadron level curves
in Sect. \ref{jetresultlab}.
The COJETS generator which uses independent fragmentation
is not considered since it was seen to describe the data badly in certain
kinematic regions.\par
By the ``parton level prediction,'' 
we understand quantities determined from the set
of particles emerging at the end of the parton shower generation, i.e.
immediately before the hadronization step, including possible final 
state photon radiation. 
The curves in the lower partitions in Figs.~\ref{asfitlab_35}, \ref{asfitlab_91} and \ref{asfitlab_189}
are the ratios $f$ of the parton level
over the hadron level predictions for the differential 2-jet fractions and
the mean jet multiplicities as a function of $y_{cut}$  at all 
c.m.s. energies and for the three generators. The correction factors 
for PYTHIA, HERWIG and ARIADNE are
shown, respectively, as solid, dashed and dotted lines.
As is expected, hadronization corrections
become notably smaller with rising c.m.s.
energy.\par
We 
transform the theoretical calculations to hadron level by dividing 
the calculation for each value of $y_{cut}$ by the factors $f$ shown
in the figures.
The result 
is then compared with the hadron
level measurement at the given $y_{cut}$. The PYTHIA generator will be
used for the quoted central results.

\subsubsection{Determination of Fit Ranges}
The ranges of $y_{cut}$ for the fits have to be chosen with 
care. 
A fundamental limitation is given by the kinematic region over
which the respective theoretical predictions can be assumed to be valid.
In particular, for pure ${\cal O}(\alpha_s^2)$ fits, the choice of the
fit range is limited to large $y_{cut}$ regions. In the fits presented
below, the fit range was adjusted separately for each type of calculation
used.\par
In order to
be as independent as 
possible of assumptions on the hadronization process, we 
require hadronization corrections to be less 
than 10\% in the case of the differential 2-jet rates and
less than
5\% in the case of the mean jet multiplicity and to be insensitive to changes
in hadronization parameters and models.
As can be seen from Figs.~\ref{asfitlab_35} to \ref{asfitlab_189}, these
conditions have to be loosened at lower energies and for observables obtained
with the Cambridge jet finder, in order to make $\alpha_s$ 
determinations possible
at all. The condition of small hadronization corrections
turns out to be the most stringent limitation of the fit ranges at low c.m.s. energies.
The size of the corrections for experimental effects is also taken into
account, but not considered as crucial because the simulation of the detectors
is believed to be rather reliable and, in most cases, these corrections
are clearly smaller than those for hadronization.\par
As a further limitation we demand that the value of $\chi^2/d.o.f.$ obtained in
the fit be not dominated by the contribution from a single bin at the boundary of the fit range.
Another necessary check concerns the stability of the fit results
under variations of the fit range.
The results can only be regarded as reliable if they do not
change significantly when the boundaries of the fit range are varied around
the chosen values.\par
Observing all conditions listed, we try to maximize the fit range in order
to give the most significant possible results.

\subsubsection{Determination of Systematic Errors}
\label{fitsystlab}
The determination of $\alpha_s$ is repeated with variations in the details
of the analysis procedure. For each variation, the absolute difference between
the obtained value for $\alpha_s$ and the central value is taken as the
systematic error from the respective source. All contributions are
added quadratically.\par
At all energies, the same systematic variations of the selection procedure
are performed as in the measurements of the jet quantities themselves (see
Sect. \ref{systlab}), and
the fit is repeated using only charged tracks. 
At the c.m.s. energies of 35 GeV, 44 GeV, 91 GeV and 189 GeV, one observes in
some cases a significant dependence of the resulting $\alpha_s$ value 
on the choice of the fit range. In order to estimate this uncertainty,
both the upper and the lower boundaries of the fit ranges are varied by
one bin in both directions, keeping the respective opposite boundary 
fixed.
All contributions mentioned, added in quadrature, define the total experimental
systematic error.\par
Uncertainties from the hadronization models used for the transformation
of the QCD prediction to hadron level are determined by varying the 
parameters and the Monte Carlo generators used.
In particular, the parameter $b$ appearing in the
Lund fragmentation function, used for the hadronization of u, d and s quarks by
the PYTHIA generator, 
and the width $\sigma_q$ of the 
transverse momentum distribution of the produced hadrons
with respect to the parent parton
are each varied in both
directions within the uncertainties allowed by 
the OPAL tunes \cite{opaljettune}.
At $\sqrt{s}=133$ GeV and below, the parameter $Q_0$ defining the end
of the parton shower cascade is also varied within its uncertainties around the
central value of 1.9 GeV
to 1.4 GeV and 2.4 GeV, 
yielding, on the average, correspondingly 
smaller and higher numbers of partons at the end of the
cascade. At the higher c.m.s. energies, these variations are replaced
by a more radical change to $Q_0=4$ GeV,
and the fits are repeated using HERWIG 5.9 and ARIADNE 4.08
as described in Sect. \ref{hadronizlab} instead of PYTHIA.\par
All theoretical predictions to be
used make the assumption of zero quark masses. 
In an attempt to estimate the effect of this, we repeat the fit
using a Monte Carlo sample containing only light
primary quarks (u, d, s and c) for the transformation to hadron level.
The quadratic sum of all contributions listed defines the overall
hadronization uncertainty of the result.\par
Finally, the uncertainty in the choice of the renormalization scale must be accounted for.
In the fits where the scale factor
$x_{\mu}$ is kept fixed at 1 for the central results,
the fit is repeated with $x_{\mu}=0.5$ and $x_{\mu}=2$. Usually,
the choice of a smaller scale will entail a decrease of the fitted $\alpha_s$,
while a larger scale will give larger values of $\alpha_s$. The different
deviations obtained from varying $x_{\mu}$ in both directions are then 
added asymmetrically to the hadronization error to yield the total
error from theory. In cases where both variations of $x_{\mu}$ let 
$\alpha_s$ change
in the same direction, the average of both deviations is taken and added as
symmetric error.

\subsection{QCD Tests at Fixed c.m.s. Energies}
\label{singtestslab}

\subsubsection{Fit Results}
\label{fixfitreslab}
\asfita{35}{\asfitatext{35}}{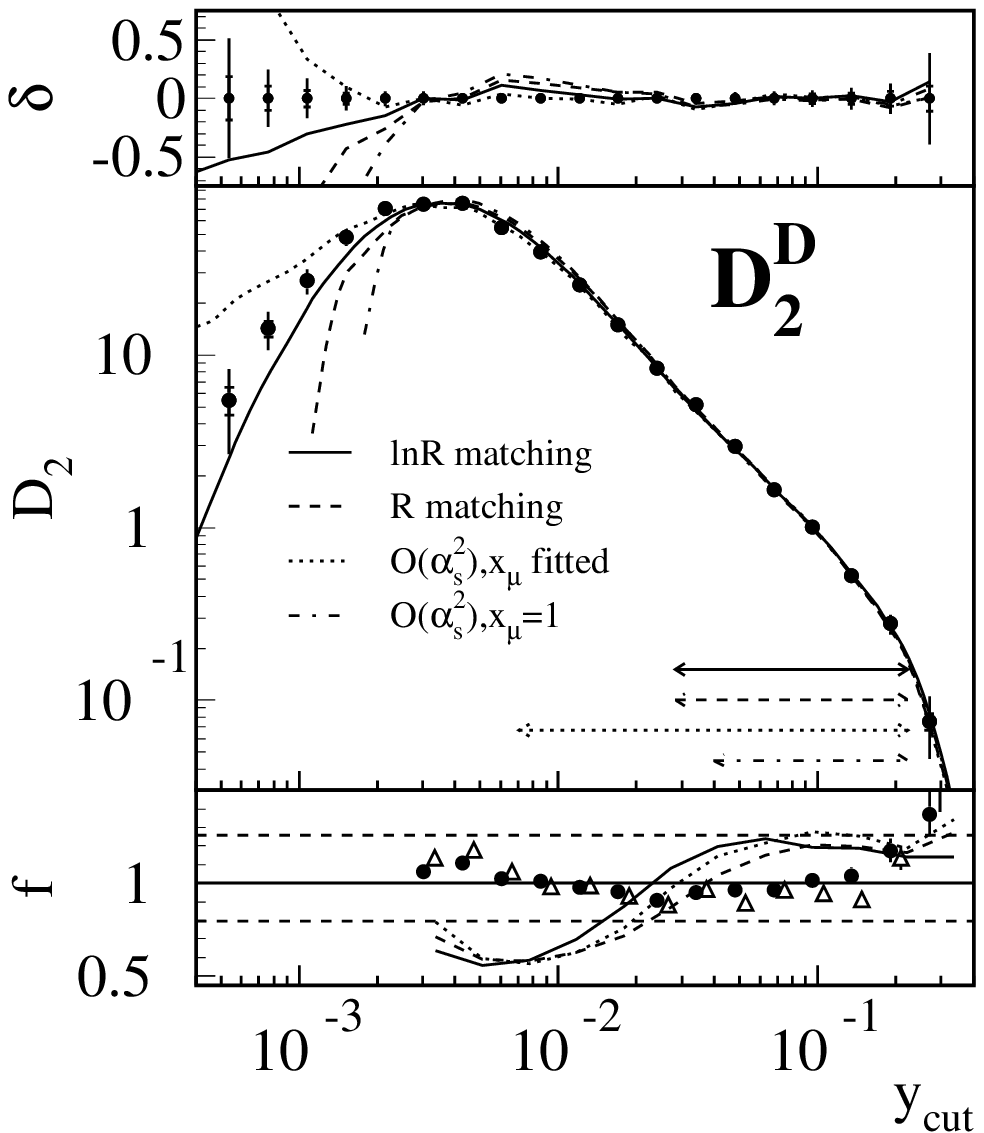}
\asfita{91}{\asfitatextt{91}}{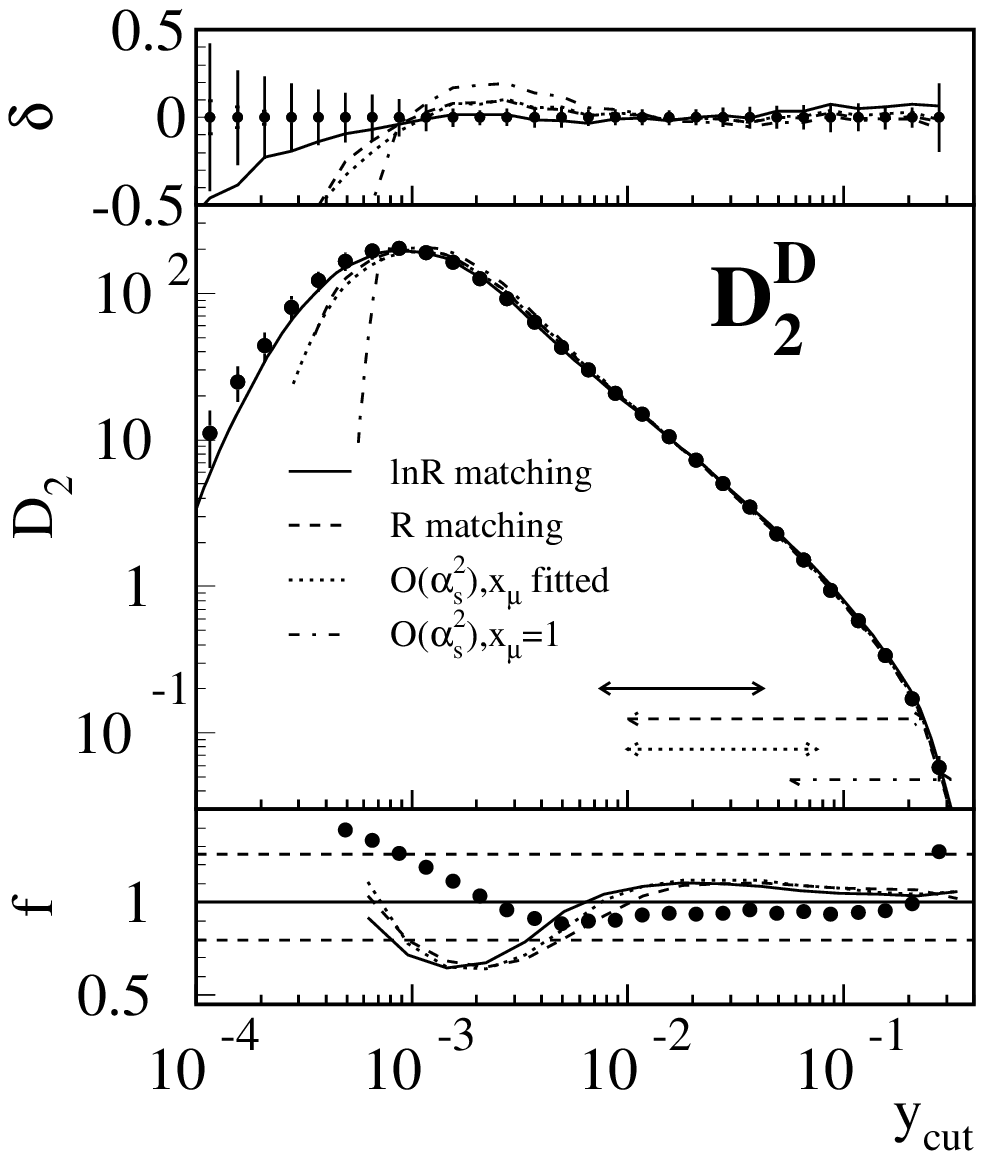}
\asfita{189}{\asfitatexttt{189}}{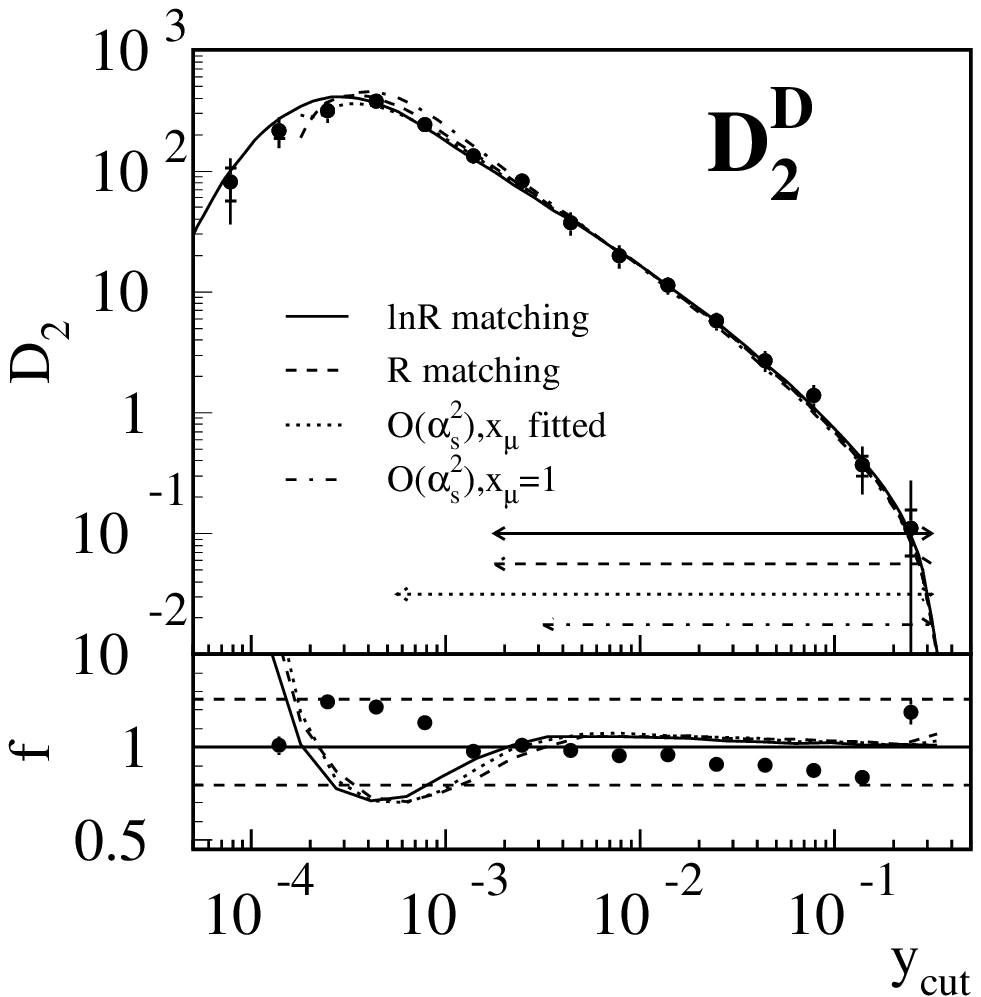}
Fits of the different matched calculations described in Appendix~\ref{matchtheolab}, i.e. the $\ln R$ matching, the $R$ matching and their modified
variants, as
well as ${\cal O}(\alpha_s^2)$ predictions have been performed for the
measured observables $D_2^D$, $D_2^C$, $N^D$ and $N^C$. In the case
of the ${\cal O}(\alpha_s^2)$ calculations, fits both with a fixed QCD
scale factor of 1 and with a variable scale $x_{\mu}$ were tried.
The fitted predictions of the various matching schemes and the
${\cal O}(\alpha_s^2)$ calculations for $\sqrt{s}=35$
GeV, 91 GeV and 189 GeV are shown in the central parts of
Figs.~\ref{asfitlab_35}, \ref{asfitlab_91} and \ref{asfitlab_189}
as smooth lines. Similar fits were done at all energies. At $\sqrt{s}=35$ GeV,
no stable fits with a fitted QCD scale could be obtained for observables
$N^D$ and $N^C$.
The small sections below each plot give the correction factors $f$ for
experimental and hadronization effects.
In the additional small section above the plots, the differences between
the fitted predictions and the data, denoted as $\delta$, is plotted, 
normalized to the measured value.
The relative total and statistical errors on the
data are shown along the $\delta=0$ line.\par
The fitted values for $\alpha_s$ are summarized in Figs.~\ref{pr299_10a_fig}
and \ref{pr299_11a_fig} with their total errors (outer bars) and
purely experimental errors (inner bars).
The $\chi^2/d.o.f.$ is given for each result to the right of the
plots.
For all observables the same
systematic pattern of the results from each calculation type
used repeats 
at all c.m.s. energies. Furthermore, 
the theoretical uncertainties in the predictions resulting
from the choice of the calculation type are at least as large as the
experimental errors, in particular at $\sqrt{s}=35$ GeV, 91 GeV, 183 GeV and
189 GeV. Further refinements of the experimental procedure and increase
of statistics will therefore not lead to significantly more precise
results for $\alpha_s$ as long as theoretical uncertainties can not
be reduced.\par
Fits of ${\cal O}(\alpha_s^2)$ calculations
with a fixed QCD scale of 1 require a limitation of the fit range to
regions of large $y_{cut}$ in order to achieve acceptable values of
$\chi^2/d.o.f.$ The resulting statistical errors are correspondingly
large.
As can be seen from Fig.~\ref{pr299_10a_fig},
the 
results for $\alpha_s$ are generally high at lower
c.m.s. energies, while, at $\sqrt{s}\ge133$ GeV, they are similar to those
obtained with the $\ln R$ and modified $R$-matchings.\par
Allowing the scale factor $x_{\mu}$ to vary allows the fit range
to be extended to 
lower $y_{cut}$. In fact, a wider fit range is required in most cases
to obtain stability of the fits. 
At the lower c.m.s. energies of 35 GeV and 44 GeV, the fit ranges have
to be extended into regions where hadronization uncertainties become
large, resulting in comparatively large theoretical errors of the results.\par
In the case of the Durham scheme,
the fitted $\alpha_s$ values obtained using ${\cal O}(\alpha_s^2)$ predictions
with a fitted scale are systematically smaller than the results from fits with a constant $x_{\mu}$. In contrast, in the case of
the Cambridge scheme they are somewhat larger.
We conclude that a suitable choice of $x_{\mu}$ is able to improve 
the agreement of the predictions with the measurements.
\matchfig{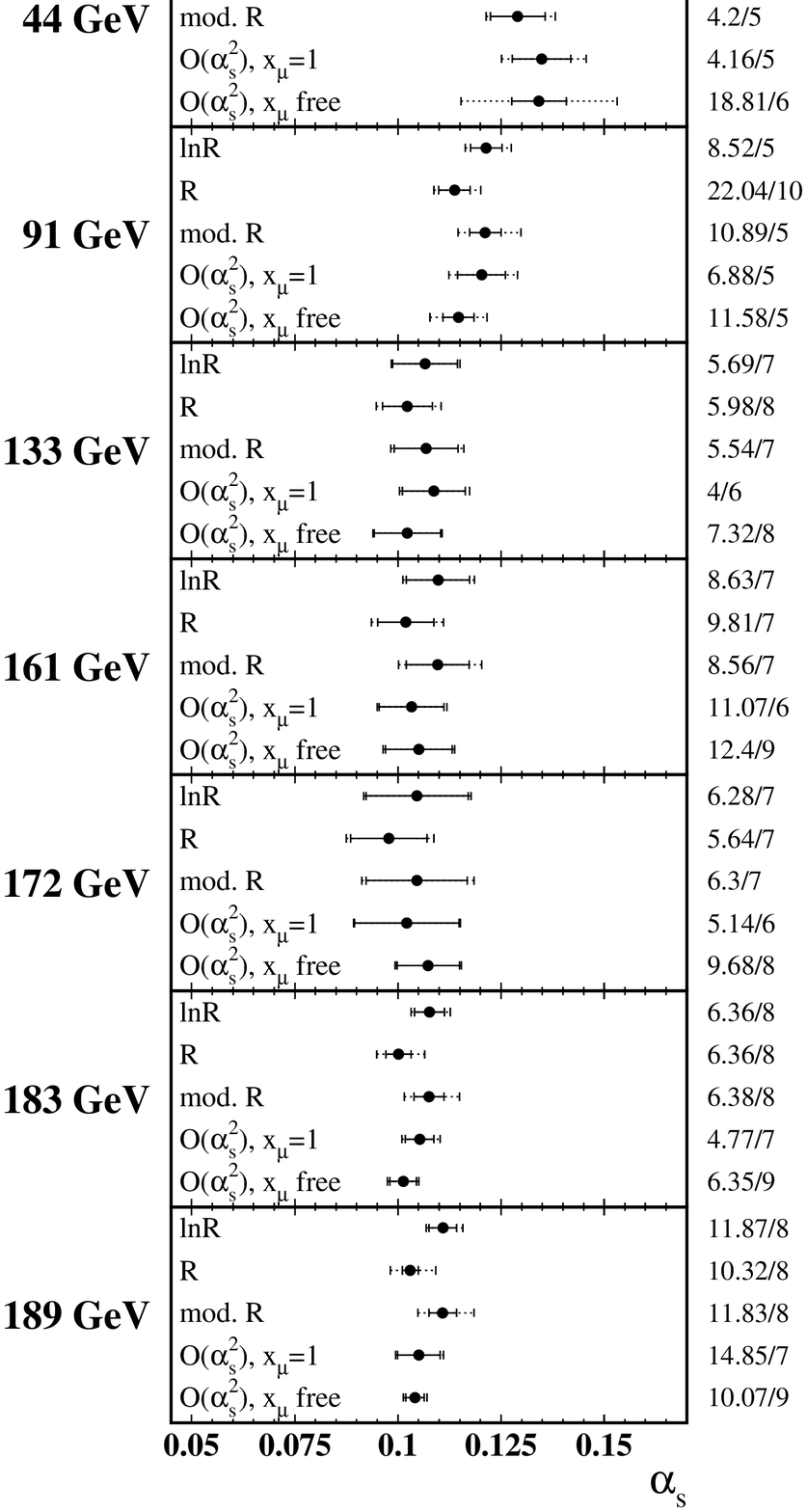}{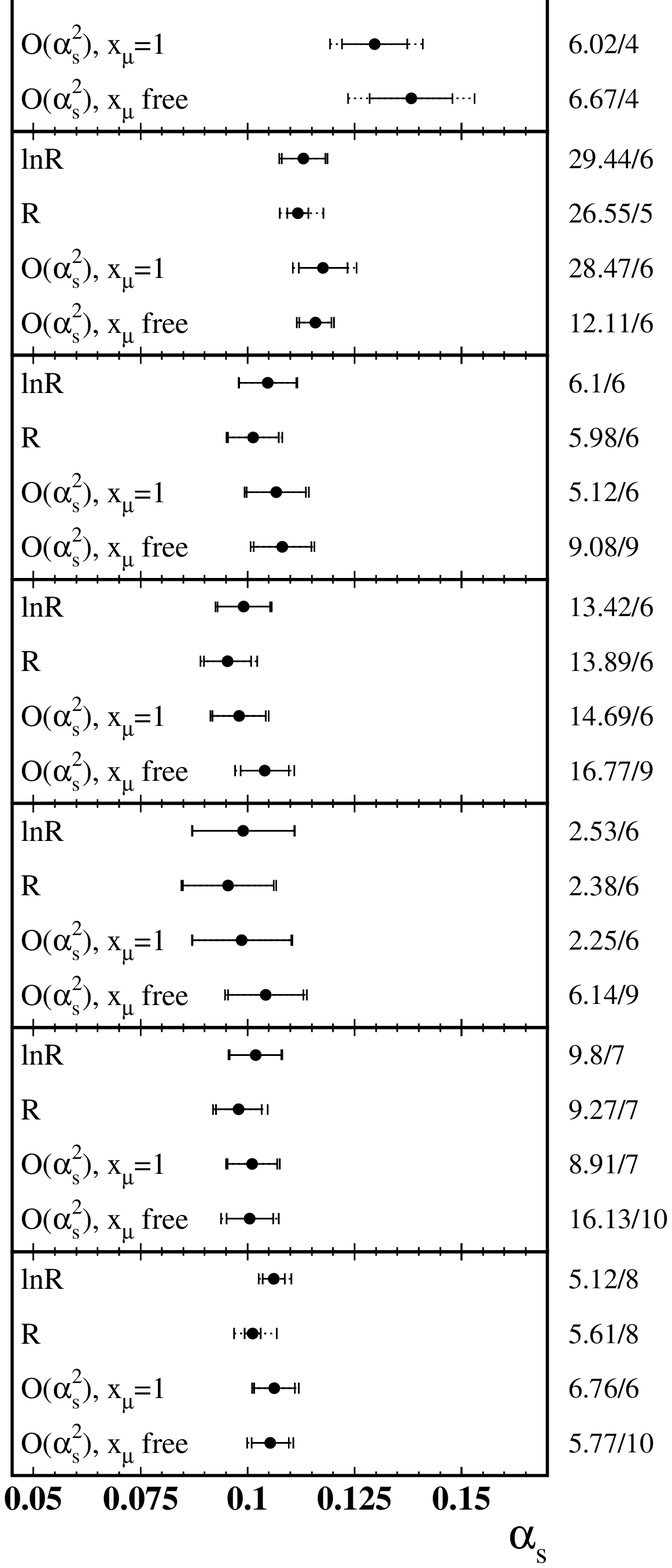}{$D_2^D$}{$D_2^C$}{-0.5}{\matchtext{$D_2^D$ and $D_2^C$}}\matchfig{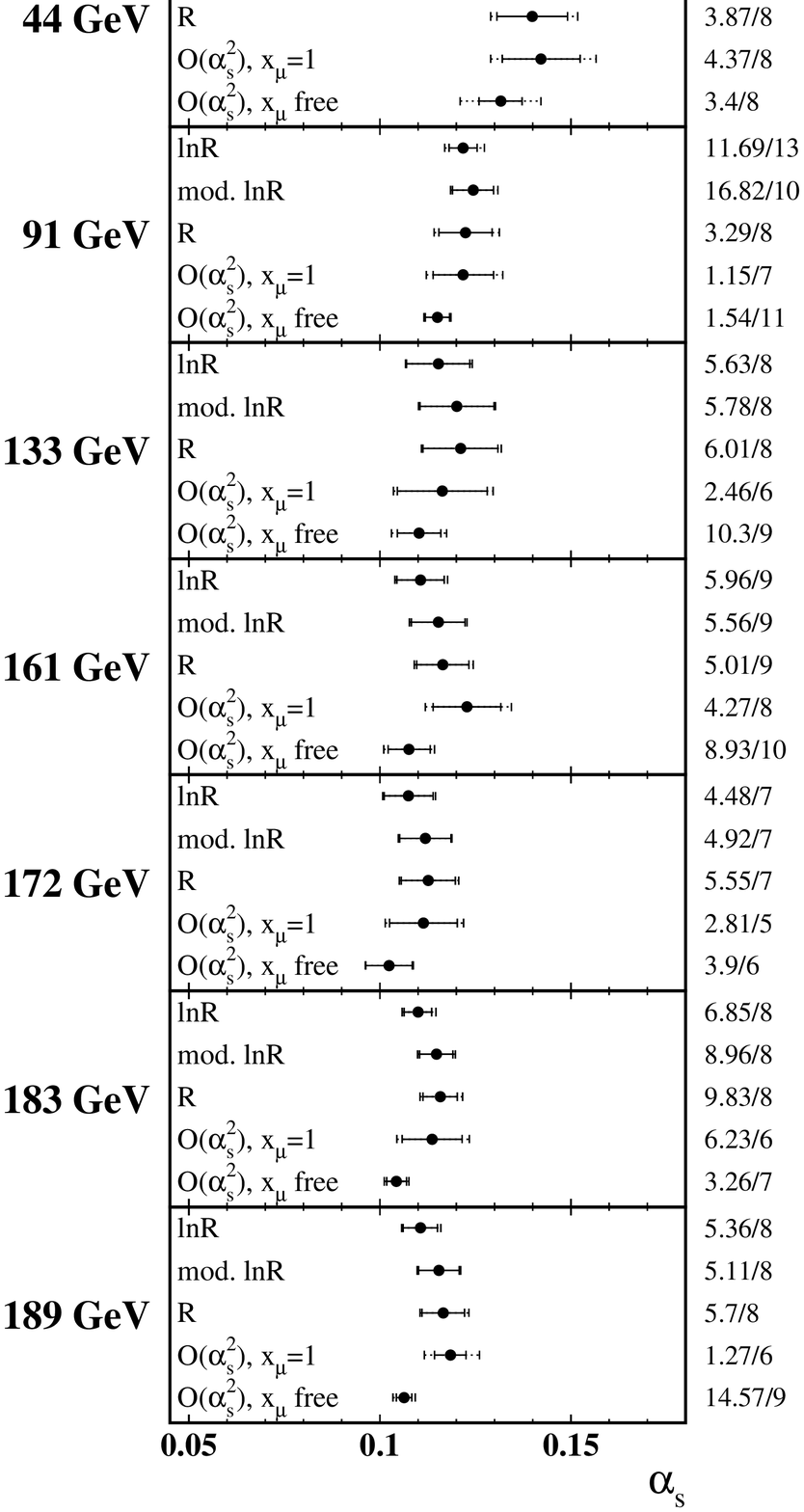}{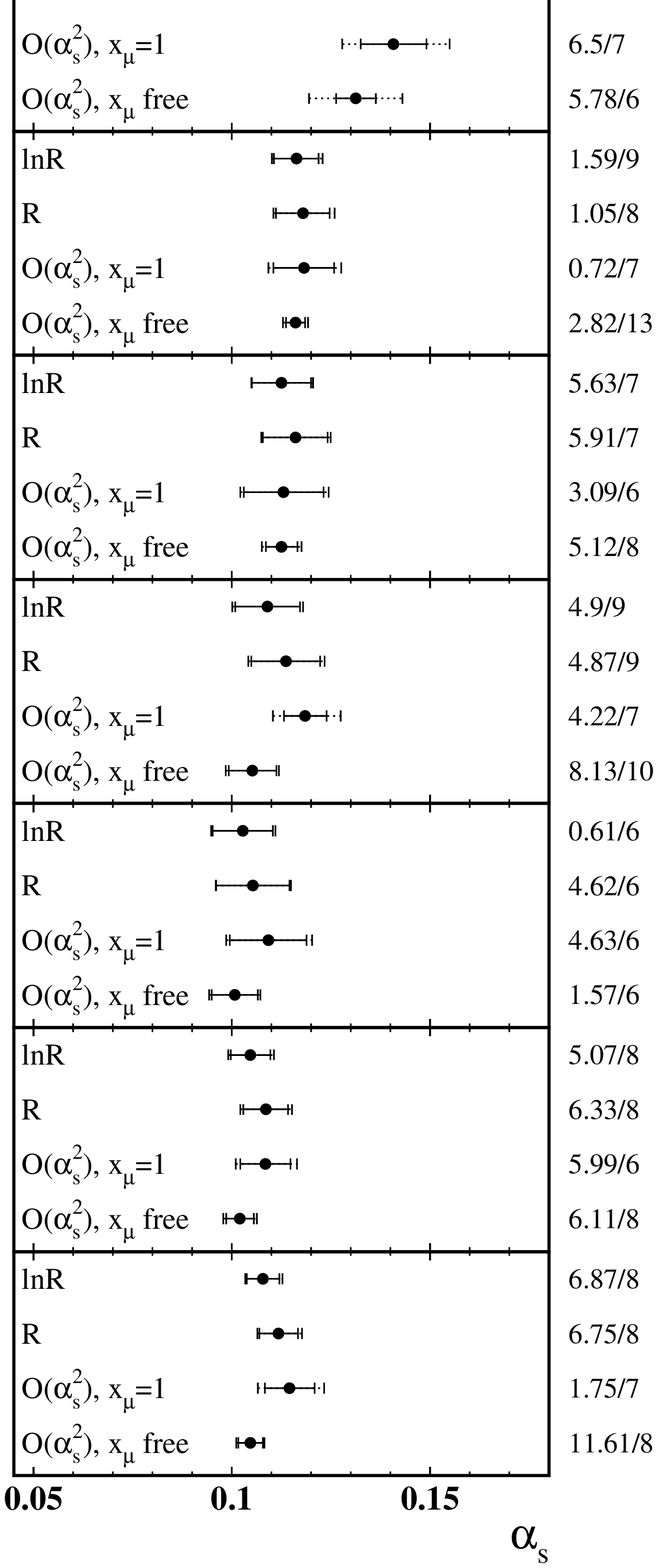}{$N^D$}{$N^C$}{-0.5}{\matchtextt{$N^D$ and $N^C$}}\par
\vspace*{-0.3cm}
\subsubsubsection{a) Fit results for the 2-jet fractions}
The comparison of the quality of the fits (i.e. the $\chi^2/d.o.f.$)
does not indicate a clear preference for one of the matching schemes 
for either one of the variables. The normalized deviations $\delta$ 
also behave similarly for all matching schemes. Only in the case of
the Cambridge scheme at $\sqrt{s}=$91 GeV, they differ somewhat
because rather distinct fit ranges had to be chosen for the different
schemes.\par
The results obtained with
the $\ln R$ and the modified $\ln R$-matching with $y_{max}=1/3$ turn
out to be virtually identical up to at least the third decimal place
of the extracted value for $\alpha_s$, which 
is in accordance with the findings of a previous OPAL 
analysis \cite{opal91}. The fit results are therefore not shown 
separately, but only for the $\ln R$-matching. The
results of the modified $R$-matching are also generally similar to those
of the $\ln R$-matching, but have slightly larger theoretical
errors. $R$-matching, however, leads to systematically low 
values of $\alpha_s$ for both jet algorithms. In fact, the $R$-matching
results are in most cases the lowest of all calculation types under
consideration. 
As is explained in Appendix~\ref{matchtheolab}, the $R$-matching may be
expected to describe the data less well than the modified variant because 
the term $G_{21}L$ is not exponentiated. The observed behaviour seems
to indicate that the inclusion of this term has more significance than
the choice between $\ln R$ and $R$-matching. We decide to follow
the practise of \cite{opal91} and use 
the $\ln R$-matching for the final results.
The difference plots $\delta$ also indicate that the $\ln R$ matching
provides a good description of the data over a wider range in $y_{cut}$ than both the $R$
matching and ${\cal O}(\alpha_s^2)$ calculations.\par
The stability of the $\ln R$-matching fits under variations of 
the fit range boundaries turns out to be generally good.\par
\subsubsubsection{b) Fit results for the mean jet multiplicities}
The fits of ${\cal O}(\alpha_s^2)$
predictions for $N^D$ and $N^C$
with the QCD scale factor fixed at $x_{\mu}=1$ describe the data significantly
less well in the regions of low $y_{cut}$ than all the other 
fits, requiring a corresponding limitation of the fit ranges.
Allowing $x_{\mu}$ to vary results in an improved agreement
between data and predictions. 
Only at $\sqrt{s}=35$ GeV, the ${\cal O}(\alpha_s^2)$
calculations with $x_{\mu}=1$ seem to follow the data well even
in low $y_{cut}$ regions. A closer inspection reveals, however, that
the agreement at higher $y_{cut}$, i.e. within the selected fit range, 
is worse.\par
The necessity of keeping hadronization uncertainties small limits the
fit ranges rather strongly and leads to somewhat larger statistical
errors than in the case of the 2-jet rates. Fig.~\ref{pr299_11a_fig}
shows that fits with a free QCD scale always result in values of $\alpha_s$
which are smaller than those from any other fits, while
those with $x_{\mu}$ fixed to 1 yield usually the largest values.\par
The agreement between data and theory, according to the normalized 
differences $\delta$, is again rather similar for
all matching schemes. All types of predictions display the 
property of overshooting the data below some $y_{cut}$, i.e. 
predicting too many jets.\par
For both jet algorithms, the fit results for $\alpha_s$ turn out to be 
less dependent on the choice of the 
matching scheme than is the case for the 2-jet rates, and the
corresponding uncertainties are currently much smaller than the experimental
errors. The modified $\ln R$-matching of the Durham scheme contains
the additional term $L H_{21}$ in the exponent (see Appendix~\ref{matchtheolab}) 
and should therefore be
a better description of the measurements than the $\ln R$-matching.
As can be seen from Fig.~\ref{pr299_11a_fig}, the results obtained by the modified
$\ln R$-matching, the $R$-matching and 
the $\ln R$-matching are always quite similar.
Considering the small dependence on the calculation
type mentioned above, one could expect to obtain rather precise
determinations of $\alpha_s$ from these variables if better hadronization
models were available which would allow to extend the fit range towards
lower $y_{cut}$.\par
The mean jet multiplicities exhibit a stronger dependence on the choice
of the
boundaries of the fit ranges than the 2-jet rates, in particular at
small c.m.s. energies.
The fit range dependence (and therefore also the corresponding systematic
error) is larger for the $R$-matching and the modified $\ln{R}$-matching
than for $\ln{R}$-matching. We conclude that the latter is a more
appropriate description, valid over a wider range in $y_{cut}$, and decide
to use it for the main quoted fit results.

\subsubsection{Investigation of the renormalization scale dependence}
\xmudepfigd
The fits of the ${\cal O}(\alpha_s^2)$ calculations show 
clearly that the data are better described
by fixed order QCD predictions if small renormalization scales are
used, with $x_{\mu}$ typically between 0.15 and 0.2 in the case of the 2-jet fractions and
between 0.03 and 0.1 in the case of the mean jet multiplicities. 
Similar results
have been obtained in many other analyses involving event shape variables
(e.g. \cite{ourjade,opal91,xmudelphi}). 
The value of the best scale depends strongly on the observable.
Although the fitted value of $x_{\mu}$ does not bear any physical
significance, its deviation from unity indicates the importance of
neglected higher order contributions in the fixed order calculation for
the given observable.\par
The inclusion of higher order terms in the matched predictions leads
one to expect that the dependence of these predictions
on the QCD scale factor $x_{\mu}$
will be reduced, as compared with the ${\cal O}(\alpha_s^2)$ calculations.
This has, in fact, been confirmed in \cite{ourjade,opal91}. As was done
in those publications, we have studied
the $x_{\mu}$ dependence of the fit results by performing fits
for $\alpha_s$ at various fixed values of $x_{\mu}$. Examples of the 
results
for $D_2^D$ and $N^D$ are shown
in Fig.~\ref{xmudep_lab} at $\sqrt{s}=35$, 91 and 189 GeV.
The solid lines demonstrate the behaviour of $\alpha_s$ 
as a function of $x_{\mu}$ for both matched and ${\cal O}(\alpha_s^2)$ 
predictions. 
The fit ranges used for the matched predictions are the
same as for the central values of $\alpha_s$ from the previous
section. In the case of the ${\cal O}(\alpha_s^2)$ predictions, the
fit ranges from the simultaneous fits for $\alpha_s$ and $x_{\mu}$ are
taken. The shape of the curves 
depends rather strongly on the fit range, which is why the circular
and square markers 
are sometimes not situated on the corresponding curves. In general, an
extension of the fit range causes the minima of the $\chi^2$ curves to
become more pronounced, though the positions in $x_{\mu}$ of the minima 
are found to be mostly unaffected by such changes.\par
At 189 GeV statistics are low, and the
fits of the ${\cal O}(\alpha_s^2)$ predictions tend to become unstable under
variations of the QCD scale for larger $x_{\mu}$.  
The corresponding curves are therefore not shown.
Stability of the
fits requires the limitation of the fit ranges to those used in the 
previous section for fits with fixed scales.
Instabilities are further encountered at $\sqrt{s}=35$ GeV in the
case of the matched
predictions of the 2-jet fractions at low $x_{\mu}$, as well as for the
two-parameter fits of the ${\cal O}(\alpha_s^2)$ calculations for
the mean jet multiplicities.\par 
As can be
seen in Fig.~\ref{xmudep_lab}, the dependence of $\alpha_s$ on the scale in the
vicinity of $x_{\mu}=1$ is generally much smaller with $\ln R$-matching
than with ${\cal O}(\alpha_s^2)$ calculations.
Furthermore, fits of the matched predictions 
with a free QCD scale result in very shallow minima in $\chi^2$ and
correspondingly large errors in the fitted parameters. Sometimes they do 
not converge at all. In some cases, convergence can be achieved by
sufficiently extending the fit range towards low $y_{cut}$. The results
for $x_{\mu}$ are then usually found to be closer to 1. We conclude that
the matched predictions indeed show little preference for any specific
choice of the QCD scale, while the second-order calculations require
rather definite values of $x_{\mu}$ to describe the data well. However,
as can be seen in the figures, the $x_{\mu}$ 
dependence of the resulting $\alpha_s$
is still sizeable. 
\begin{figure}[!t]

\centerline{
\hspace*{-0.5cm}
\epsfig{file=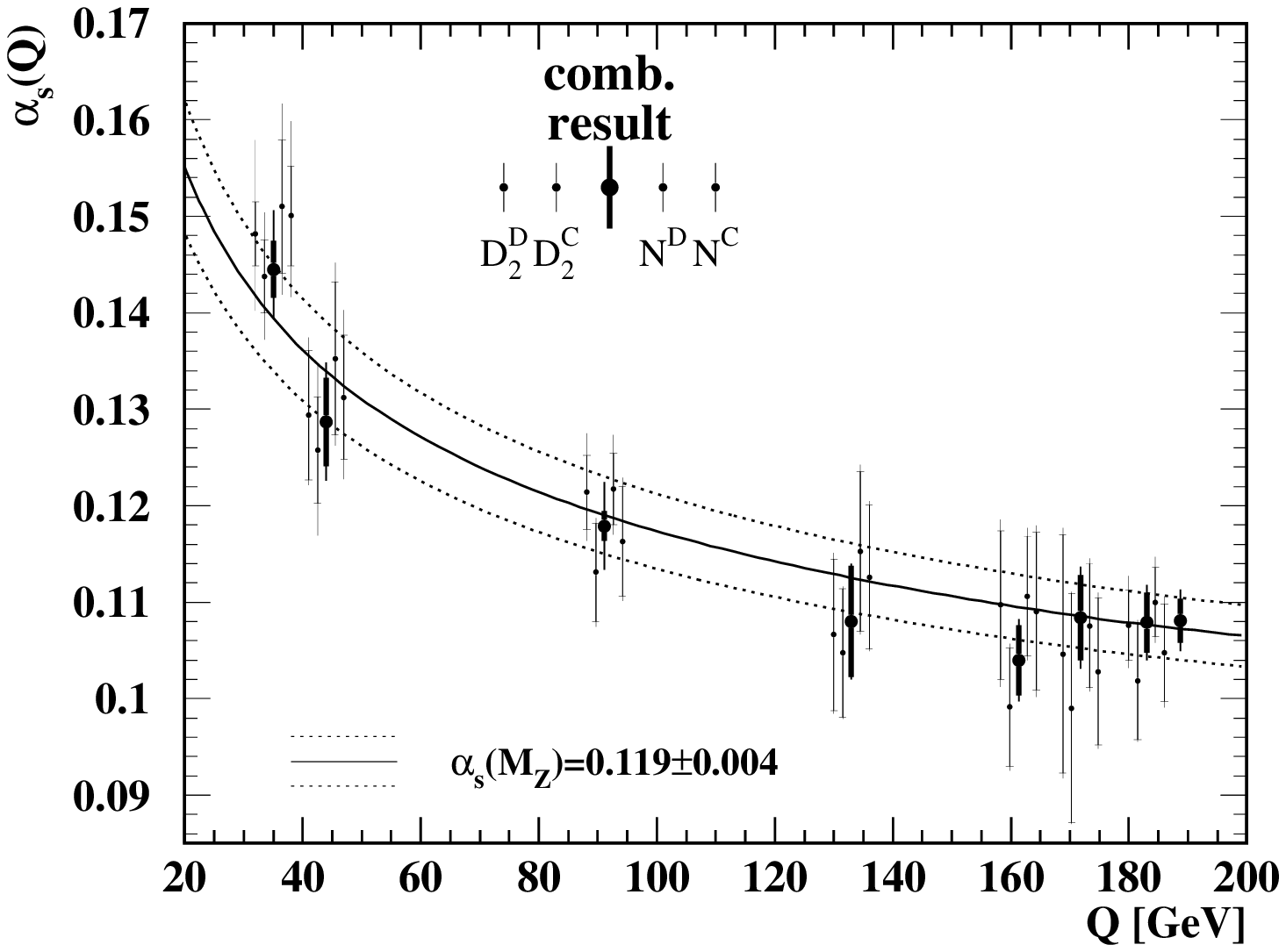,width=15cm}
}

\vskip-11.1cm
\centerline{
\hspace*{9.1cm}
\epsfig{file=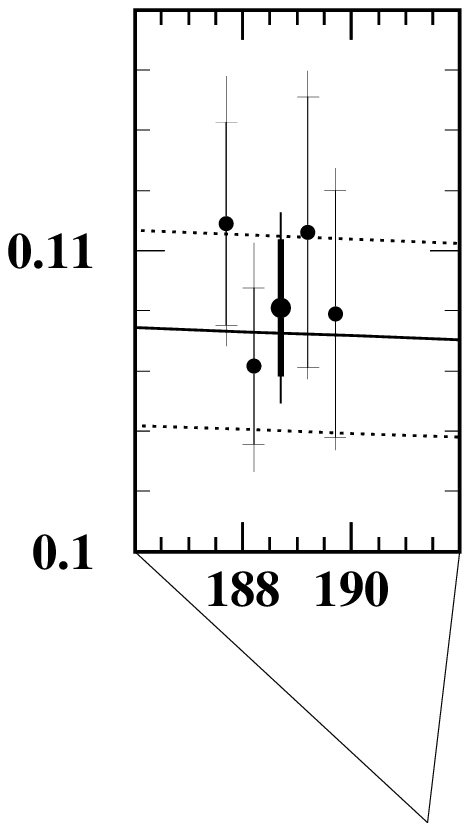,width=4cm}
}

\vspace*{2.5cm}

\caption[Final results for $\alpha_s(\protect\sqrt{s})$ versus $\protect\sqrt{s}$]
{The combined results for $\alpha_s(Q)$ ($Q=\sqrt{s}$) 
from fits of the matched predictions
plotted versus the c.m.s. energy. Arranged around the combined values
are the contributing separate results for each of the four observables under
consideration. All results are shown with their total errors (outer
error bars) and the purely experimental (inner bars) component of the errors.
A three-loop evolution of the current world average \cite{worldav} is overlaid as solid and
dotted lines.\label{asqfiglab}}
\end{figure}

\subsubsection{Systematic Errors of the Main Result}
Based on the investigations carried out in the previous sections, we 
use the
$\ln{R}$-matching scheme for our final results for all observables.
Tables \ref{astablab_35} through \ref{astablab_189} in Appendix~\ref{astableslab} list the final results for $\alpha_s$ for
all variables and c.m.s. energies, along with the composition of their
errors. We discuss the systematic errors for this scheme.\par
At all c.m.s. energies, the use of only charged tracks rather than
both tracks and electromagnetic clusters has a large effect on the
result. At JADE energies, variations of the selections generally
contribute little to the systematic errors. At c.m.s. energies of 133 GeV
and above, however, the influence of variations in the selections becomes
sizeable. This is to be expected since the selections themselves play a
more important role here than at the lower energies. The restriction
of the cut on the thrust axis, which is applied at all c.m.s. energies,
induces mostly small effects.\par
The hadronization errors of the results are, in particular at lower
c.m.s. energies, dominated by differences between the
generators (i.e.
HERWIG or ARIADNE). This could be anticipated considering the size of the
hadronization corrections shown in Figs.~\ref{asfitlab_35} to
\ref{asfitlab_189}. A similar
behaviour is seen for the dependence of the results on $x_{\mu}$ which 
is largest at low c.m.s. energies. The influence of
the omission of b quarks in the hadronization model
and of varying the termination point of the parton
shower is sizeable and relatively independent of the c.m.s.\
energy. Variations of the hadronization parameters $b$ and $\sigma_q$
have very small effects throughout.\par 
\subsubsection{Combination of the Fit Results}
\label{combreslab}
We combine the results obtained using the four observables, separately
at each c.m.s. energy by forming the mean value,
taking into account the covariance matrix corresponding
to the total errors of the individual values. The 
statistical correlations are determined from the data/Monte Carlo
subsamples of Sect. \ref{genfixlab}. The systematic uncertainties
are treated as fully correlated, aside from the variations 
of $x_{\mu}$ and those of the bin boundaries which
are considered as uncorrelated and
added to the diagonal of the matrix only. To this end, the scale uncertainties
are made symmetric by taking the mean value of the
two separate deviations. In order to break up the total error into its
single components, the mean is determined using, respectively, 
the covariance matrix corresponding to only statistical, only
experimental, and only experimental and hadronization uncertainties.
Fig.~\ref{asqfiglab} gives an overview of the combined results,
as well as the results from each single observable with their experimental
and total errors\footnote{At $\sqrt{s}=172$ GeV the mean value lies above
all contributing individual values. Here, relatively
large positive correlations and the fact that the highest individual value
(from $N^D$) has the smallest error have the combined effect that the
lowest $\chi^2$ is obtained when all individual values lie on the
same, rather than on opposite sides of the mean.}. 
The combined results are given
in Table \ref{asqtablab}.
The values listed in the right column of the table are 
three-loop evolutions to $\sqrt{s}=M_{Z^0}$. 
The results
are compatible within their errors.\par 
\begin{table}[!t]

{\footnotesize
\begin{center}
\begin{tabular}{l|c@{$\pm$}c@{$\pm$}c@{$\pm$}c@{$\pm$}c|c@{$\pm$}c@{$\pm$}c@{$\pm$}c@{$\pm$}c}
\multicolumn{1}{c}{$\sqrt{s}$} & \multicolumn{5}{|c|}{$\alpha_s(\sqrt{s})$} & \multicolumn{5}{|c}{$\alpha_s(M_{Z^0})$}\\ 
 \multicolumn{1}{c|}{[GeV]} & \multicolumn{1}{c}{} & \multicolumn{1}{c}{(stat.)} & \multicolumn{1}{c}{(exp. sy.)} & \multicolumn{1}{c}{(hadron.)} & \multicolumn{1}{c|}{(scale)} & \multicolumn{1}{c}{} & \multicolumn{1}{c}{(stat.)} & \multicolumn{1}{c}{(exp. sy.)} & \multicolumn{1}{c}{(hadron.)} & (scale) \\  \hline

35 & 0.1445 & 0.0007 & 0.0029 & 0.0042 & 0.0033 
   & 0.1226 & 0.0005 & 0.0021 & 0.0030 & 0.0024   \\[0.2cm]

44 & 0.1287 & 0.0010 & 0.0044 & 0.0039 & 0.0012
   & 0.1149 & 0.0008 & 0.0035 & 0.0031 & 0.0009 \\[0.2cm]

91 & 0.1179 & 0.0001 & 0.0016 & 0.0038 & 0.0021
   & 0.1179 & 0.0001 & 0.0016 & 0.0038 & 0.0021 \\[0.2cm]

133 & 0.1080 & 0.0019 & 0.0054 & 0.0014 & 0.0010 
   & 0.1139  & 0.0021 & 0.0061 & 0.0016 & 0.0011 \\[0.2cm]

161 & 0.1040 & 0.0020 & 0.0030 & 0.0019 & 0.0012
   & 0.1125  & 0.0023 & 0.0035 & 0.0022 & 0.0014 \\[0.2cm]

172 & 0.1084 & 0.0037 & 0.0025 & 0.0016 & 0.0024
   & 0.1188  & 0.0045 & 0.0031 & 0.0020 & 0.0030 \\[0.2cm]

183 & 0.1079 & 0.0020 & 0.0024 & 0.0020 & 0.0011 
   & 0.1193  & 0.0024 & 0.0029 & 0.0024 & 0.0013  \\[0.2cm]

189 & 0.1081 & 0.0012 & 0.0019 & 0.0021 & 0.0007
   & 0.1201  & 0.0015 & 0.0023 & 0.0026 & 0.0008   \\
\end{tabular}
\end{center}
}

\caption[Final results for $\alpha_s(\protect\sqrt{s})$ versus $\protect\sqrt{s}$]
{Combined results from fits of matched predictions, along with the 
statistical, experimental systematic, hadronization and QCD scale
errors (left column). The values in the right column are results of a three-loop
evolution of each result to $Q=M_{Z^0}$.\label{asqtablab}}
\end{table}
As can be seen in the Fig.~\ref{asqfiglab}, the separate results are distributed in roughly
the same pattern 
at each c.m.s. energy. The solid and dotted
lines represent a three-loop evolution of the current world average of 
$\alpha_s(M_{Z^0})=0.119\pm0.004$ 
\cite{worldav} with its total error. No significant discrepancies between the curve and
any of the fit results can be noted. Our values agree with the
world average within their experimental errors.
Our results using the 2-jet fraction of the Durham scheme
are also in good agreement with previous determinations using
resummed calculations at these energies
(see, e.g., \cite{ourjade,opal161,opal133,opal91,alephmean,otherexp}).\par
\begin{figure}[!t]

\centerline{
\hspace*{-0.5cm}
\epsfig{file=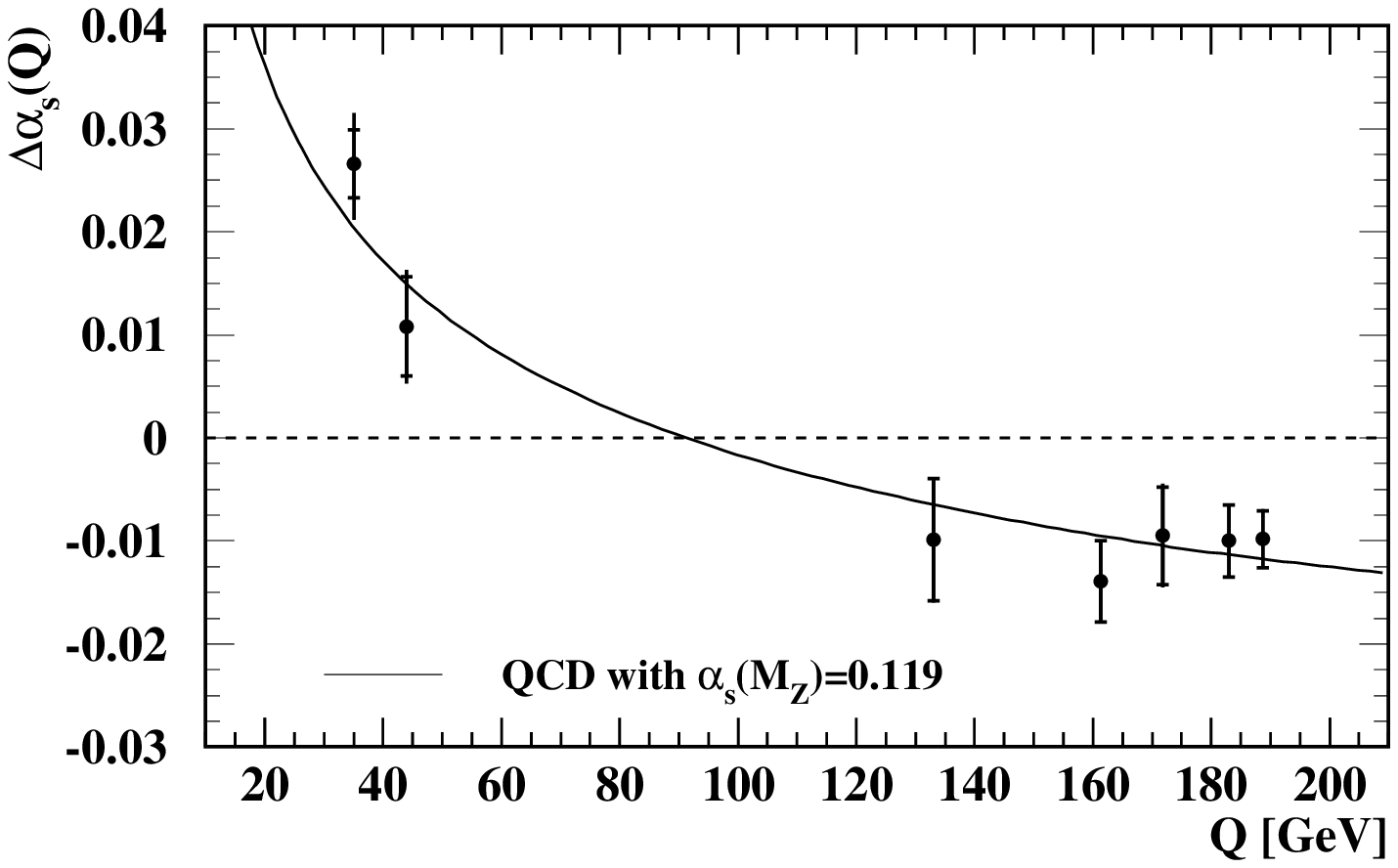,width=12cm}
}

\vskip -1.cm

\caption[$\Delta\alpha_s$ versus $\protect\sqrt{s}$]
{The difference
$\Delta\alpha_s\equiv\alpha_s(Q)-\alpha_s(M_{Z^0})$ 
versus $Q=\sqrt{s}$ from $\alpha_s$ fits of matched predictions. 
The definition of the error bars is as in Fig.~\ref{asqfiglab}. The solid line represents the current world average.\label{asdiffreslab}}
\end{figure}

The difference $\Delta\alpha_s\equiv\alpha_s(Q)-\alpha_s(M_{Z^0})$ is plotted
in Fig.~\ref{asdiffreslab}.
Again, the inner error bars in the plot represent
the total experimental errors while the outer bars include also hadronization
and QCD scale uncertainties. The statistical and experimental systematic errors
are obtained from those quoted in Table \ref{asqtablab} 
assuming that correlations between different energies can be neglected. 
In order to take correlations of the theoretical errors into account, each
difference $\Delta\alpha_s$ was calculated separately under each variation
of the hadronization model and of the QCD scale. 
As usual, the difference between
the value of $\Delta\alpha_s$ obtained after each such variation and 
its central value
was counted as a contribution to the systematic 
error. The contributions to the 
hadronization error were combined in quadrature.\par
Finally, in order to obtain a single result for $\alpha_s$, we perform a fit
of the three-loop running of $\alpha_s$ to the values of Fig.~\ref{asqfiglab}, the fitted variable being $\alpha_s(M_{Z^0})$. 
We assume that correlations between the total experimental errors can be neglected
and use these for the determination of $\chi^2$. The resulting fit error is taken
as the total experimental error of the final result. The statistical component
is assessed by repeating the fit using only statistical errors. 
The total hadronization and scale errors are then
calculated as for $\Delta\alpha_s$, i.e. by repeating the fits under each
variation. The final result of 0.1187\dderr{0.0034}{0.0019} is given with its individual error
contributions and the value of $\chi^2/d.o.f.$ in Table \ref{ascombtablab}.

\begin{table}

{\footnotesize
\begin{center}
\begin{tabular}{|r||c|} \hline
$\alpha_s(M_{Z^0})$ &  {\bf 0.1187}\\
$\chi^2/d.o.f.$ & 7.85/7 \\ \hline\hline
Stat. error & $\pm0.0001$ \\ \hline\hline
Total exp. &  $\pm0.0010$ \\ \hline\hline
$b+1\sigma$ & $-0.0001$ \\\hline
$b-1\sigma$ & $+0.0002$ \\\hline
$\sigma_q+1\sigma$ & $+0.0001$ \\\hline
$\sigma_q-1\sigma$ & $<0.0001$ \\\hline
$Q_0=2.4$ GeV &  $+0.0007$ \\\hline
$Q_0=1.4$ GeV &  $-0.0004$ \\\hline
udsc only &   $+0.0010$ \\\hline
HERWIG &   $-0.0010$ \\\hline
ARIADNE &   $-0.0002$ \\  \hline\hline
Total hadronization & $\pm0.0016$\\ \hline\hline
$x_{\mu}=0.5$ &   $-0.0001$ \\\hline
$x_{\mu}=2$ &  $+0.0028$ \\  \hline\hline
 & {\bf +0.0034} \\
\rb{Total error} & {\bf -0.0019} \\ \hline
\end{tabular}
\end{center}
}

\caption[Global result for $\alpha_s(M_{Z^0})$ from separate energy measurements]{
\label{ascombtablab}
The result for $\alpha_s(M_{Z^0})$ obtained from a fit of the three-loop running
of the strong coupling to the combined values determined using matched QCD calculations
at single c.m.s. energies 
(Table \ref{asqtablab}), 
with the individual error contributions. 
For each systematic variation of Monte Carlo parameters and of
the QCD scale, the deviation with respect to the central value is given with a sign
indicating the direction of the deviation.

}

\end{table}

\subsection{$\alpha_s(M_{Z^0})$ from the Energy Evolution of Jet Observables}
\label{enevollab}
The availability of measurements
performed over a wide range of c.m.s. energy with very similar experimental
conditions suggests investigations of 
the c.m.s. energy dependence of jet-related observables rather than analyses
at fixed energies. In Sect. 
\ref{jetmeaslab}, 
the energy evolution of jet fractions was already 
qualitatively observed, and the extracted $\alpha_s$
values shown in Fig.~\ref{asqfiglab} were seen to be in accordance with 
the QCD prediction of the running of the strong coupling.\par
Similar analyses have been done based on data
from many ${\rm e}^+{\rm e}^-$ collision experiments, present and past. For
an early test of the running of $\alpha_s$, the three-jet fraction was
used (e.g. \cite{bethkejade}). This is, 
in lowest order, simply proportional to $\alpha_s$.
The JADE scheme was employed for these analyses
because it has the smallest and least energy-dependent
hadronization corrections of all clustering
schemes at low c.m.s. energies. More recent
measurements have been mainly concerned with the measurements of the energy
evolution of event shape moments and of the mean jet multiplicity (e.g. \cite{ourjade,mymontp,alephmean,othermean}).\par
In this analysis, the energy evolution of the three-jet fraction $R_3$ at a fixed 
value of $y_{cut}$ and that of the mean values of the observable $y_2$
are compared with the predictions of QCD for various clustering 
algorithms.

\subsubsection{Analysis Procedure}
\begin{figure}[!t]
\vskip \momuup

\vskip \momup
\hspace*{\momleft}
\epsfig{file=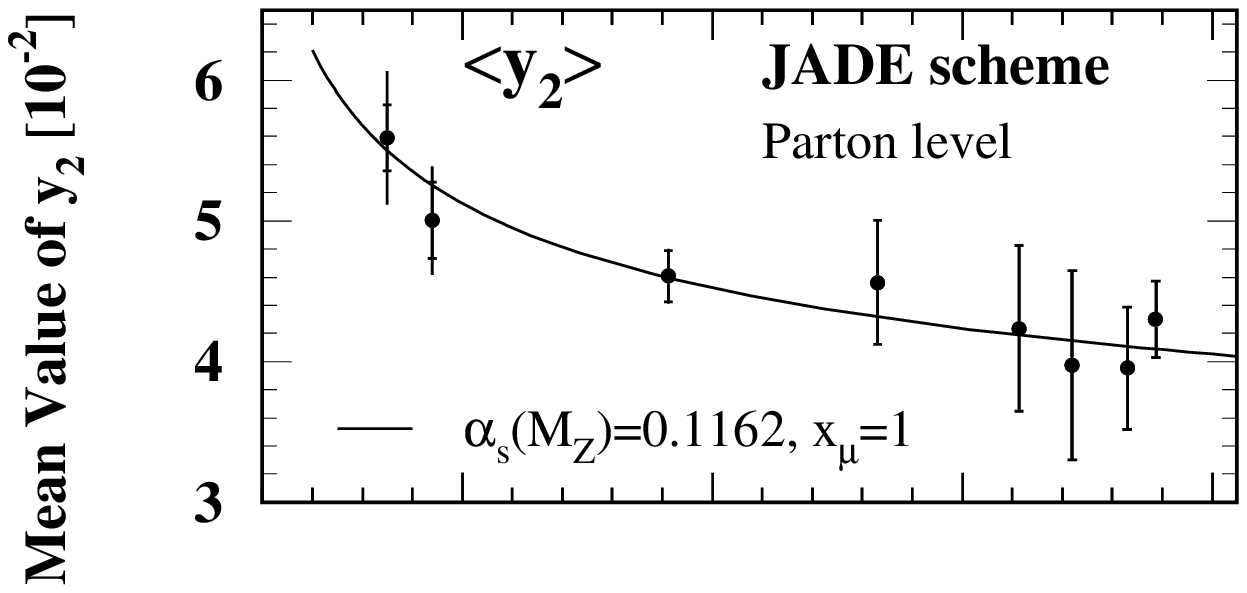,width=\momwid}
\vskip\mommid
\hspace*{\momleft}
\epsfig{file=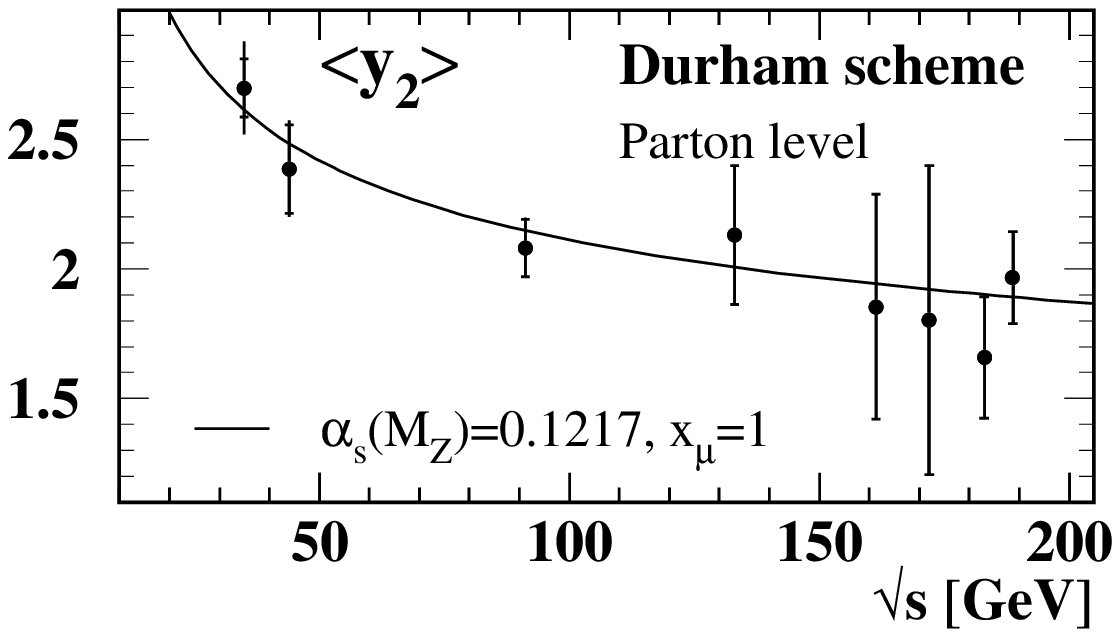,width=\momwid}

\vspace*{\momdown}

\caption[Mean values of $y_n$ vs. $\protect\sqrt{s}$]
{
Mean values of $y_2$ as obtained 
with the JADE and Durham schemes at parton 
level versus the c.m.s. energy $\protect\sqrt{s}$. 
The solid lines represent the fitted ${\cal O}(\alpha_s^2)$ predictions at a fixed
QCD scale of 1.\label{momfitfig}}
\end{figure}

The comparisons between measurement and
predictions 
are done by making $\chi^2$ fits to the respective
observables over the entire range of c.m.s. energies considered in 
our analysis, taking $\alpha_s(M_{Z^0})$ as the fitted variable.
The expressions
for the ${\cal O}(\alpha_s)$ and
${\cal O}(\alpha_s^2)$ coefficients in the prediction 
are taken from \cite{galgo} for the
JADE scheme and from \cite{calgo} for the Durham and Cambridge schemes.
The expression to be fitted to the mean value of $y_2$ is given by
(\ref{oas2mom2}) of Appendix~\ref{theocalclab}.
The coefficients $A$ and $B$ were 
obtained from runs of the EVENT2 program \cite{event2,kluthpriv}.\par
In the case of
the fits to the three-jet fractions, a specific value of $y_{cut}$ has
to be selected at which the jet fraction is given. 
Here, we select a value in a region where the sensitivity to changes in the c.m.s.
energy is large (see Figs.~\ref{e0ratefig}, \ref{dratefig} and \ref{cratefig}) and
hadronization corrections are small.
In the case of the JADE scheme, the value of $y_{cut}=0.08$
from \cite{opalr3} is used. For the Durham and Cambridge schemes, we
take $y_{cut}=0.01$.\par 
For the moments of some event shape variables, there exist non-perturbative corrections
(also called ``power corrections'') to the perturbative calculations \cite{powcorr}.
It was found in \cite{ourjade}
that corrections of order $1/Q$ as well as $1/Q^2$ to the mean values of $y_2$ as obtained
with the Durham scheme can be neglected at the c.m.s. energies
considered in this analysis.
Hadronization corrections for the three-jet fractions at the chosen
value of $y_{cut}$
are found to be small (between 2\% and 11\%) for the JADE and
Durham schemes,
but large (about 25\%) for
the Cambridge scheme at low c.m.s. energies.
We 
perform the fits for all observables
at parton level. The transformation of the 
measurements from hadron to parton level is done separately at each c.m.s.
energy by means of multiplicative factors which are obtained from
the same Monte Carlo samples as were used for the $\alpha_s$ fits. 
The factors predicted by the PYTHIA 5.722 Monte Carlo
at standard OPAL tune are used 
to correct the results.\par
The systematic errors can not be treated
by repeating the fits with purely statistical errors for each 
systematic variation, because the systematic variations applied are essentially
different at each c.m.s. energy. We therefore use the
total experimental errors of each measurement for the calculation of
the $\chi^2$ and take the resulting fit error as the overall
experimental error of $\alpha_s$. Its statistical component is determined
by repeating the fit using only statistical errors. Hadronization 
uncertainties are determined by repeating the fits with variations
of the Monte Carlo predictions for the correction factors from hadron
to parton level. We apply only
the variations leading to the predominant error
contributions in the fits of the previous section, i.e. changes to
HERWIG and ARIADNE and the removal of the b quark.
The absolute differences from the central result are again added
in quadrature to get the total hadronization error.\par
Fits are carried out both with a constant QCD scale factor, fixed
at unity, and with $x_{\mu}$ taken as an additional fit parameter. In
the first case, systematic variations of $x_{\mu}$ to 0.5 and 2
are performed and added asymmetrically to the hadronization
error.

\subsubsection{Results from the Energy Evolution Fits}
\begin{figure}[!t]
\vspace*{-1.5cm}

\centerline{
\hspace*{1cm}
\epsfig{file=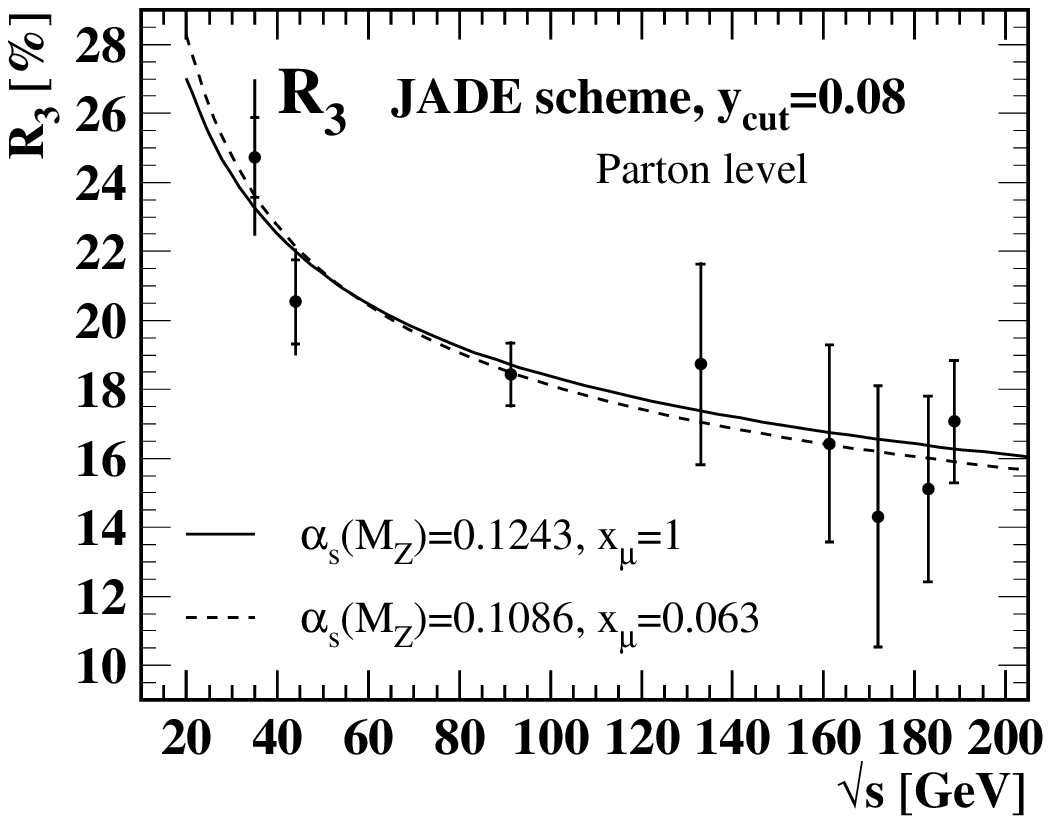,width=11cm}
\hspace*{-2.5cm}
\epsfig{file=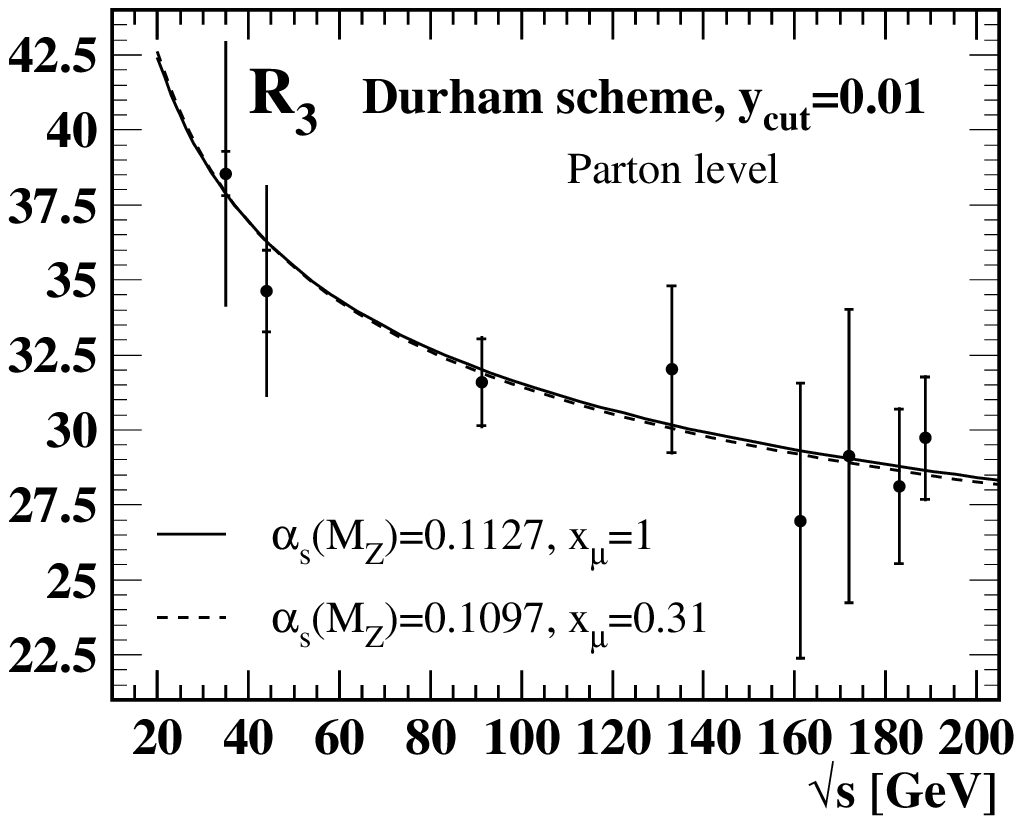,width=11cm}
}
\vskip -1.8cm
\centerline{
\hspace*{1cm}
\epsfig{file=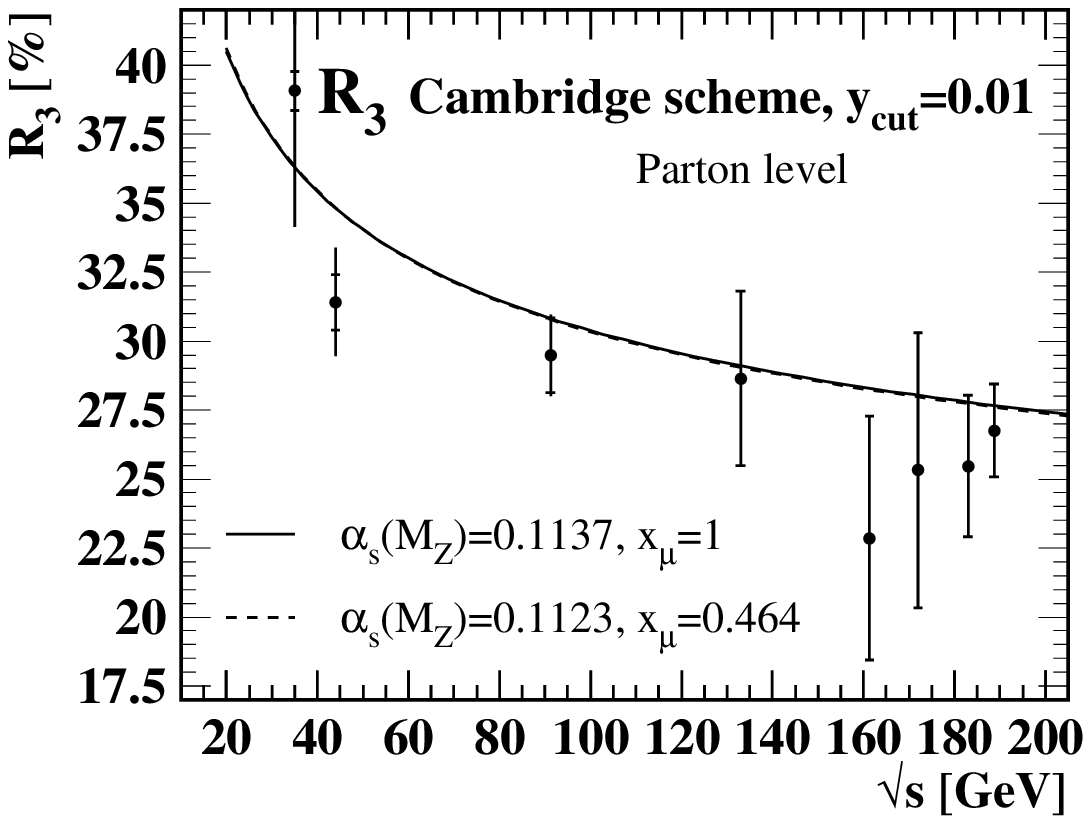,width=11cm}
}

\vskip -1cm

\caption[$R_3$ versus $\protect\sqrt{s}$]
{
The 3-jet fraction as obtained 
with the JADE, Durham and Cambridge schemes at parton 
level versus the c.m.s. energy $\protect\sqrt{s}$. 
The solid lines represent the fitted ${\cal O}(\alpha_s^2)$ predictions at a fixed
QCD scale of 1, the dashed lines are the results obtained with a fitted scale.\label{r3_fig}}
\end{figure}

\noindent Fig.~\ref{momfitfig} 
shows the measurements of $\langle y_2\rangle$ for
the JADE and the Durham scheme at parton level with the result of the
fit at fixed $x_{\mu}$. The inner error bars denote the size of the purely
experimental errors, the outer bars include also hadronization errors.
The obtained values for $\alpha_s(M_{Z^0})$ from the fits shown are
listed in 
Table \ref{momrestablab} of Appendix~\ref{astableslab} with the deviations induced by each
systematic variation, as before.
The total errors are of the same order of magnitude as from
the fits at separate c.m.s. energies. 
Fits with a free QCD scale $x_{\mu}$ lead to errors of
almost 100\% in the scale for both observables. 
We therefore present the
results obtained at a fixed scale of 1, performing the usual variations
to estimate the error from the scale uncertainty. The somewhat small values of
$\chi^2/d.o.f.$ indicate correlations in the systematic uncertainties.\par
The results of the fits to the three-jet fractions are summarized in 
Table \ref{r3restablab} of Appendix~\ref{astableslab} 
and are displayed in Fig.~\ref{r3_fig}. Unlike 
the previous fits to
the mean values of $y_2$, the fits with a free QCD scale lead to
smaller errors in $x_{\mu}$.
The corresponding prediction is added as a dashed line in the plots.
The optimized
scales again turn out to be significantly smaller than $x_{\mu}=1$. The fit
results for $\alpha_s$ with a fitted scale are systematically smaller than with a fixed 
scale of 1. In order to be able to estimate the error induced by
scale uncertainties, we quote central results for
$x_{\mu}=1$ and vary the scale to 0.5 and 2.\par
We combine
the separate results
with $x_{\mu}=1$ from the five observables
into one value for $\alpha_s(M_{Z^0})$.
The combined result is calculated 
by taking 
the average of the remaining five
single values, weighted with their respective total errors. We assume 
total experimental errors to be largely unaffected by correlations
and determine these also by taking the weighted mean of the single values. The
error contributions from each hadronization and QCD scale variation are determined
in the same way separately for each variation, resulting
in the 
overall hadronization and asymmetric scale error of the combined value. Table
\ref{combevreslab} shows the combined result 0.1181\dderr{0.0066}{0.0056} 
with its single error components. 
\begin{table}[!t]

{\footnotesize
\begin{center}
\begin{tabular}{|r||c|} \hline
$\alpha_s(M_{Z^0})$ &  {\bf 0.1181}\\ \hline\hline
Stat. error & $\pm0.0006$ \\ \hline\hline
Total exp. &  $\pm0.0020$ \\ \hline\hline
udsc only &   $+0.0015$ \\\hline
HERWIG &   $-0.0035$ \\\hline
ARIADNE &   $-0.0011$ \\  \hline\hline
Total hadronization & $\pm0.0039$ \\ \hline\hline
$x_{\mu}=0.5$ &   $-0.0035$ \\\hline
$x_{\mu}=2$ &  $+0.0048$ \\  \hline\hline
 & {\bf +0.0066} \\
\rb{Total error} & {\bf -0.0056} \\ \hline
\end{tabular}
\end{center}
}

\caption[Global result for {\boldmath $\alpha_s(M_{Z^0})$} from energy evolution measurements]{
\label{combevreslab}
The weighted mean of $\alpha_s(M_{Z^0})$ as determined from the energy evolution
of observables $\langle y_2^J\rangle$, $\langle y_2^D\rangle$, $R_3^J$, $R_3^D$ and 
$R_3^C$,
with the single error contributions. 
For each systematic variation of Monte Carlo parameters and of
the QCD scale, the deviation with respect to the central value is given with a sign
indicating the direction of the deviation.

}

\end{table}

\subsection{MLLA Prediction for the Mean Jet Multiplicity}
\begin{figure}[tbhp]
\vskip \fituup

\centerline{
\hskip \fitleft
\epsfig{file=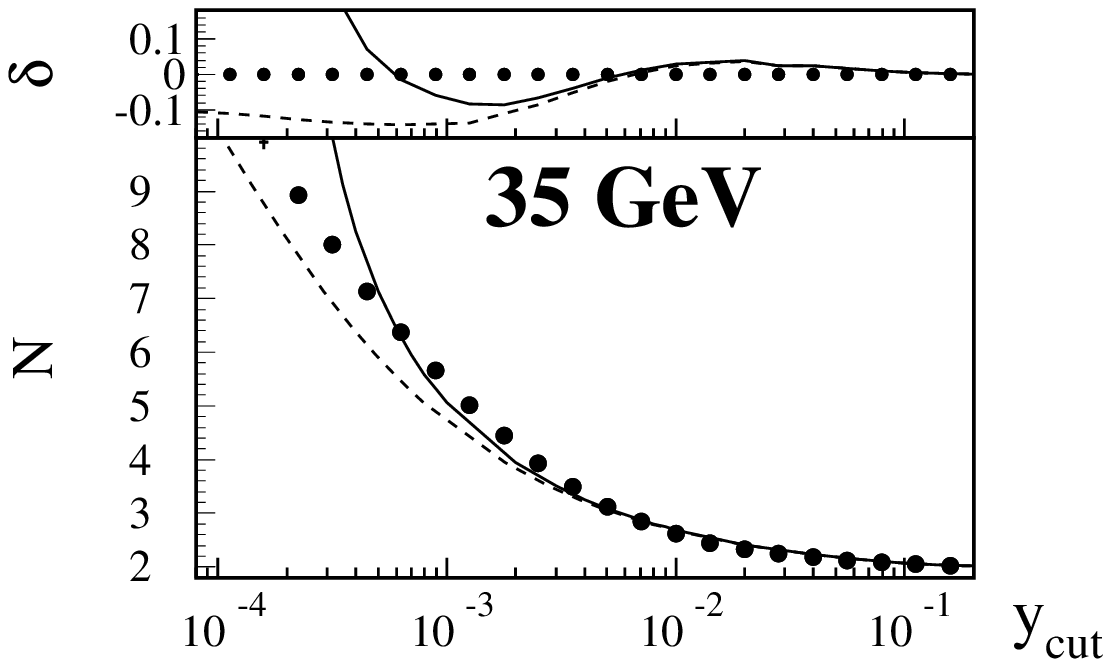,width=\fitwid}
\hskip \fitcent
\epsfig{file=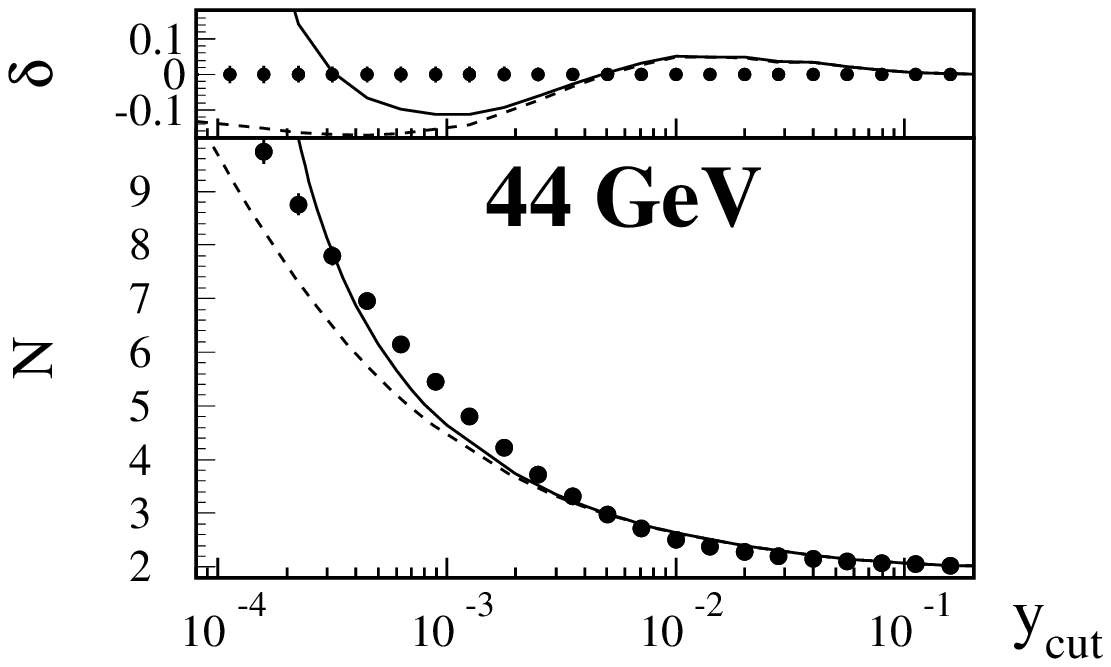,width=\fitwid}
}

\vspace*{\fitmid}

\centerline{
\hskip \fitleft
\epsfig{file=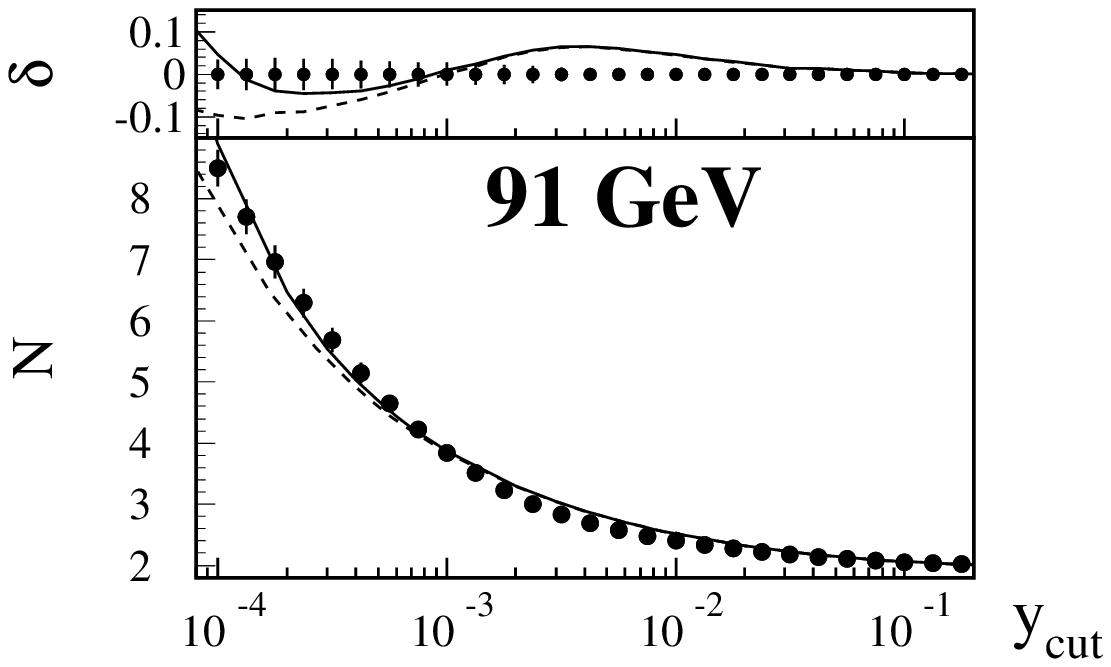,width=\fitwid}
\hskip \fitcent
\epsfig{file=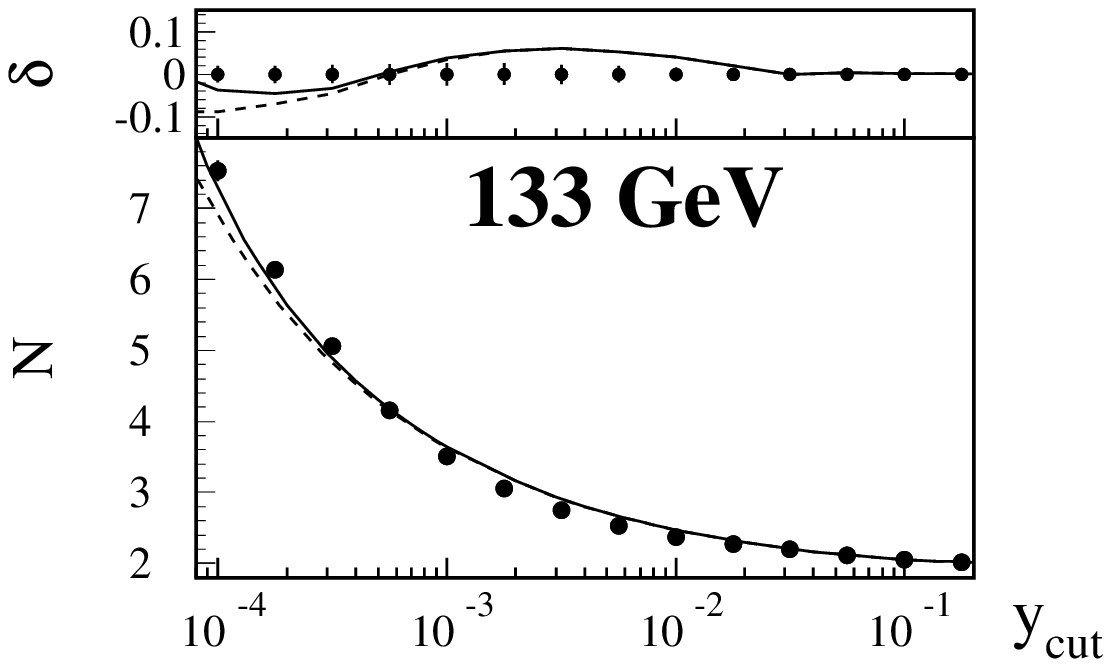,width=\fitwid}
}

\vspace*{\fitmid}

\centerline{
\hskip \fitleft
\epsfig{file=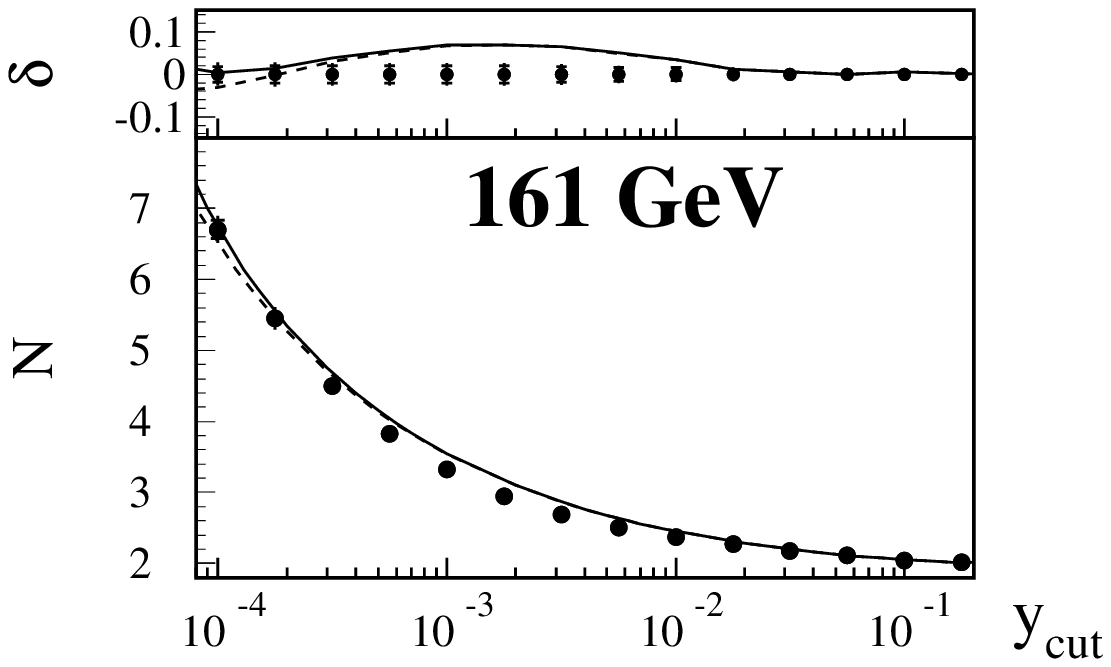,width=\fitwid}
\hskip \fitcent
\epsfig{file=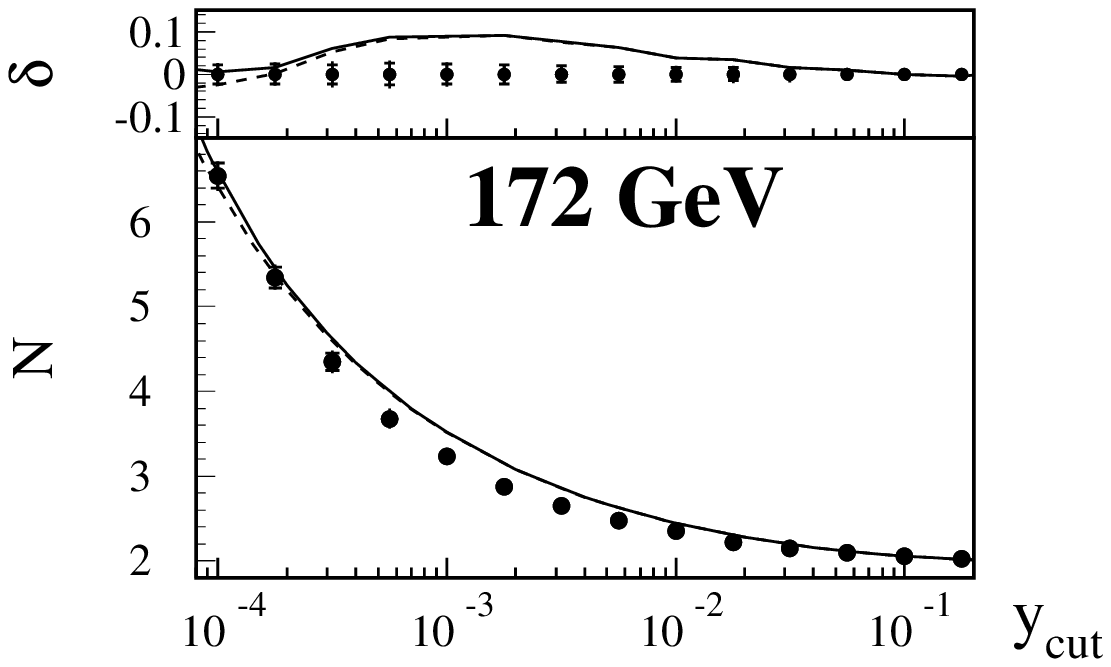,width=\fitwid}
}

\vspace*{\fitmid}

\centerline{
\hskip \fitleft
\epsfig{file=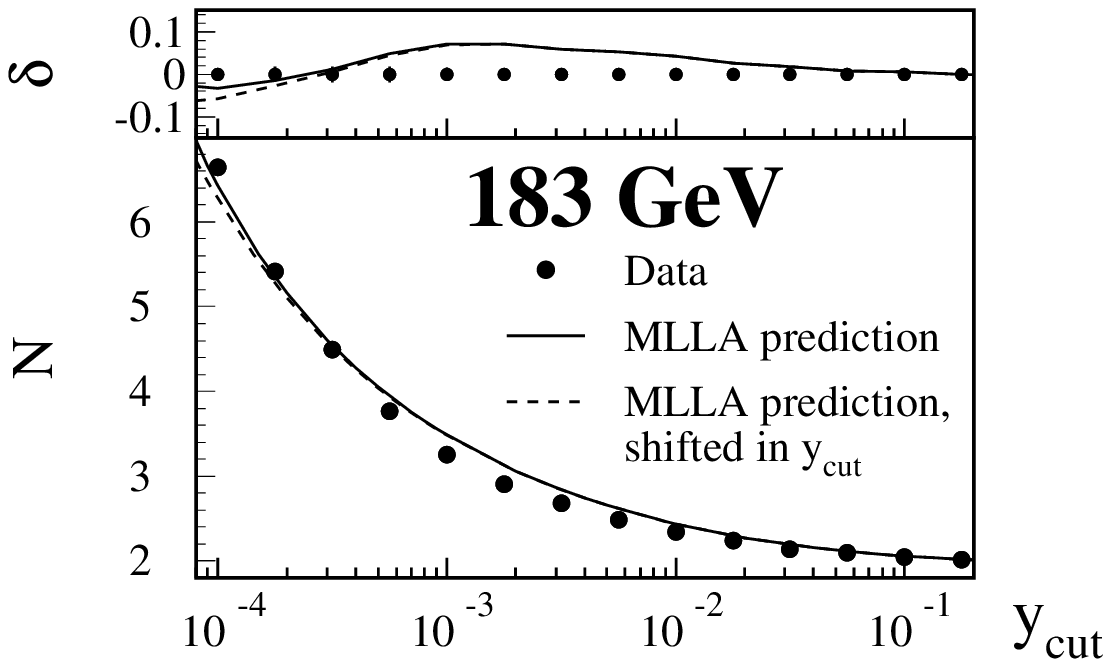,width=\fitwid}
\hskip \fitcent
\epsfig{file=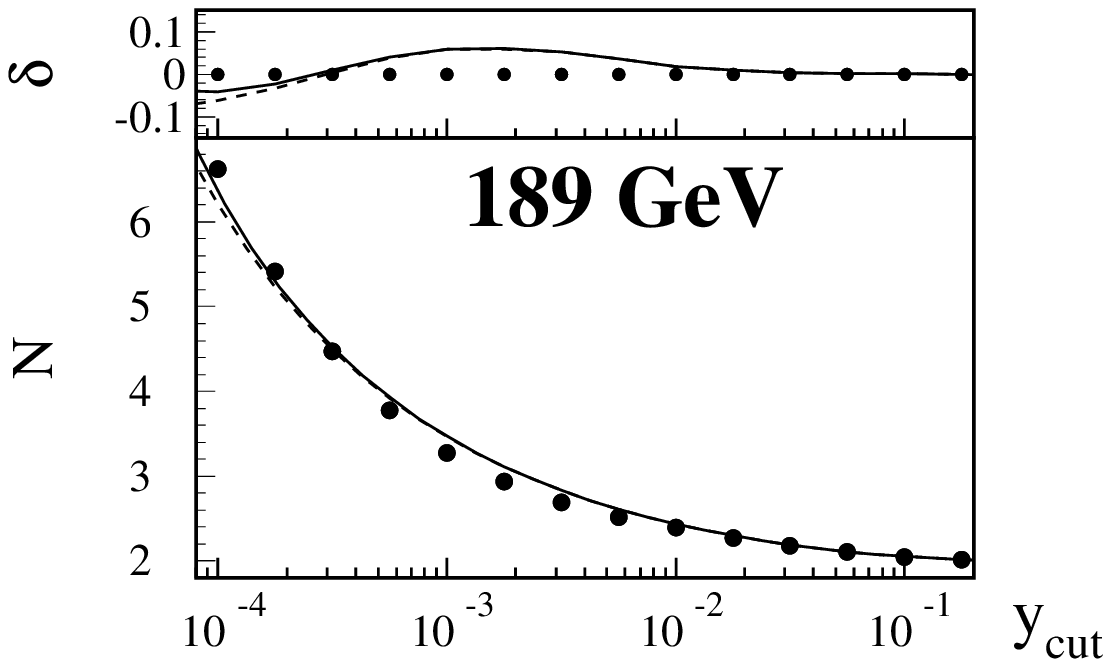,width=\fitwid}
}

\vspace*{\fitdown}

\caption[Mean jet multiplicity (Durham scheme) and MLLA prediction]
{The mean jet multiplicity as obtained with the Durham scheme plotted
versus parameter $y_{cut}$ at all energies. The 
MLLA prediction according to \cite{ochs} for each energy is the solid line
and the same prediction shifted by $Q_0^2/s$ in $y_{cut}$ as described in the text is the
dashed line.\label{mllafig}}
\end{figure}

Lastly, we compare our measurements of the mean jet multiplicity of the
Durham scheme with a recent hadron level prediction \cite{ochs}. The calculation
was carried out in the framework of the modified leading logarithmic approximation
(MLLA) in the form of a cascade of successive parton branchings which is continued until
the relative transverse momentum between the two partons emerging from the branching
falls below a cut-off $Q_0$. Assuming ``local parton-hadron duality''
to be valid 
and setting $Q_0$ to a value of typical hadron masses, the prediction may
be compared with hadron level measurements.
The lower portions of each part of Fig.~\ref{mllafig}
show the hadron level 
measurement of $N^D$ at all c.m.s. energies with their errors as in Fig.~\ref{jetmeanfig} and the MLLA prediction 
\cite{ochspriv} 
as a solid line. 
The cut-off $Q_0$ is fixed by $\ln{(Q_0/\Lambda_{QCD})}=0.015$ with 
$\Lambda_{QCD}=500$ MeV \cite{ochs}. This value was obtained from
fits to measured hadron multiplicities over a c.m.s. energy range from
1.5 GeV to 91 GeV.
The smaller captions above each
plot show the normalized difference $\delta$ between measurements and predictions,
defined as in Sect. \ref{fixfitreslab}. 
A first observation is that the MLLA curves
rise more rapidly than the data below some $y_{cut}$ as a result of 
the singularity in the strong coupling as the transverse parton
momentum comes close to $\Lambda_{QCD}$. The difference seen at
low values of $y_{cut}$ becomes smaller
for higher c.m.s. energies.\par
The MLLA prediction assumes
quarks to be massless. 
In order to artificially introduce a hadron mass in the calculation, it is 
suggested in \cite{ochs} that one may compare the parton transverse momenta
$k_{\perp}$ with the transverse {\it energies}
of the hadrons given by $E_{\perp}=\sqrt{k_{\perp}^2+Q_0^2}$.
In the application of the predictions to the Durham jet
multiplicities, this amounts to comparing the measurement at some $y_{cut}$
with the prediction $ $ at $y_{cut}+Q_0^2/s$, i.e. a relative shift by
a fixed amount along the $y_{cut}$ axis. The dashed lines in the plots 
represent the prediction after this shift.
At $\sqrt{s}=35$ GeV and 44 GeV,
the predictions somewhat
undershoot the data after the shift, showing that such a crude method for
introducing a hadron mass into the calculations works less well at such low energies.\par
At all c.m.s. energies, the data fall below the MLLA predictions in the region
of medium $y_{cut}$. This behaviour is also observed in \cite{ochs} for
$\sqrt{s}=91$ GeV and is there attributed
to the omission of higher loops in the definition of $\alpha_s$.

\section{Summary}
\label{outrolab}
We have performed QCD related measurements at c.m.s. energies of
35, 44, 91.2, 133, 161, 172, 183 and 189 GeV using data from the JADE and
OPAL experiments which are very similar
in their components used in the analysis.
The same measuring techniques were
applied in the two experiments, including
details of the selections and the method of reconstructing single
particle four-momenta.
The results of the measurements display the same systematic behaviour
over the entire c.m.s. energy range, suggesting that the
desired homogeneity of the analysis is indeed realized.\par
Hadron level measurements of $n$-jet fractions, differential jet fractions,
distributions in the variables $y_n$ and mean jet multiplicities have been presented
up to large jet multiplicities and down to very low values of $y_{cut}$
using the JADE, Durham and Cambridge jet finders. Measurements of jet
fractions as obtained with a cone algorithm, varying both of its
parameters $R$ and $\epsilon$, have also been performed. The
mean values of the $y_n$ distributions have been measured. The numerical values
will appear in the Durham data base.
The
measured values were compared qualitatively with the predictions of
four Monte Carlo generators, PYTHIA, HERWIG, ARIADNE and COJETS, representing
the major currently available
models for parton shower evolution and hadronization. All
generators except for COJETS were found to be in agreement with the data. COJETS
was seen to predict too many jets, in particular in regions
of high jet multiplicities (high jet resolution). The discrepancy rises with
c.m.s. energy and becomes significant at $\sqrt{s}\gtrsim91$ GeV. 
It can
be explained by the omission of gluon coherence effects in the generator
which will lead to an excess of soft gluons.
Qualitatively, 
all observables based on the clustering schemes display a clear scaling violation
with c.m.s. energy, as expected from QCD.\par
The 2-jet fractions and mean jet multiplicities as obtained with the Durham
and Cambridge schemes were used for quantitative tests of QCD. Matched
${\cal O}(\alpha_s^2)$ and NLLA predictions for these observables were fitted
to the data at each separate c.m.s. energy over an appropriate range of $y_{cut}$,
the fitted parameter being the strong coupling at the respective energy. The
$\ln R$ and the $R$-matching schemes, as well as their ``modified'' variants were
used in the fits.
In addition, ${\cal O}(\alpha_s^2)$ predictions were fitted, where the QCD
scale factor $x_{\mu}$ was taken as an additional free parameter or kept fixed at
unity. The fit quality in terms of $\chi^2$ was found to be 
reasonable, confirming
the general validity of QCD within the errors of the results. None of the
different calculation types could be clearly disqualified on the basis of the
$\chi^2$ values. To obtain an acceptable
$\chi^2$, the fits
of the ${\cal O}(\alpha_s^2)$ calculations at $x_{\mu}=1$ required a
limitation of the fit range to regions of large $y_{cut}$, which is in accordance with
the expectation that the omission of higher orders of $\alpha_s$ becomes noticeable
in regions of high jet multiplicities (low $y_{cut}$). The range of validity of
the ${\cal O}(\alpha_s^2)$ calculations
can be extended towards lower $y_{cut}$ by choosing
an appropriate scale $x_{\mu}$. In agreement with other analyses, the 
optimized values for $x_{\mu}$ turn out to be significantly lower than 1. At
all c.m.s. energies, the
resulting $\alpha_s$ values display the same relative shifts as a function of the type of prediction chosen. The results using ${\cal O}(\alpha_s^2)$
calculations and $x_{\mu}=1$ tend to be systematically larger and the
corresponding results with a fitted scale tend to be smaller than those from
the matched predictions. In the case of the 2-jet fractions, the differences between
the results using simple and modified $\ln R$-matching are negligible. The
results of the modified $R$-matching are also generally comparable with those of the
$\ln R$-matching, while the $R$-matching, which is known to be less complete, 
leads to systematically small values of $\alpha_s$. In the case of
the mean jet multiplicities, the differences between the matching schemes are less
significant. Generally, we observe that at c.m.s. energies with sufficiently
high statistics (35 GeV, 91 GeV, 183 GeV and 189 GeV) the precision of the obtained
$\alpha_s$ is currently limited by the theoretical uncertainties.\par
The dependence
of the results on the QCD scale $x_{\mu}$ has been investigated.
Simultaneous fits of $\alpha_s$ and $x_{\mu}$, using matched predictions and the
same fit ranges as with a fixed scale either do not converge at all or result in
values for $x_{\mu}$ around unity affected by very large fit errors. The $\alpha_s$
results, however, still depend on the scale, which leads to sizeable contributions to
the theoretical errors.\par
Combinations of the $\ln R$-matching results from the four observables, separately 
at each c.m.s. energy, are found to be in agreement with a three-loop QCD evolution of the
current world average for $\alpha_s(M_{Z^0})$ of $0.119\pm0.004$. A fit of the
three-loop running expression for 
$\alpha_s$ to the combined results over all c.m.s. energies returned 
a final value of 
$$\alpha_s(M_{Z^0})=0.1187\derr{0.0034}{0.0019} $$
with a $\chi^2/d.o.f.$ of 7.85/7.\par
We have also carried out $\alpha_s$ fits to the energy evolution of the 
mean values $\langle y_2\rangle$ as obtained with the JADE and Durham schemes and
of the 3-jet fraction of the JADE, Durham and Cambridge schemes, each evaluated
at a fixed $y_{cut}$. In all cases, $\alpha_s(M_{Z^0})$ was fitted for, and
predictions of ${\cal O}(\alpha_s^2)$ were used with a fixed QCD scale of unity. 
The results found are again in agreement with the world average with reasonable
$\chi^2/d.o.f.$, except for the 3-jet fraction of the Cambridge scheme where
a somewhat high value of about $\chi^2/d.o.f.=31/7$ is obtained. The differences
between the results from different observables turn out to be larger than the 
overall experimental error. Attempts to fit the QCD scale simultaneously resulted
in very large fit errors in the case of the quantities $\langle y_2\rangle$,
precluding any reasonable fixing of the scale. In the case
of the 3-jet fractions, optimized QCD scales were again found smaller than unity
and led to systematically smaller results for $\alpha_s(M_{Z^0})$. The weighted
mean value of the $\alpha_s$ results of the five observables at $x_{\mu}=1$
is $\alpha_s(M_{Z^0})=0.1181\dderr{0.0066}{0.0056}.$\par
Finally, we have tested a hadron level MLLA prediction for the mean jet multiplicity in 
the Durham scheme, separately at each c.m.s. energy.
A qualitative comparison with the data showed that the
prediction overshoots the data in regions of medium $y_{cut}$ which may be attributed
to the fact that $\alpha_s$ is included only in one-loop accuracy. Another 
significant deviation is seen towards very low
$y_{cut}$, where the prediction begins to rise 
significantly faster than the data due to the
singularity in the $\alpha_s$ running. This deviation is 
largest at low c.m.s. energies. Better 
agreement between the data and the prediction in these regions can be achieved if
the prediction is shifted in $y_{cut}$ by an amount 
corresponding to typical hadron masses.\par
    In summary, this analysis presents a unique investigation 
  of the running of $\alpha_s$ in a large c.m.s. energy range
  of 35 through 189 GeV based on a consistent treatment of the
  data and employing up-to-date theoretical predictions. The
  numerical value of $\alpha_s$ obtained from our study of jet 
  rates and jet multiplicities is found to be in agreement
  with the world average, which has been obtained from a large 
  variety of observables and processes. It is of
  comparable precision.

\vspace*{2cm}
\par
\noindent Acknowledgements:
\par
\noindent We particularly wish to thank the SL Division for the efficient operation
of the LEP accelerator at all energies
 and for their continuing close cooperation with
our experimental group.  We thank our colleagues from CEA, DAPNIA/SPP,
CE-Saclay for their efforts over the years on the time-of-flight and trigger
systems which we continue to use.  In addition to the support staff at our own
institutions we are pleased to acknowledge the  \\
Department of Energy, USA, \\
National Science Foundation, USA, \\
Particle Physics and Astronomy Research Council, UK, \\
Natural Sciences and Engineering Research Council, Canada, \\
Israel Science Foundation, administered by the Israel
Academy of Science and Humanities, \\
Minerva Gesellschaft, \\
Benoziyo Center for High Energy Physics,\\
Japanese Ministry of Education, Science and Culture (the
Monbusho) and a grant under the Monbusho International
Science Research Program,\\
Japanese Society for the Promotion of Science (JSPS),\\
German Israeli Bi-national Science Foundation (GIF), \\
Bundesministerium f\"ur Bildung, Wissenschaft,
Forschung und Technologie, Germany, \\
National Research Council of Canada, \\
Research Corporation, USA,\\
Hungarian Foundation for Scientific Research, OTKA T-029328, 
T023793 and OTKA F-023259.\\

\begin{appendix}
\section{Quality Criteria for Charged Tracks and Electromagnetic Clusters}
\label{qualitylab}
The quality cuts defining acceptable charged tracks are listed for the JADE and
OPAL experiments in Table \ref{trackqualtablab}. 
For the JADE part of the analysis, two different types of 
tracks, ``long'' and ``central'' tracks, are defined. In the case of OPAL the
criteria used in the preselection 
differ from
those applied in the actual analysis. $n_{hits}$ denotes the number of hits in the
respective central jet chamber, and $n_{exp}$ the number of hits which is to be
expected taking into account track direction and detector geometry. 
For the OPAL preselection hits in the entire
central detector were counted. 
The symbol
$p$ denotes the reconstructed three-momentum of a track, $p_{\perp}$ its
projection onto the $xy$ plane and $\theta$ its polar angle. The beam energy
dependent upper limit
$p_{max}(E_{beam})$ is taken to be 
$(1+6\sqrt{0.02^2+(0.0015 E_{beam})^2})E_{beam}$.
The upper limits on $p$ are motivated by the occurrence of left-right ambiguities for tracks
running close to and parallel to a wire plane. In some cases, this situation leads
to the reconstruction of very straight tracks with extremely high momenta. The
expression for $p_{max}(E_{beam})$ is the beam energy, augmented by six times the track
momentum resolution.
$z_0$ is the $z$ coordinate of the point of closest
approach (p.c.a.) of the fitted track helix to the origin of the coordinate system, and
$d_0$ and $R_1$ are the transverse distances from, respectively, the p.c.a. and
the first hit to the origin.\\ 
\begin{table}[!t]

\begin{center}
\begin{tabular}{||c||c|c||c|c|c||} \hline\hline
 & \multicolumn{2}{c||}{JADE} &  \multicolumn{3}{c||}{OPAL}\\
Variable & ``long & ``central & 
preselection & analysis & analysis \\ 
&  tracks'' &  tracks'' & & (tracks only) & (tracks \& ECAL)\\ \hline \hline
$n_{hits}$ & $\ge25$ & -- & $\ge20$ & $\ge40$ & $\ge20$  \\ 
$n_{hits}/n_{exp}$ & -- & -- & -- & -- & $\ge0.5$  \\ 
$p$ [GeV] & -- & $>0.1$ & -- &  $<90$ & $<p_{max}(E_{beam})$ \\ 
$p_{\perp}$ [GeV] & $>0.5$ & -- & $>0.05$ & $>0.15$ & $>0.15$ \\ 
$|z_0|$ [cm] & -- & $<35$ & $<40$ & $<25$ & $<30$ \\ 
$d_0$ [cm] & -- & $<3$ & $<2$ & $<2$ & $<2.5$ \\ 
$|\cos{\theta}|$ & -- & -- & $<0.995$ & -- & $<0.9622$ \\
$R_1$ [cm] & -- & -- & $<60$ & -- & -- \\ \hline\hline
\end{tabular}
\end{center}
\caption{
Quality criteria for charged tracks at JADE and OPAL
\label{trackqualtablab}}
\end{table}The quality criteria for electromagnetic calorimeter clusters are
given in Table \ref{clustqualtablab}. 
In the case of OPAL, the
criteria used in the preselection 
differ again from
those applied in the actual analysis. $E_{clust}$ and $n_{bl}$ are, respectively,
cluster energy and number of lead-glass blocks contained in the cluster.
\begin{table}[!t]

\begin{center}
\begin{tabular}{||c|c||c||c|c||} \hline\hline
Detector & Variable & JADE &  \multicolumn{2}{c||}{OPAL}\\
region &  &  & preselection & analysis \\ \hline \hline
barrel & $E_{clust}$ [GeV] & $>0.15$ & $>0.1$ & $>0.1$ \\
       & $n_{bl}$ & -- & $\ge1$ & $\ge1$  \\ \hline
endcap & $E_{clust}$ [GeV] & $>0.15$ & $>0.2$ & $>0.25$ \\
       & $n_{bl}$ & -- & $\ge2$ &  $\ge2$ \\ \hline\hline
\end{tabular}
\end{center}

\caption[Quality criteria for electromagnetic calorimeter 
clusters at JADE and OPAL]{
Quality criteria for electromagnetic calorimeter clusters at JADE and OPAL
\label{clustqualtablab}

}

\end{table}
\section{Theoretical Calculations for Jet-Multiplicity Related Observables}
\label{theocalclab}
\subsection{Second-Order Approximations}
Predictions up to fixed second order in $\alpha_s$ are available
for the jet fractions and mean jet multiplicities of all jet
finders presented in Sect. \ref{jetmeaslab}. 
The perturbative power series
for the 2-jet fractions and the mean jet multiplicity can be written in terms of
$\bar{\alpha}_s\equiv\alpha_s/(2\pi)$ as 
\mathlab{R_2^{{\cal O}(\alpha_s^2)}(y_{cut})=1+{\cal A}_R(y_{cut})\bar{\alpha}_s
+{\cal B}_R(y_{cut})\bar{\alpha}_s^2}{oas2r2}
and
\mathlab{N^{{\cal O}(\alpha_s^2)}(y_{cut})=2+{\cal A}_N(y_{cut})\bar{\alpha}_s
+{\cal B}_N(y_{cut})\bar{\alpha}_s^2}{oas2mean}
where indices ``{\it R}'' and ``{\it N}'' on the coefficient functions indicate
the 2-jet fraction and the mean jet multiplicity, respectively.
The coefficient functions ${\cal A}$ and ${\cal B}$ may be calculated by
integration of the ${\cal O}(\alpha_s^2)$ matrix elements.
We employ second-order predictions for the 2-jet
fractions and mean jet multiplicities as obtained with the Durham
and Cambridge schemes for tests of perturbative QCD. For the Durham
scheme, we use values for the respective coefficients obtained 
for a wide range in $y_{cut}$ \cite{kluthpriv} from numerical matrix
element integration using the program EVENT2 \cite{event2}. In the case of
the Cambridge scheme, the calculations from \cite{calgo},
valid for $0.001\le y_{cut}\le 0.2$, are used.\par
${\cal O}(\alpha_s^2)$ predictions for the mean values of $y_n$ can be obtained
for $n=2$ and $n=3$, being, respectively, of next-to-leading and leading
order:
\mathlab{\langle y_2\rangle^{{\cal O}(\alpha_s^2)}=A\bar{\alpha}_s+B\bar{\alpha}_s^2}{oas2mom2}
\mathlab{\langle y_3\rangle^{{\cal O}(\alpha_s^2)}=C\bar{\alpha}_s^2}{oas2mom3}
Very recently, calculations of four-jet observables in 
next-to-leading
fixed order (i.e. ${\cal O}(\alpha_s^3)$) 
approximation have been
presented \cite{zoltan}. 
We do not use these for QCD tests, but we present
measurements
of the relevant quantities in Sect. \ref{jetmeaslab}.

\subsubsection{Next-To-Leading Logarithmic Approximations}
\label{nllatheolab}
The truncation of the perturbative series 
after a fixed order is meaningful as long
as the omitted terms can be assumed to be small. 
For the jet-related observables under consideration, 
as well as for cumulative cross
sections of event shape variables, the coefficients
can themselves be written as
series in 
$L\equiv\ln(1/y_{cut})$, which
becomes large in
the region of small $y_{cut}$. For some observables, reasonable predictions of
cross-sections
may nevertheless be obtained in these kinematic regions of low $y_{cut}$, 
because they allow the ``resummation'' of
the largest logarithmic components of the coefficients
to all orders in $\alpha_s$ \cite{nllatheo}. 
Both the differential 2-jet fraction and the mean
jet multiplicities as obtained with the Durham and Cambridge schemes
belong to this group of observables, which is why we chose to use them
to test QCD.\par
In the case of the differential 2-jet rates, the feasibility of
resummation is closely connected to the fact that the
complete prediction can be written in the ``exponentiated'' form
\mathlab{R_2(y_{cut})=C(\alpha_s)\exp G(\alpha_s,L)+D(\alpha_s,L)}{expo}
where
\mathlab{C(\alpha_s)=1+\sum_{n=1}^\infty C_n\bar{\alpha}_s^n}{cfunc}
and
\begin{eqnarray}
G(\alpha_s,L) & = & \sum_{n=1}^{\infty}\sum_{m=1}^{n+1}G_{nm}\bar{\alpha}_s^n L^m\nonumber\\
& \equiv & Lg_1(\alpha_s L)+g_2(\alpha_s L)+\alpha_s g_3(\alpha_s L)+\alpha_s^2 g_4(\alpha_s L)+...
\label{gfunc}
\end{eqnarray}
and the remainder function $D(\alpha_s,L)$ is assumed to vanish 
for $y_{cut}\rightarrow0$. Exponentiation implies that
terms with $m>n+1$ are absent in the sum (\ref{gfunc}). 
As can be seen from (\ref{gfunc}), each
of the functions $g_i$ is defined as an infinite power series in $\alpha_s L$.
The 
two functions, $g_1$ and $g_2$, contributing, respectively, the 
largest and second-largest logarithmic
parts of the coefficients, have been calculated for the 2-jet
fractions of the Durham and Cambridge schemes \cite{schmelling}, 
yielding a prediction in
``next-to-leading logarithmic'' approximation (NLLA). Up to this order,
the predictions are identical for both schemes \cite{calgo}. The coefficient
$C_1$ can be obtained from the ${\cal O}(\alpha_s)$ matrix element
\cite{schmelling}, and $C_2$ from integration of the 
${\cal O}(\alpha_s^2)$ matrix 
element and comparison of the result with (\ref{expo}), 
e.g. by fitting the latter
to the first \cite{schmelling}.\par
An improvement of the NLLA prediction suggested in
\cite{nllatheo} concerns the subleading logarithmic term $G_{21}\bar{\alpha}_s^2 L$,
which is included in the
second-order prediction
and may therefore be determined from a comparison of the NLLA with the
${\cal O}(\alpha_s^2)$ prediction
as was done for $C_2$. A result is again quoted in \cite{schmelling}.
The overall NLLA prediction is then\par
\mathlab{R_2^{NLLA}(y_{cut})=(1+C_1\bar{\alpha}_2+C_2\bar{\alpha}_s^2)
\exp\left[Lg_1(\alpha_s L)+g_2(\alpha_s L)\right].}{nllar2}
NLLA predictions have also been derived for the mean jet multiplicities
of the Durham scheme \cite{nllameantheo}. 
Here, however, the complete prediction does not
exponentiate as in (\ref{expo}). 
One therefore has to include powers of $L$ above $n+1$ and obtains
\begin{eqnarray}
N(y_{cut}) & = & 2+\sum_{n=1}^{\infty}\sum_{m=0}^{2n}H_{nm}\bar{\alpha}_s^n L^m\nonumber\\
 & \equiv & 2+h_1(L\alpha_s^2)+L^{-1}h_2(L\alpha_s^2)+L^{-2}h_3(L\alpha_s^2)+...,
\label{nllameantot}
\end{eqnarray}
where the ``leading'' and ``next-to-leading''
logarithmic terms are now those with $m=2n$ and $m=2n-1$, respectively. Each
function $h_i$ is an infinite series in $L\alpha_s^2$.
The complete NLLA predictions are given in \cite{nllameantheo}.\par
As in the case of the 2-jet fraction, one may additionally 
obtain the coefficients of subleading logarithmic terms of 
${\cal O}(\alpha_s)$ by comparing them with the fixed order prediction.
Here, the terms in question are $H_{22}L^2$ and $H_{21}L$. In fact,
from expansions of the calculations in \cite{nllameantheo}, an analytic 
expression for $H_{22}$ can be derived \cite{webberpriv}.
Determinations of both $H_{22}$ and $H_{21}$ have been performed 
using fits to ${\cal B}_N$ \cite{mymontp,kluthpriv}.
Including these two terms, the overall NLLA prediction takes the form
\mathlab{N^{NLLA}(y_{cut})=2+h_1(L\alpha_s^2)+L^{-1}h_2(L\alpha_s^2)+H_{22}L^2+H_{21}L.}{nllamean}
Again, the prediction turns out identical for the Cambridge scheme.

\subsubsection{Matched Predictions}
\label{matchtheolab}
Various techniques have
been devised to combine fixed-order with logarithmic predictions
in order to obtain
a prediction which is valid over a wide range of $y_{cut}$.
In the most obvious of such combination schemes, generally called the 
``$R$-matching'' scheme,
the exponential function in (\ref{nllar2}) is expanded in its power series and
the coefficients $c_1$ and $c_2$ of the $\alpha_s$ and $\alpha_s^2$ terms
are read off. 
The matched prediction $R_2(y_{cut})$ is then formed by subtracting the  
${\cal O}(\alpha_s)$ and  ${\cal O}(\alpha_s^2)$ terms from $R^{NLLA}_2$ and 
adding the second-order prediction,
yielding
\mathlab{R_2(y_{cut})=R^{NLLA}_2(y_{cut})-c_1\bar{\alpha}_s-c_2\bar{\alpha}_s^2+R^{{\cal O}(\alpha_s^2)}_2(y_{cut}).}{bla}
The procedure requires the explicit knowledge of 
$G_{11}$, $G_{12}$, $G_{22}$ and $G_{23}$ whose values can be
obtained by expanding function $G(\alpha_s,L)$ (with the known functions
$g_1$ and $g_2$) in a power series and reading off the appropriate
coefficients. 
We shall follow the convention of \cite{opal91} and use the expression
``$R$-matching'' if the $G_{21}$ term
is left out in $R_2^{NLLA}$ (and consequently also in $c_2$). If the
term is kept, we shall speak of ``modified $R$-matching''. Because of the
additionally included subleading term, one may expect
the modified variant to be the more precise prediction.\par
In the corresponding matching procedure for the mean jet multiplicity 
the ${\cal O}(\alpha_s)$ and  ${\cal O}(\alpha_s^2)$ coefficients which
appear in both types of predictions can be
read off directly from (\ref{nllameantot})
resulting in the matched prediction
\mathlab{N(y_{cut})=N^{NLLA}(y_{cut})-c'_1\bar{\alpha}_s-c'_2\bar{\alpha}_s^2+{\cal A}_N(y_{cut})\bar{\alpha}_s
+{\cal B}_N(y_{cut})\bar{\alpha}_s^2.}{blaa}
It does not make any difference in this case whether the
subleading terms $H_{22}L^2$ and $H_{21}L$ are included or not, 
because they are subtracted out in any case.
The required coefficients $H_{11}$, $H_{12}$, $H_{23}$ and $H_{24}$ have
been calculated analytically and are given in equation (8) of
\cite{nllameantheo}.\par
An alternative way of matching the two types of calculations, called
``$\ln R$-matching'' in \cite{nllatheo}, applies a similar procedure as
before, but now to the {\it logarithms} of the predictions.
To this end, the logarithms of the two predictions are 
expanded in their power series 
and terms of ${\cal O}(\alpha_s^3)$ 
as well as non-logarithmic terms 
are omitted.
The ${\cal O}(\alpha_s)$ and ${\cal O}(\alpha_s^2)$ terms are then
subtracted from the resulting NLLA expression,
and the two logarithms are added to
yield the matched prediction
\begin{eqnarray}
\ln R_2(y_{cut}) & = & Lg_1(\alpha_s L)+g_2(\alpha_s L)\nonumber\\
   & & -(G_{11}L+G_{12}L^2)\bar{\alpha}_s-(G_{22}L^2+G_{23}L^3)\bar{\alpha}_s^2\label{blaaa}\\
   & & +{\cal A}_R(y_{cut})\bar{\alpha}_s+\left[{\cal B}_R(y_{cut})-{\scriptsize\frac{1}{2}}{\cal A}_R(y_{cut})^2\right]\bar{\alpha}_s^2.\nonumber
\end{eqnarray}
The subleading term $G_{21}\bar{\alpha}_s^2 L$ in
the NLLA part of the prediction can be ignored since it is removed in the
process, being implicitly contained in ${\cal B}_R$.\par
It has been pointed out in \cite{nllatheo} that neither the 
$R$-matching nor the $\ln R$-matching scheme ensures that the resulting prediction
for $R_2$ tends to 1 in the kinematic limit of $y_{cut}=y_{max}$ where all events are resolved
in two jets. In the case
of the $\ln R$-matching, this drawback can simply be cured by
replacing $L$ with $L'\equiv\ln(1/y_{cut}-1/y_{max}+1)$. As in \cite{opal91},
we refer to this variant as ``modified $\ln R$-matching''.\par
The mean jet multiplicity may be subjected to a procedure analogous
to the $\ln R$-matching where 
the logarithm is taken of $N-1$ rather
than $N$ itself. 
The expansion of $\ln(N^{NLLA}-1)$
makes it possible 
to read off the common ${\cal O}(\alpha_s)$ and ${\cal O}(\alpha_s^2)$
coefficients $d_1$ and $d_2$. 
Combining
both predictions and subtracting double terms leads then to the matched
result
\begin{eqnarray}
\ln[N(y_{cut})-1] & = & \ln[N^{NLLA}(y_{cut})-1]\nonumber\\
& & -d_1\bar{\alpha}_s-d_2\bar{\alpha}_s^2
+{\cal A}_N(y_{cut})\bar{\alpha}_s+\left[{\cal B}_N(y_{cut})-\frac{1}{2}{\cal A}_N(y_{cut})^2\right]\bar{\alpha}_s^2
\label{blaaaa}
\end{eqnarray}
Here, it does make a difference whether the subleading terms $H_{22}L^2$ and 
$H_{21}L$ are left out. We shall speak of ``$\ln R$-matching'' if only
the analytic expression for $H_{22}$ 
is included, and of a ``modified $\ln R$-matching'' if the fitted values for both
$H_{22}$ and $H_{21}$ are used.\par
$R$ and $\ln R$-matching differ generally in their assumptions
on the unknown subleading and non-logarithmic terms of ${\cal O}(\alpha_s^3)$.
In the case of the $R$-matching, these terms are ignored in the predictions
for $R_2$ and $N$, while, in the case of the $\ln R$-matching,
they are assumed to vanish in the logarithms.\par

\newpage
\section{Fit Results for $\protect\alpha_s$}
\label{astableslab}
The tables on the subsequent pages summarize 
the values obtained for $\alpha_s$ in Sects. \ref{fixasfitslab}
and \ref{enevollab} and the detailed composition of the systematic
errors. For each systematic variation, the deviation with
respect to the central result is given with a sign indicating
the direction of the deviation.
\begin{table}[!b]

\begin{center}

{
\footnotesize

\begin{tabular}{|r||r|r|r|r|}\hline
& \multicolumn{1}{c|}{$D_2^D$} & \multicolumn{1}{c|}{$D_2^C$} & \multicolumn{1}{c|}{$N^D$} & \multicolumn{1}{c|}{$N^C$} \\ \hline\hline
$\alpha_s$(35 GeV)&\bf{0.1482}&\bf{0.1438}&\bf{0.1510}&\bf{0.1500}\\ \hline \hline
Fit range [$\log_{10}(y_{cut})$]&-1.55 \ \ ---\ \ -0.65&-1.55 \ \ ---\ \ -0.65&-1.4 \ \ ---\ \ -0.5&-1.55 \ \ ---\ \ -0.65\\ \hline \hline
$\chi^2$/d.o.f.&5.25/5&7.66/5&0.27/6&0.66/6\\ \hline \hline
Statistical error& $\pm$0.0011&$\pm$0.0013&$\pm$0.0039&$\pm$0.0029\\ \hline \hline
Experimental syst.&$\pm$0.0031&$\pm$0.0036&$\pm$0.0057&$\pm$0.0043\\ \hline \hline
Total hadronization&$\pm$0.0073&$\pm$0.0047&$\pm$0.0046&$\pm$0.0063\\ \hline \hline
$x_{\mu}=0.5$&-0.0003&+0.0007&-0.0038&-0.0019\\ \hline
$x_{\mu}=2$&+0.0055&+0.0044&+0.0068&+0.0054\\ \hline \hline
&{\bf+0.0097}&{\bf+0.0066}&{\bf+0.0107}&{\bf+0.0098}\\
\rb{Total error}&{\bf-0.008}&{\bf-0.0066}&{\bf-0.0091}&{\bf-0.0084}\\

\hline
\end{tabular}

\vspace*{0.5cm}

\begin{tabular}{|r||r|r|r|r|}\hline
& \multicolumn{1}{c|}{$D_2^D$} & \multicolumn{1}{c|}{$D_2^C$} & \multicolumn{1}{c|}{$N^D$} & \multicolumn{1}{c|}{$N^C$} \\ \hline\hline
$\alpha_s$(44 GeV)&\bf{0.1294}&\bf{0.1257}&\bf{0.1353}&\bf{0.1313}\\ \hline \hline
Fit range [$\log_{10}(y_{cut})$]&-1.55 \ \ ---\ \ -0.65&-1.85 \ \ ---\ \ -0.95&-1.7 \ \ ---\ \ -0.5&-1.85 \ \ ---\ \ -0.65\\ \hline \hline
$\chi^2$/d.o.f.&4.21/5&12.36/5&3.56/8&4.58/8\\ \hline \hline
Statistical error& $\pm$0.0013&$\pm$0.0016&$\pm$0.0028&$\pm$0.0017\\ \hline \hline
Experimental syst.&$\pm$0.0066&$\pm$0.0053&$\pm$0.0074&$\pm$0.0062\\ \hline \hline
Total hadronization&$\pm$0.0028&$\pm$0.0067&$\pm$0.0036&$\pm$0.0056\\ \hline \hline
$x_{\mu}=0.5$&-0.0001&+0.0004&-0.0024&-0.0004\\ \hline
$x_{\mu}=2$&+0.0036&+0.0030&+0.0047&+0.0031\\ \hline \hline
&{\bf+0.0081}&{\bf+0.0088}&{\bf+0.0099}&{\bf+0.0091}\\
\rb{Total error}&{\bf-0.0073}&{\bf-0.0088}&{\bf-0.0090}&{\bf-0.0085}\\

\hline
\end{tabular}

\vspace*{0.5cm}

\begin{tabular}{|r||r|r|r|r|}\hline
& \multicolumn{1}{c|}{$D_2^D$} & \multicolumn{1}{c|}{$D_2^C$} & \multicolumn{1}{c|}{$N^D$} & \multicolumn{1}{c|}{$N^C$} \\ \hline\hline
$\alpha_s$(91.14 GeV)&\bf{0.1214}&\bf{0.1131}&\bf{0.1218}&\bf{0.1163}\\ \hline \hline
Fit range [$\log_{10}(y_{cut})$]&-2.125 \ \ ---\ \ -1.375&-1.5 \ \ ---\ \ -0.625&-2.25 \ \ ---\ \ -0.625&-1.625 \ \ ---\ \ -0.5\\ \hline \hline
$\chi^2$/d.o.f.&8.52/5&29.44/6&11.69/13&1.59/9\\ \hline \hline
Statistical error& $\pm$0.0002&$\pm$0.0002&$\pm$0.0003&$\pm$0.0008\\ \hline \hline
Experimental syst.&$\pm$0.0038&$\pm$0.0051&$\pm$0.0037&$\pm$0.0056\\ \hline \hline
Total hadronization&$\pm$0.0032&$\pm$0.0021&$\pm$0.0027&$\pm$0.0022\\ \hline \hline
$x_{\mu}=0.5$&-0.0006&+0.0007&-0.0013&-0.0009\\ \hline
$x_{\mu}=2$&+0.0035&+0.0019&+0.0032&+0.0026\\ \hline \hline
&{\bf+0.0061}&{\bf+0.0056}&{\bf+0.0056}&{\bf+0.0066}\\
\rb{Total error}&{\bf-0.0051}&{\bf-0.0056}&{\bf-0.0047}&{\bf-0.0061}\\

\hline
\end{tabular}

}

\vspace*{0.5cm}

\caption[$\alpha_s$ results from fits using $\ln{R}$-matching at $\protect\sqrt{s}$=44, 91 and 133 GeV]{
\label{astablab_35}
\label{astablab_44}
\label{astablab_91}
Results for $\alpha_s$ from fits of the $\ln{R}$-matching predictions 
for the four observables $D_2^D$, $D_2^C$, $N^D$ and $N^C$ at
$\sqrt{s}=35$, 44 and 91 GeV
}

\end{center}

\end{table}

\clearpage\newpage
\begin{table}[!h]

\begin{center}

{
\footnotesize

\begin{tabular}{|r||r|r|r|r|}\hline
& \multicolumn{1}{c|}{$D_2^D$} & \multicolumn{1}{c|}{$D_2^C$} & \multicolumn{1}{c|}{$N^D$} & \multicolumn{1}{c|}{$N^C$} \\ \hline\hline
$\alpha_s$(133 GeV)&\bf{0.1066}&\bf{0.1047}&\bf{0.1153}&\bf{0.1126}\\ \hline \hline
Fit range [$\log_{10}(y_{cut})$]&-2.5 \ \ ---\ \ -0.5&-2.25 \ \ ---\ \ -0.5&-2.5 \ \ ---\ \ -0.5&-2.25 \ \ ---\ \ -0.5\\ \hline \hline
$\chi^2$/d.o.f.&5.69/7&6.1/6&5.63/8&5.63/7\\ \hline \hline
Statistical error& $\pm$0.0021&$\pm$0.0035&$\pm$0.0028&$\pm$0.0042\\ \hline \hline
Experimental syst.&$\pm$0.0076&$\pm$0.0057&$\pm$0.0078&$\pm$0.0062\\ \hline \hline
Total hadronization&$\pm$0.0024&$\pm$0.0011&$\pm$0.0021&$\pm$0.0017\\ \hline \hline
$x_{\mu}=0.5$&$<$0.0001&+0.0006&-0.0010&-0.0003\\ \hline
$x_{\mu}=2$&+0.0021&+0.0015&+0.0027&+0.0020\\ \hline \hline
&{\bf+0.0085}&{\bf+0.0068}&{\bf+0.0090}&{\bf+0.0079}\\
\rb{Total error}&{\bf-0.0082}&{\bf-0.0068}&{\bf-0.0086}&{\bf-0.0077}\\

\hline
\end{tabular}

\vspace*{1cm}

\begin{tabular}{|r||r|r|r|r|}\hline
& \multicolumn{1}{c|}{$D_2^D$} & \multicolumn{1}{c|}{$D_2^C$} & \multicolumn{1}{c|}{$N^D$} & \multicolumn{1}{c|}{$N^C$} \\ \hline\hline
$\alpha_s$(161 GeV)&\bf{0.1097}&\bf{0.0991}&\bf{0.1106}&\bf{0.1090}\\ \hline \hline
Fit range [$\log_{10}(y_{cut})$]&-2.75 \ \ ---\ \ -0.75&-2.5 \ \ ---\ \ -0.75&-2.75 \ \ ---\ \ -0.5&-2.75 \ \ ---\ \ -0.5\\ \hline \hline
$\chi^2$/d.o.f.&8.63/7&13.42/6&5.96/9&4.9/9\\ \hline \hline
Statistical error& $\pm$0.0038&$\pm$0.0039&$\pm$0.0028&$\pm$0.0021\\ \hline \hline
Experimental syst.&$\pm$0.0067&$\pm$0.0047&$\pm$0.0056&$\pm$0.0079\\ \hline \hline
Total hadronization&$\pm$0.0035&$\pm$0.0022&$\pm$0.0023&$\pm$0.0034\\ \hline \hline
$x_{\mu}=0.5$&-0.0006&+0.0001&-0.0009&$<$0.0001\\ \hline
$x_{\mu}=2$&+0.0027&+0.0016&+0.0023&+0.0016\\ \hline \hline
&{\bf+0.0089}&{\bf+0.0066}&{\bf+0.0070}&{\bf+0.0089}\\
\rb{Total error}&{\bf-0.0085}&{\bf-0.0066}&{\bf-0.0067}&{\bf-0.0089}\\

\hline
\end{tabular}

\vspace*{1cm}

\begin{tabular}{|r||r|r|r|r|}\hline
& \multicolumn{1}{c|}{$D_2^D$} & \multicolumn{1}{c|}{$D_2^C$} & \multicolumn{1}{c|}{$N^D$} & \multicolumn{1}{c|}{$N^C$} \\ \hline\hline
$\alpha_s$(172 GeV)&\bf{0.1046}&\bf{0.0990}&\bf{0.1075}&\bf{0.1028}\\ \hline \hline
Fit range [$\log_{10}(y_{cut})$]&-2.75 \ \ ---\ \ -0.75&-2.5 \ \ ---\ \ -0.75&-2.75 \ \ ---\ \ -1&-2.75 \ \ ---\ \ -1.25\\ \hline \hline
$\chi^2$/d.o.f.&6.28/7&2.53/6&4.48/7&0.61/6\\ \hline \hline
Statistical error& $\pm$0.0046&$\pm$0.0052&$\pm$0.0037&$\pm$0.0044\\ \hline \hline
Experimental syst.&$\pm$0.0115&$\pm$0.0107&$\pm$0.0053&$\pm$0.0062\\ \hline \hline
Total hadronization&$\pm$0.0037&$\pm$0.0016&$\pm$0.0019&$\pm$0.0026\\ \hline \hline
$x_{\mu}=0.5$&-0.0004&+0.0002&-0.0008&-0.0001\\ \hline
$x_{\mu}=2$&+0.0023&+0.0015&+0.0021&+0.0015\\ \hline \hline
&{\bf+0.0131}&{\bf+0.0120}&{\bf+0.0070}&{\bf+0.0082}\\
\rb{Total error}&{\bf-0.0129}&{\bf-0.0120}&{\bf-0.0068}&{\bf-0.0081}\\

\hline
\end{tabular}

}

\vspace*{1cm}

\caption[$\alpha_s$ results from fits using $\ln{R}$-matching at $\protect\sqrt{s}$=44, 91 and 133 GeV]{
\label{astablab_133}
\label{astablab_161}
\label{astablab_172}
Results for $\alpha_s$ from fits of the $\ln{R}$-matching predictions 
for the four observables $D_2^D$, $D_2^C$, $N^D$ and $N^C$ at
$\sqrt{s}=133$, 161 and 172 GeV
}

\end{center}

\end{table}

\clearpage\newpage
\begin{table}[!h]

\begin{center}

{
\footnotesize

\begin{tabular}{|r||r|r|r|r|}\hline
& \multicolumn{1}{c|}{$D_2^D$} & \multicolumn{1}{c|}{$D_2^C$} & \multicolumn{1}{c|}{$N^D$} & \multicolumn{1}{c|}{$N^C$} \\ \hline\hline
$\alpha_s$(183 GeV)&\bf{0.1076}&\bf{0.1019}&\bf{0.1100}&\bf{0.1048}\\ \hline \hline
Fit range [$\log_{10}(y_{cut})$]&-2.75 \ \ ---\ \ -0.5&-2.5 \ \ ---\ \ -0.5&-2.75 \ \ ---\ \ -0.75&-2.75 \ \ ---\ \ -0.75\\ \hline \hline
$\chi^2$/d.o.f.&6.36/8&9.8/7&6.85/8&5.07/8\\ \hline \hline
Statistical error& $\pm$0.0030&$\pm$0.0032&$\pm$0.0021&$\pm$0.0022\\ \hline \hline
Experimental syst.&$\pm$0.0020&$\pm$0.0052&$\pm$0.0029&$\pm$0.0046\\ \hline \hline
Total hadronization&$\pm$0.0025&$\pm$0.0014&$\pm$0.002&$\pm$0.0026\\ \hline \hline
$x_{\mu}=0.5$&-0.0005&+0.0002&-0.0008&-0.0001\\ \hline
$x_{\mu}=2$&+0.0025&+0.0017&+0.0023&+0.0015\\ \hline \hline
&{\bf+0.0050}&{\bf+0.0063}&{\bf+0.0047}&{\bf+0.0059}\\
\rb{Total error}&{\bf-0.0044}&{\bf-0.0063}&{\bf-0.0042}&{\bf-0.0057}\\

\hline
\end{tabular}

\vspace*{1cm}

\begin{tabular}{|r||r|r|r|r|}\hline
& \multicolumn{1}{c|}{$D_2^D$} & \multicolumn{1}{c|}{$D_2^C$} & \multicolumn{1}{c|}{$N^D$} & \multicolumn{1}{c|}{$N^C$} \\ \hline\hline
$\alpha_s$(189 GeV)&\bf{0.1109}&\bf{0.1062}&\bf{0.1106}&\bf{0.1079}\\ \hline \hline
Fit range [$\log_{10}(y_{cut})$]&-2.75 \ \ ---\ \ -0.5&-2.75 \ \ ---\ \ -0.5&-2.75 \ \ ---\ \ -0.75&-2.75 \ \ ---\ \ -0.75\\ \hline \hline
$\chi^2$/d.o.f.&11.87/8&5.12/8&5.36/8&6.87/8\\ \hline \hline
Statistical error& $\pm$0.0016&$\pm$0.0015&$\pm$0.0013&$\pm$0.0015\\ \hline \hline
Experimental syst.&$\pm$0.0029&$\pm$0.0021&$\pm$0.0043&$\pm$0.0038\\ \hline \hline
Total hadronization&$\pm$0.0022&$\pm$0.0024&$\pm$0.0017&$\pm$0.0019\\ \hline \hline
$x_{\mu}=0.5$&-0.0005&-0.0001&-0.0008&-0.0001\\ \hline
$x_{\mu}=2$&+0.0028&+0.0022&+0.0023&+0.0017\\ \hline \hline
&{\bf+0.0049}&{\bf+0.0041}&{\bf+0.0053}&{\bf+0.0049}\\
\rb{Total error}&{\bf-0.0041}&{\bf-0.0035}&{\bf-0.0049}&{\bf-0.0045}\\

\hline
\end{tabular}

}

\vspace*{1cm}

\caption[$\alpha_s$ results from fits using $\ln{R}$-matching at $\protect\sqrt{s}$=161, 172 and 183 GeV]{
\label{astablab_183}
\label{astablab_189}
Results for $\alpha_s$ from fits of the $\ln{R}$-matching predictions 
for the four observables $D_2^D$, $D_2^C$, $N^D$ and $N^C$ at
$\sqrt{s}=183$ and 189 GeV
}

\end{center}

\end{table}

\begin{table}[!h]

\centering{{\footnotesize

\begin{tabular}{|r||c|c|} \hline
               & $\langle y_2\rangle^D$ & $\langle y_2\rangle^J$  \\ \hline \hline
$\alpha_s(M_{Z^0})$ &  {\bf 0.1217} &  {\bf 0.1162}  \\
$\chi^2/d.o.f.$  & 2.77/7     & 1.66/7    \\  \hline\hline
Stat. error & $\pm0.0009$    & $\pm0.0006$ \\ \hline \hline
Total exp. & $\pm0.0025$ &  $\pm0.0021$  \\ \hline\hline
Total hadronization &  $\pm0.0024$&$\pm0.0037$\\
\hline \hline
 $x_{\mu}=0.5$ &   $-0.0042$ &  $-0.0031$ \\ \hline
  $x_{\mu}=2$ &   $+0.0056$ &  $+0.0043$ \\ \hline
\hline
                 & {\bf +0.0066}  & {\bf +0.0060} \\ 
\rb{Total error} & {\bf -0.0054}  &  \bf{-0.0052}  \\ \hline
\end{tabular}

}}

\caption[$\alpha_s(M_{Z^0})$ results from c.m.s. energy evolution of $\langle y_n\rangle$ and $R_3$]{
Results for $\alpha_s(M_{Z^0})$ from fits of the ${\cal O}(\alpha_s^2)$ predictions 
for the energy evolution of $\langle y_2\rangle^D$ and $\langle y_2\rangle^J$
\label{momrestablab}

}

\end{table}
\begin{table}[!h]

\centering{{\footnotesize

\begin{tabular}{|r||c|c|c|} \hline
               & $R_3^J(y_{cut}=0.08)$ &$R_3^D(y_{cut}=0.01)$  & $R_3^C(y_{cut}=0.01)$ \\ \hline \hline
$\alpha_s(M_{Z^0})$ &  {\bf 0.1243}    &  {\bf 0.1127}    &  \bf{0.1137} \\
\rb{$x_{\mu}=1$} \hspace*{\fill} $\chi^2/d.o.f.$  &  3.63/7    &    5.18/7     &   31.16/7 \\ \hline 
$\alpha_s(M_{Z^0})$ &  0.1086    &  0.1097    &  0.1123 \\
fitted  \hspace*{\fill}  $x_{\mu}$ &   0.063   &  0.310    &  0.464 \\
 $x_{\mu}$ \hspace*{\fill} stat. error &   $\pm0.025$   &  $\pm0.148$    &  $\pm0.172$ \\
$\chi^2/d.o.f.$  &  2.98/6     &   5.03/6     &  30.70/6 \\ \hline \hline
Stat. error & $\pm0.0003$ &  $\pm0.0002$ &  $\pm0.0003$ \\ \hline \hline
Total exp. &   $\pm0.0024$  &  $\pm0.0011$   &  $\pm0.0014$ \\ \hline \hline 
Total hadronization & $\pm0.0032$ & $\pm0.0082$ & $\pm0.0072$ \\ \hline \hline
 $x_{\mu}=0.5$ & $-0.0061$  &  $-0.0024$  &  $-0.0014$ \\ \hline
  $x_{\mu}=2$ &  $+0.0074$ &  $+0.0037$ & $+0.0030$ \\ \hline \hline
                 &{\bf +0.0085}  &  \bf{+0.0091} &  \bf{+0.0079} \\
\rb{Total error} & \bf{-0.0073}  &  \bf{-0.0086} &  \bf{-0.0075} \\ \hline
\end{tabular}

}}

\caption[$\alpha_s(M_{Z^0})$ results from c.m.s. energy evolution of $\langle y_n\rangle$ and $R_3$]{
Results for $\alpha_s(M_{Z^0})$ from fits of the ${\cal O}(\alpha_s^2)$ predictions 
for the energy evolution of
$R_3^J$, $R_3^D$ and $R_3^C$
\label{r3restablab}
}

\end{table}
\end{appendix}

\end{document}